\documentstyle[eqsecnum,prd,aps,epsf]{revtex}

\draft
\begin{document}

\def\agt{\mathrel{\raise.3ex\hbox{$>$}\mkern-14mu\lower0.6ex\hbox{$\sim$}}}
\def\alt{\mathrel{\raise.3ex\hbox{$<$}\mkern-14mu\lower0.6ex\hbox{$\sim$}}}

\newcommand{\beq}{\begin{equation}}
\newcommand{\eeq}{\end{equation}}
\newcommand{\beqn}{\begin{eqnarray}}
\newcommand{\eeqn}{\end{eqnarray}}
\newcommand{\pa}{\partial}
\newcommand{\vp}{\varphi}
\newcommand{\varep}{\varepsilon}
\newcommand{\ep}{\epsilon}
\newcommand{\comp}{(M/R)_\infty} 

%@twocolumnfalse\endcsname

\begin{center}
{\large\bf{Merger of binary neutron stars of unequal mass 
in full general relativity}}
~~\\
~~\\
Masaru Shibata$^1$, Keisuke Taniguchi$^1$, and K\=oji Ury\=u$^2$
~~\\
~~\\
{\em $^1$ Graduate School of Arts and Sciences, 
University of Tokyo, Komaba, Meguro, Tokyo 153-8902, Japan \\
$^2$ Astrophysical Sector, SISSA, via Beirut 2/4, Trieste 34013, Italy}
\end{center}

\begin{abstract}
~\\
We present results of three dimensional numerical simulations of
the merger of 
unequal-mass binary neutron stars in full general relativity.  
A $\Gamma$-law equation of state 
$P=(\Gamma-1)\rho\varepsilon$ is adopted, where 
$P$, $\rho$, $\varep$, and $\Gamma$ 
are the pressure, rest mass density, specific internal energy, 
and the adiabatic constant, respectively.  We take $\Gamma=2$ 
and the baryon rest-mass ratio $Q_M$ to be in the range $0.85$--1. 
The typical grid size is $(633,633,317)$ for $(x,y,z)$ . 
We improve several implementations since the latest work. 
In the present code, the radiation reaction of gravitational waves 
is taken into account with a good accuracy. This fact enables us to 
follow the coalescence all the way from the late inspiral phase
through the merger phase for which the transition is 
triggered by the radiation reaction. It is found that 
if the total rest-mass of the system is more than $\sim 1.7$ times 
of the maximum allowed rest-mass of spherical neutron
stars, a black hole is formed after the merger irrespective
of the mass ratios. The gravitational waveforms and outcomes 
in the merger of unequal-mass binaries are compared with those in 
equal-mass binaries. It is found that the disk mass around the so 
formed black holes increases with decreasing rest-mass ratios 
and decreases with increasing compactness of neutron stars.
The merger process and the gravitational waveforms also depend
strongly on the rest-mass ratios even for the range $Q_M= 0.85$--1. 
\end{abstract}

\section{Introduction}

Binary neutron stars such as the Hulse-Taylor binary pulsar \cite{HT} 
adiabatically inspiral as a result of the radiation reaction of
gravitational waves, and eventually merge. 
In the most optimistic scenario, 
the latest statistical study suggests that such mergers may occur 
approximately once per year within a 
distance of about 30 Mpc \cite{BNST}. 
Even the most conservative scenario predicts an 
event rate approximately 
one per year within a distance of about 400 Mpc \cite{BNST}. 
This implies that the merger of binary neutron stars is 
one of the promising sources for kilometer-size laser interferometric 
detectors, such as LIGO, TAMA, GEO600, and VIRGO \cite{KIP,Ando}.  

Interest has also been stimulated by a hypothesis about 
the central engine of $\gamma$-ray bursts (GRBs) \cite{piran}.
Recently, it has been found that many GRBs are of cosmological 
origin \cite{piran}. In cosmological GRBs, the central sources must
supply a large amount of the energy $\agt 10^{51}$ ergs in a 
very short time scale (order of milliseconds to minutes). 
Most GRB models involve a stellar system resulting in a
stellar-mass rotating black hole and a massive disk of mass
$\sim 0.1$--1$M_{\odot}$, which could supply a large amount of
energy by neutrino processes or by extracting the rotational energy
of the black hole. GRBs may be classified into two classes. 
One is a long burst for which the duration of the bursts is longer than
$\sim 1$ sec and typically $\sim 10$ sec,
and the other is a short burst for which the duration is
typically $\sim 100$ msec. It has been recently suggested that 
the merger of binary neutron stars is a possible progenitor 
to producing short bursts. 

Hydrodynamic simulations employing full general relativity 
provide the best approach for studying the merger of binary neutron 
stars. Over the last few years, numerical methods for solving coupled 
equations of the Einstein and hydrodynamic equations have been 
developed \cite{gr3d,bina,bina2,other,Font} 
and now such simulations are feasible. 
In previous papers \cite{bina,bina2},
we focused on the binary neutron stars of equal mass, and 
have found the following results: 
(i) the final outcome (of either a neutron star or a black hole)
depends on the compactness of each neutron star and 
on the equation of state. Even if the total mass of the system is 
$\sim 1.5$ times larger than the maximum allowed rest-mass of
a spherical star for a given equation of state, a differentially
rotating neutron star supported by a significant centrifugal
force may be formed; 
(ii) in the case of neutron star formation, nonspherical 
oscillation modes of the formed neutron star 
are excited and, as a result, gravitational waves with 
characteristic frequency $\sim 2$--3 kHz are emitted; 
(iii) in the case of black hole formation, 
the disk mass around the formed black hole is negligible 
because the specific angular momentum of all the mass elements
in equal-mass binary neutron stars is too small and also because 
the angular momentum transfer is not effective during the merger. 

So far, all the simulations in general relativity have been 
performed assuming that two neutron stars are identical 
\cite{bina,bina2,marks,illinois}, 
since they are indeed approximately identical in the observed systems 
of binary neutron stars \cite{PK}. For example, 
mass ratio of the Hulse-Taylor binary is about 0.963 \cite{MT}.
However, it seems 
there is no theoretical reason that nature should produce 
only binary neutron stars of nearly equal mass. 
Allowed mass range of neutron stars may fall in a fairly broad range 
$\sim 1$--$2M_{\odot}$ according to theories of neutron stars \cite{ST,GLEN}.
From the theoretical point of view, it is reasonable and an interesting
subject to investigate the merger of two unequal-mass neutron stars. 

One of the most important findings in the previous works 
\cite{bina,bina2} is that black hole formation is not accompanied by 
disks with a large mass. The mass of the disk is found to be
less than $0.01M_{\odot}$. 
This result suggests that binary neutron stars 
of equal mass may not be good progenitors for the central engine of 
GRBs. On the other hand, the disk mass may be much larger 
in the merger of binary neutron stars of unequal mass,
because the smaller-mass neutron star would be tidally disrupted by
the more massive primary before contact and would subsequently 
form a tidal tail around the primary in which angular
momentum transfer is likely to be efficient to form disks 
around the central object. In addition, 
in association with the change of the merger process, 
gravitational waveforms may be significantly modified. 
Actually, Newtonian and post-Newtonian simulations indicate such 
significant changes \cite{RS,C,FR,FR2}.

From a computational point of view, 
we have substantially improved our implementation
for a solution of Einstein and hydrodynamic equations 
from our previous approach \cite{bina,bina2}. Primarily, 
a modified numerical scheme for solving hydrodynamic equations 
by adopting the so-called high-resolution
shock-capturing scheme \cite{Font,shiba2d} provides better accuracy.
The spatial gauge condition is changed 
from a minimal distortion type \cite{SY,gw3p2} to
a dynamical one in which a hyperbolic type 
equation is adopted for determining the shift vector \cite{ALC,LS}.
This has resulted in saving a substantial amount of computation time.  
%As a result, the computational time is significantly saved.
Finally, we have modified the treatment for the transport terms 
in the evolution equations of geometric variables. 
This improves the accuracies for a solution of the geometric
quantities and for conservation of the total ADM mass and 
angular momentum significantly. 

The paper is organized as follows. In Sec. II, we review 
basic equations, gauge conditions, and methods for 
setting initial conditions currently adopted 
in fully general relativistic simulations of binary neutron star mergers. 
In Sec. III, methods used for analysis of gravitational waves are summarized. 
In Sec. IV, the numerical results are presented, paying particular attention 
to merger process, disk mass, and gravitational waveforms.
Section V is devoted to a summary. 
Throughout this paper, we adopt the geometrical 
units in which $G=c=1$ where $G$ and $c$ 
are the gravitational constant and the speed of light.  
Latin and Greek indices denote spatial components ($x, y, z$) 
and space-time components ($t, x, y, z$), respectively. 
$\delta_{ij}(=\delta^{ij})$ denotes the Kronecker delta. 

\section{Formulation}

\subsection{Basic equations}

In our numerical simulation, 
the Einstein and general relativistic hydrodynamic equations are
solved without any approximation. 
The formulation for a numerical solution of these coupled equations 
is based on those described in a previous work \cite{bina2}.
However, we have improved several numerical implementations 
since then, and the form of the basic equations adopted in numerical 
simulation has also been modified.
A summary of the current formulation is, therefore, in order here.

The line element is written in the form 
\beqn
ds^2=g_{\mu\nu}dx^{\mu}dx^{\nu} =(-\alpha^2+\beta_k\beta^k)dt^2
+2\beta_i dx^i dt+\gamma_{ij}dx^i dx^j ,
\eeqn
where $g_{\mu\nu}$, $\alpha$, $\beta^i~(\beta_i=\gamma_{ij}\beta^j)$, 
and $\gamma_{ij}$ are the four dimensional spacetime metric, 
the lapse function, the shift vector, and the
three dimensional spatial metric, respectively. 
Following \cite{SN,gw3p2,gr3d}, we define the quantities as 
\beqn
&& \gamma={\rm det}(\gamma_{ij}) \equiv e^{12\phi},\\
&& \tilde \gamma_{ij} \equiv e^{-4\phi}\gamma_{ij}, \\
&& \tilde A_{ij} \equiv 
e^{-4\phi} \Bigl(K_{ij}-{1 \over 3} \gamma_{ij} K \Bigr),
\eeqn
where $K_{ij}$ is the extrinsic curvature, and $K$ its trace. 
In the Cartesian coordinates adopted in our simulation,
${\rm det}(\tilde \gamma_{ij})$ should be unity. 
In the numerical computations, 
$\phi$, $\tilde \gamma_{ij}$, $K$, and $\tilde A_{ij}$ are evolved in time, 
instead of $\gamma_{ij}$ and $K_{ij}$. 
Note that the indices of $\tilde A_{ij}$ ($\tilde A^{ij}$) are 
raised (lowered) in terms of $\tilde \gamma^{ij}$ ($\tilde \gamma_{ij}$). 
Hereafter, $D_i$ and $\tilde D_i$ are used 
as covariant derivatives with respect to $\gamma_{ij}$ and 
$\tilde \gamma_{ij}$, respectively.  In addition, the Laplacians are
defined as $\Delta \equiv D^i D_i$ and  
$\tilde \Delta \equiv \tilde D^i \tilde D_i$. 

As the matter source of the Einstein equation, 
a perfect fluid is adopted. Then, the energy-momentum tensor is written as 
\beq
T_{\mu\nu}=(\rho+\rho\varep+P)u_{\mu}u_{\nu}+P g_{\mu\nu},
\eeq
where $\rho$, $\varep$, $P$, and $u_{\mu}$ are 
the baryon rest-mass density, the specific internal energy density, 
the pressure, and the four-velocity, respectively. 
Initial conditions are given using a polytropic equation of state as
\beq
P=\kappa \rho^{\Gamma},~~~~~~~\Gamma=1 + {1 \over n},
\eeq
where $\kappa$ and $n$ are a polytropic constant and a polytropic index. 
During the time evolution, 
we adopt a $\Gamma$-law equation of state of the form 
\beq
P=(\Gamma-1)\rho\varep. \label{GEOS}
\eeq
In the absence of shocks, the polytropic form of the equation of state 
is preserved even if Eq. (\ref{GEOS}) is used. Thus the quantity 
$\kappa'\equiv P/\rho^{\Gamma}[=\varep/(\Gamma-1)\rho^{\Gamma-1}]$ 
measures the efficiency of the shock heating. 
In this paper, we set $n=1$ ($\Gamma=2$) as a reasonable qualitative 
approximation to moderately stiff equations of state for neutron 
stars \cite{ST}. 

The hydrodynamic equations (continuity, Euler, and energy equations) 
are written in the forms 
\beqn
&&\pa_t \rho_* + \pa_i (\rho_* v^i )=0,\label{eqrho}\\
&& \pa_t (\rho_*  \hat u_j)
+ \pa_i (\rho_*  v^i \hat u_j
+ P\alpha e^{6\phi} \delta^i_{~j}) 
=P\pa_j(\alpha e^{6\phi})-\rho_* \Big[w h\pa_j \alpha - \hat u_i
\pa_j \beta^{i}+{1 \over 2 u^t h}\hat u_k \hat u_l \pa_j \gamma^{kl}\Big],
\label{euler}\\
&& \pa_t (\rho_* \hat e )+ \pa_i [\rho_* 
\hat e v^i + P e^{6\phi} (v^i+\beta^i)] =\alpha e^{6\phi} P K
+{\rho_* \over u^t h}\hat u_i \hat u_j K^{ij}
-\rho_* \hat u_i \gamma^{ij} D_j \alpha, \label{energy}
\eeqn
where 
\beqn
&& \rho_*=\rho w e^{6\phi},\\
&& h=1+\varep+{P \over \rho},\\
&& w=\alpha u^t, \\
&& \hat u_k=h u_k, \\
&& \hat e \equiv {e^{6\phi} \over \rho_*}
T_{\mu\nu} n^{\mu}n{^\nu} =  h w-{P \over \rho w},\\
&&v^i \equiv {u^i \over u^t} = -\beta^i + \gamma^{ij}{u_j \over u^t}=
-\beta^i + {\alpha \tilde \gamma^{ij} \hat u_j 
\over w h e^{4\phi}}. \label{eqvelo}
\eeqn
Here, the conservative form of the energy equation is adopted in contrast
with the previous works \cite{bina,bina2}. 
In numerical simulations, Eqs. (\ref{eqrho})--(\ref{energy}) are solved  
to evolve $\rho_*$, $\hat u_k$, and $\hat e$. 
Once $\hat u_i$ is obtained, $w$ is determined from the normalization
relation of the four-velocity, $u^{\mu}u_{\mu}=-1$, which can be written as 
\beq
w^2 = 1 +\gamma^{ij} u_i u_j 
=1 +\gamma^{ij} \hat u_i \hat u_j \biggl({\hat e \over w}+{P \over \rho w^2}
\biggr)^{-2},\label{wweq}
\eeq
where $P$ and $\rho$ are related to $\rho_*$, $\hat e$, and $w$ as 
$P=P(\rho,\varepsilon)=P[\rho_*/(w e^{6\phi}),\hat e]$
and $\rho=\rho_*/(w e^{6\phi})$. 

The Einstein equation is split into the constraint and evolution equations. 
The Hamiltonian and momentum constraint equations are written in the form 
\beqn
&& \tilde \Delta \psi = {\psi \over 8}\tilde R_k^{~k} 
- 2\pi \rho_{\rm H} \psi^5 
-{\psi^5 \over 8} \Bigl(\tilde A_{ij} \tilde A^{ij}
-{2 \over 3}K^2\Bigr), \label{hameq} \\
&& \tilde D_i (\psi^6  \tilde A^i_{~j}) - {2 \over 3} \psi^6 
\tilde D_j K = 8\pi J_j \psi^6, \label{momeq}
\eeqn
where $\psi\equiv e^{\phi}$, 
$\rho_{\rm H} \equiv T^{\mu\nu}n_{\mu}n_{\nu}$, and 
$J_i \equiv -T^{\mu\nu} n_{\mu}\gamma_{\nu i}$ with 
$n_{\mu}=(-\alpha, 0)$. Here, 
$R_{ij}$ ($\tilde R_{ij}$) denotes the Ricci tensor with respect 
to $\gamma_{ij}$ ($\tilde \gamma_{ij}$), 
and $R_k^{~k}=R_{ij}\gamma^{ij}$
($\tilde R_k^{~k}=\tilde R_{ij}\tilde \gamma^{ij}$).
These constraint equations are 
solved only at $t=0$ to set initial conditions (see Sec. II D) and 
for $t>0$, they are used to monitor the accuracy of numerical solutions. 

Following \cite{SN,gw3p2,gr3d,bina,bina2}, 
evolution equations for the geometric variables are written as 
\beqn
&&(\pa_t - \beta^l \pa_l) \tilde \gamma_{ij} 
=-2\alpha \tilde A_{ij} 
+\tilde \gamma_{ik} \beta^k_{~,j}+\tilde \gamma_{jk} \beta^k_{~,i}
-{2 \over 3}\tilde \gamma_{ij} \beta^k_{~,k}, \label{heq} \\
%%%%%%%%%%%%
&&(\pa_t - \beta^l \pa_l) \tilde A_{ij} 
= e^{ -4\phi } \biggl[ \alpha \Bigl(R_{ij}
-{1 \over 3}e^{4\phi}\tilde \gamma_{ij} R_k^{~k} \Bigr) 
-\Bigl( D_i D_j \alpha - {1 \over 3}e^{4\phi}
\tilde \gamma_{ij} \Delta \alpha \Bigr)
\biggr] \nonumber \\
&& \hskip 2.5cm +\alpha (K \tilde A_{ij} 
- 2 \tilde A_{ik} \tilde A_j^{~k}) 
+\beta^k_{~,i} \tilde A_{kj}+\beta^k_{~,j} 
\tilde A_{ki}
-{2 \over 3} \beta^k_{~,k} \tilde A_{ij} \nonumber \\
&& \hskip 2.5cm-8\pi\alpha \Bigl( 
e^{-4\phi} S_{ij}-{1 \over 3} \tilde \gamma_{ij} S_k^{~k}
\Bigr), \label{aijeq} \\
%%%%%%%%%%%
&&(\pa_t - \beta^l \pa_l) K 
=\alpha \Bigl[ \tilde A_{ij} \tilde A^{ij}+{1 \over 3}K^2
\Bigr] 
-\Delta \alpha +4\pi \alpha (\rho_{\rm H}+ S_k^{~k}), 
\label{keq}
\eeqn
where $S_{ij} \equiv T^{\mu\nu}\gamma_{\mu i}\gamma_{\nu j}$. 
Equations (\ref{heq})--(\ref{keq}) are solved to evolve 
$\tilde \gamma_{ij}$, $\tilde A_{ij}$, and $K$. 

In the previous works \cite{SN,gw3p2,gr3d,bina,bina2}, 
the evolution equation for $\phi$ is written in the form 
\beq
(\pa_t - \beta^l \pa_l) \phi = {1 \over 6}\Bigl( 
-\alpha K + \beta^k_{~,k} \Bigr). \label{peq} 
\eeq
Instead of this form, in the present work,
a conservative form is adopted as
\beq
\pa_t e^{6\phi} -\pa_i (\beta^i e^{6\phi})= 
-\alpha K e^{6\phi}. \label{peq2} 
\eeq

As in the previous works, we introduce an auxiliary variable 
$F_i=\delta^{jl}\pa_l \tilde \gamma_{ij}$ \cite{SN}, 
which evolves according to the evolution equation 
\beqn
(\pa_t - \beta^l \pa_l)F_i& =&-16\pi \alpha J_i+2\alpha 
\Bigl\{ f^{kj} \tilde A_{ik,j}
+f^{kj}_{~~,j} \tilde A_{ik} 
-{1 \over 2} \tilde A^{jl} h_{jl,i} 
+6\phi_{,k} \tilde A^k_{~i}-{2\over 3}K_{,i} \Bigr\} 
\nonumber \\
&+&\delta^{jk} \Bigl\{ -2\alpha_{,k} \tilde A_{ij} 
+ \beta^l_{~,k}h_{ij,l} 
+\Bigl(\tilde \gamma_{il}\beta^l_{~,j}+\tilde \gamma_{jl}\beta^l_{~,i}
-{2\over 3}\tilde \gamma_{ij} \beta^l_{~,l}\Bigr)_{,k}\Bigr\}, 
\label{eqF}
\eeqn
where $\tilde \gamma_{ij}$ and $\tilde \gamma^{ij}$ are split into 
$\delta_{ij}+h_{ij}$ and $\delta^{ij}+f^{ij}$. 
In the numerical simulations, a term 
$\delta^{kl}\tilde \gamma_{ik,lj}$ which appears 
in the expression of $R_{ij}$ in Eq. (\ref{aijeq}) is
evaluated using $F_i$ as $F_{i,j}$. 
This replacement is crucial to enable a stable and longterm simulation. 

\subsection{Improvements for numerical implementations}

In this section, we report our improvement to several numerical 
implementations since the latest work \cite{bina2}. 
Firstly, the hydrodynamic part has been improved 
adopting the so-called high-resolution shock-capturing scheme 
for computation of the transport terms of
Eqs. (\ref{eqrho})--(\ref{energy}) as described in \cite{shiba2d}. 
With this scheme, shocks are captured with a much better accuracy than
in the previous implementation \cite{gr3d,bina,bina2}. 
Although the shocks generated during the merger of binary neutron stars are
not very strong, it is promising to use such high-resolution schemes 
to accurately compute peak densities and to evaluate effects 
of the shock heating. 

Numerical treatments for transport terms in the evolution 
equation for geometric variables have been also improved.
For the transport terms in Eqs. (\ref{heq})--(\ref{keq}),
a second-order upwind scheme has been adopted \cite{gw3p2}.
To avoid numerical instabilities, 
we incorporate a limiter $f$ by which the order of the finite
differencing for the numerical flux is lowered from the second order
to the first order at a point of steep gradient as 
\beq
F_{\rm num}= F_{1} f + F_2 (1-f), 
\eeq
where $F_{\rm num}$, $F_1$, and $F_2$ denote the total, 
first-order, and second-order fluxes. Since the previous choice of $f$
is found to be too dissipative \cite{gw3p2}, we have changed 
the functional form of $f$ to 
\beq
f=1-{2 \sqrt{|\delta Q_u \delta Q_d|} \over |\delta Q_u| + |\delta Q_d|},
\eeq
where $Q$ denotes one of the variables among 
$\tilde A_{ij}$, $K$, and $\tilde \gamma_{ij}$.  
$\delta Q_u$ and $\delta Q_d$ denote the difference of
$Q$ for two neighboring grid points as
$\delta Q_u=Q_{I+1}-Q_I$ and $\delta Q_d=Q_{I}-Q_{I-1}$ where  
$Q_I$ denotes the value of $Q$ at the $I$-th grid point.

For the evolution of $e^{6\phi}$,
we have also changed the finite differencing
scheme. As shown in Eq. (\ref{peq2}), the
evolution equation for $e^{6\phi}$ has the same conservative form
as that of hydrodynamic equations.
Thus the numerical flux is computed using the 
third-order upwind scheme with an appropriate min-mod
limiter as done in the hydrodynamic equations \cite{shiba2d}. 
This change plays a significant role for enforcing the conservation
of the total ADM mass and angular momentum. 

The outer boundary condition for $\phi$ has been also improved. 
In our previous works, we simply imposed 
\beqn
\phi = O(r^{-1}). 
\eeqn
It is replaced with a better condition as 
\beqn
\phi ={M \over 2r} + O(r^{-2}), \label{phibnd}
\eeqn
where $M$ is the ADM mass which will be defined in Sec. II E.

\subsection{Change in gauge conditions}

As the time slicing condition, 
an approximate maximal slice (AMS) condition $K \approx 0$ 
is adopted following previous papers \cite{bina2}.
On the other hand, the spatial gauge condition has been changed. 

Our previous simulations were performed adopting an approximately 
minimal distortion (AMD) gauge condition \cite{gw3p2}. 
The equation in this condition is schematically written as 
\beq
\delta_{ij} \Delta_{\rm f} \beta^j + {1 \over 3}\pa_i \pa_j \beta^j =S_i,
\eeq
where $S_i$ denotes the source term composed of geometric variables
and matter sources, and $\Delta_f$ the flat Laplacian. In this 
gauge condition, a vector elliptic-type equation has to be solved. 
A serious drawback of this is the long computational time needed 
to obtain its numerical solution. Typically, $\sim 50 \%$ of
total computational time is consumed in solving this equation. 

\begin{figure}[t]
\begin{center}
\epsfxsize=3.in
\leavevmode
\epsffile{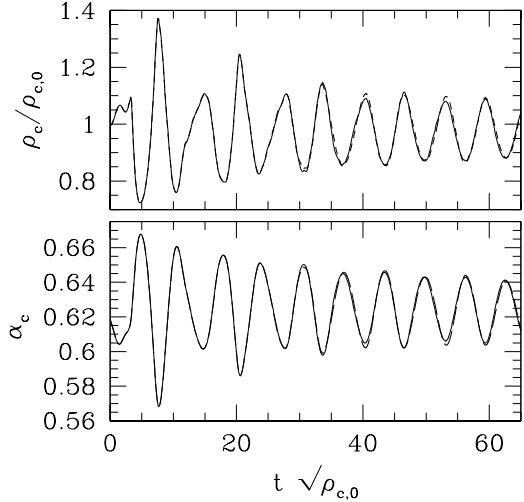}
\end{center}
\vspace{-4mm}
\caption{Evolution of central density $\rho_c$
in units of the initial value $\rho_{c,0}$ and 
central value of the lapse function $\alpha_c$ 
for an oscillating and rapidly rotating neutron star 
with $n=1$. The baryon rest-mass and the ADM mass in units of
$\kappa=1$ (see Sec. II E) are 0.186 and 0.172, respectively. 
The compactness measured by the equatorial (polar) radius is 0.129 (0.207).
The angular velocity is constant and equal to the Kepler velocity
at the equatorial surface. The time is shown in units of $\rho_{c,0}^{-1/2}$.
The solid and dashed curves denote the results by the 
dynamical and AMD gauge conditions, respectively. 
\label{FIG1} }
\end{figure}

To overcome this drawback, we adopt a dynamical spatial gauge condition, 
e.g., in \cite{ALC,LS}. Following \cite{S03}, the equation for the
shift vector is chosen to be 
\beq
\pa_t \beta^k = \tilde \gamma^{kl} [F_l +\Delta t (\pa_t F_l)],\label{dyn}
\eeq
where $\Delta t$ denotes a time step in numerical computation.
The second term in the right-hand side of Eq. (\ref{dyn})
is introduced to stabilize the numerical computation. 
In this choice, $\beta^k$ obeys a hyperbolic type equation
(for a sufficiently small value of $\Delta t$) as
\beq
\pa_t^2 \beta^j = 
\Delta_{\rm f} \beta^j + {1 \over 3}\tilde \gamma^{jk}\pa_k \pa_i \beta^i
- \tilde S^j,
\eeq
where $\tilde S^j$ denotes the source term. With this gauge condition, 
the fraction of the computational time occupied
for imposing the spatial gauge is negligible. 
Furthermore, we have confirmed that the numerical solution in this
gauge condition agrees well with that in the AMD gauge condition
for collapse of neutron stars to a black hole \cite{S03}
and for oscillating and rapidly rotating neutron stars 
(cf. Fig. \ref{FIG1}). This indicates that the present dynamical gauge
is physically as good as the AMD gauge. 

Since $\beta^k$ obeys a hyperbolic type equation 
in this gauge condition, so should $F_l$. 
Thus we impose an outgoing boundary condition for $F_l$ as 
\beqn
F_l = {f_l(t-r) \over r},
\eeqn
where $f_l(t-r)$ is a function and its value at outer boundaries 
is determined from the values on the eight nearby 
grid points at a previous time step.
The same type of boundary condition is imposed for
$\tilde A_{ij}$ and $\tilde \gamma_{ij}$ \cite{SN}.

\subsection{Initial conditions}

Binary neutron stars with a moderate compactness of orbits
as $a/M \agt 6$ where $a$ denotes an orbital separation 
are in a quasiequilibrium state even just before the merger because 
the time scale of gravitational radiation reaction at Newtonian order 
$\sim 5/\{64\Omega(M_{\rm N}\Omega)^{5/3}\}$ \cite{ST} 
(where $M_{\rm N}$ and $\Omega$ denote the Newtonian total mass of system 
and the orbital angular velocity of binary neutron stars) 
is several times longer than the orbital period. Thus 
a quasiequilibrium state should be prepared as the initial condition
for a realistic simulation of the merger. 
Such quasiequilibrium states are obtained by solving 
coupled equations of gravitational field and 
hydrostatic equations. For the gravitational field,
we adopted the conformal flatness formulation \cite{WM} 
in which the three geometry is assumed to be 
conformally flat and the selected components of 
the Einstein equation are solved. Specifically, the
selected components are the Hamiltonian 
and momentum constraints, and the trace of spatial projection of the
Einstein equation with the maximal slicing condition $K=0$.
The solution in this formalism is fully general relativistic
in the sense that they satisfy the constraints. 

It is expected that most of the close binary neutron stars in 
quasiequilibrium circular orbits have 
irrotational velocity fields approximately since 
the viscous time scale is much longer than the gravitational
radiation time scale and the orbital period $\sim 2$ msec is much shorter
than the typical spin period of neutron stars \cite{CBS}. 
Assuming the irrotational velocity field and 
the presence of a helical Killing vector as 
\beq
\ell^{\mu}=\biggl({\pa \over \pa t}\biggr)^{\mu}
+\Omega \biggl({\pa \over \pa \varphi}\biggr)^{\mu},
\eeq
the hydrodynamic equations are written into 
a first integral of the Euler equation and 
an elliptic-type equation for a velocity potential \cite{irre}.
%%%These two equations constitute the hydrostatic equations. 

The coupled equations of the selected
Einstein and hydrostatic equations are solved
by a pseudospectral method developed by
Bonazzola, Gourgoulhon, and Marck \cite{GBM}.
Detailed numerical calculations have been done by Taniguchi
and part of the numerical results are presented in \cite{TG}. 

Quasiequilibrium solutions are given as the initial conditions for 
simulations without any modification. 
In a realistic system of binary neutron stars, 
the orbit is not strictly circular because of the presence of 
the approaching velocity due to gravitational radiation reaction. 
As pointed out by Miller \cite{mark}, neglecting the 
effect of the approaching velocity yields a systematic error 
in waveforms and the merger process, if we choose a quasiequilibrium
with a very small orbital separation as the initial condition. 
It is likely that gravitational waveforms, in particular
the wave phase, obtained below contain a small systematic error.
However, it has been studied in \cite{bina2}
that the merger process and the final outcome depend very weakly on 
an artificial approaching velocity of $\approx 10\%$
of the orbital velocity.

\subsection{Definitions of quantities}

In numerical simulations, we refer to the total baryon rest-mass, 
the ADM mass, and the angular momentum of the system, which are given by 
\beqn
M_* &&\equiv \int \rho_* d^3x, \\
M &&\equiv -{1 \over 2\pi} 
\oint_{r\rightarrow\infty} D^i \psi dS_i \nonumber \\
&&=\int \biggl[ \rho_{\rm H} e^{5\phi} +{e^{5\phi} \over 16\pi}
\biggl(\tilde A_{ij} \tilde A^{ij}-{2 \over 3}K^2 -\tilde R_k^{~k} 
e^{-4\phi}\biggr)\biggr]d^3x, \label{eqm00}\\
J &&\equiv {1 \over 8\pi}\oint_{r\rightarrow\infty} 
\varphi^ i \tilde A_i^{~j} e^{6\phi} dS_j \nonumber \\
&&=\int e^{6\phi}\biggl[J_i \varphi^i  
+{1 \over 8\pi}\biggl( \tilde A_i^{~j} \pa_j \varphi^i 
-{1 \over 2}\tilde A_{ij}\varphi^k\pa_k \tilde \gamma^{ij}
+{2 \over 3}\varphi^j \pa_j K \biggr) \biggr]d^3x,~~~
\label{eqj00}
\eeqn
where $dS_j=r^2 D_j r d(\cos\theta)d\varphi$ 
and $\varphi^j=-y(\pa_x)^j + x(\pa_y)^j$. 
To rewrite the expressions for $M$ and $J$, the Gauss law is used. 
Here, $M_*$ is a conserved quantity, and it uniquely specifies a model 
of a stable neutron star for a given value of $\Gamma$.

$M$ and $J$ are computed using the volume integral
shown in Eqs. (\ref{eqm00}) and (\ref{eqj00}).
Since the computational domain is finite, 
they are not constant and decrease
after gravitational waves propagate to the outside of the
computational domain during time evolution.
Therefore, in the following, they are referred to as 
the ADM mass and the angular momentum computed in the finite domain 
(or simply as $M$ and $J$, which decrease with time).
As easily predicted from the calculation 
using the quadrupole formula, $M$ decreases at most by 0.5\% and
may be regarded as an approximately conserved quantity, while
$J$ decreases by $\sim 5$--10\%.

A model of each neutron star is specified 
using the compactness $\comp$ which is defined as the ratio of 
the ADM mass to the circumferential radius of 
a spherical neutron star in isolation (see Tables I and II). 
To indicate how massive the system is, 
we also introduce the ratio of the total baryon rest-mass 
of the system to the maximum allowed mass of spherical 
neutron stars for a given equation of state $M_{*\rm max}^{\rm sph}$, 
\beq
Q_{*} \equiv {M_* \over M_{*\rm max}^{\rm sph}}.
\eeq

%%In addition to the above quantities, 
%%we use the relation between the baryon rest-mass and 
%%specific angular momentum that we define as 
%%\beq
%%M_*(j)=\int_{j' > j} \rho_*(x') d^3x',
%%\eeq
%%where $j$ denotes the specific angular momentum 
%%$j\equiv h u_i \varphi^i$. Here, $M_*(j)/M_*$ denotes a baryon 
%%rest-mass fraction whose specific angular momentum 
%%is larger than a given value $j$ (i.e., $M_*(j=0)/M_*=1$). 

Physical units enter the problem through the polytropic 
constant $\kappa$ initially chosen, which can be completely 
scaled out of the problem. Since $\kappa^{n/2}$
has the dimension of length, time, and mass
in the geometrical units $c=G=1$, 
dimensionless variables can be constructed as 
\beqn
&& \bar M_* = M_* \kappa^{-n/2}, 
~~~\bar M = M \kappa^{-n/2}, ~~~
\bar R = R \kappa^{-n/2},  \nonumber \\
&& \bar J = J \kappa^{-n}, ~~~
\bar \rho = \rho \kappa^{n}, ~~~{\rm and}~~~
\bar \Omega = \Omega \kappa^{n/2}.
\eeqn
In the following, only these dimensionless quantities are presented 
(namely units of $\kappa=1$ are adopted) and, hence, the bar is omitted. 

Nondimensional quantities may be converted to dimensional 
ones for a value of $\kappa$.
For $\kappa = 1.455 \times 10^{5}$ cgs which is chosen in \cite{other},  
the mass, the density, and time in the dimensional units are written as 
\beqn
&& M_{\rm dim}=1.80 M_{\odot}
\biggr({\kappa \over 1.455\times 10^5~{\rm cgs}}\biggr)^{1/2}
\biggl({M \over 0.180}\biggr),\\
&& \rho_{\rm dim}=1.86 \times 10^{15}~{\rm g/cm^3} 
\biggr({\kappa \over 1.455\times 10^5~{\rm cgs}}\biggr)^{-1}
\biggl({\rho \over 0.300}\biggr),\\
&& T_{\rm dim}=4.93~{\rm msec} 
\biggr({\kappa \over 1.455\times 10^5~{\rm cgs}}\biggr)^{1/2}
\biggl({T \over 100}\biggr). 
\eeqn

\subsection{Calibration}

Several test simulations, including spherical collapse of dust,
stability of spherical and rotating neutron stars, 
comparison of eigenoscillation modes of spherical stars with
the known results, 
and longterm evolution of rotating stars, have been performed 
to check the reliability of numerical results obtained in the
new implementation. A list of these test simulations and
some of their results are described in \cite{shiba2d}. 

During the simulations, we monitored the violation of the
Hamiltonian constraint, and 
the conservation of the baryon rest-mass, the ADM mass, 
and the angular momentum. 
Because of the emission of gravitational waves, 
$M$ and $J$ computed in the finite volume by 
Eqs. (\ref{eqm00}) and (\ref{eqj00}) decrease with time. 
However, the sum of $M$ and accumulated radiated energy of
gravitational waves, and the sum of $J$ and accumulated radiated
angular momentum of gravitational 
waves should be conserved (at least approximately)
in numerical computation as \cite{foot00}
\beqn
&&M(t) + \Delta E(t) =M_0,\label{eqm01}\\
&&J(t) + \Delta J(t) =J_0,\label{eqj01}
\eeqn
where $\Delta E(t)$ and $\Delta J(t)$ denote the total radiated 
energy and angular momentum by gravitational waves until a time $t$, 
for which the definitions are described in Sec. III. 
$M_0$ and $J_0$ denote the initial values of $M$ and $J$. 

The violation of the Hamiltonian constraint 
is locally measured by the equation as 
\beq
\displaystyle
f_{\psi} \equiv 
{\Bigl|\tilde \Delta \psi - {\psi \over 8}\tilde R_k^{~k} 
+ 2\pi \rho_{\rm H} \psi^5 
+{\psi^5 \over 8} \Bigl(\tilde A_{ij} \tilde A^{ij}
-{2 \over 3}K^2\Bigr)\Bigr| \over
|\tilde \Delta \psi | + |{\psi \over 8}\tilde R_k^{~k}| 
+ |2\pi \rho_{\rm H} \psi^5| 
+{\psi^5 \over 8} \Bigl(|\tilde A_{ij} \tilde A^{ij}|+
{2 \over 3}K^2\Bigr)}. 
\eeq
Following \cite{shiba2d}, we define and monitor a global quantity as 
\beq
H \equiv {1 \over M_*} \int \rho_* f_{\psi} d^3x. \label{vioham}
\eeq
Hereafter, this quantity will be referred to as the averaged violation
of the Hamiltonian constraint. 

\section{Analysis of gravitational waves} 

Gravitational waves are measured in terms of the gauge-invariant 
Moncrief variables in a flat spacetime \cite{moncrief}.
To compute them, first, we perform a coordinate transformation for
the three-metric from the Cartesian coordinates 
to the spherical polar coordinates, and then split $\gamma_{ij}$ 
into $\eta_{ij}+\sum_{lm} \zeta_{ij}^{lm}$, where 
$\eta_{ij}$ is the flat metric in the spherical polar coordinates 
and $\zeta_{ij}^{lm}$ is given by 
\beqn
\zeta_{ij}^{lm}=&& \left(
\begin{array}{lll}
\displaystyle 
H_{2lm} Y_{lm} & h_{1lm} Y_{lm,\theta} & h_{1lm} Y_{lm,\varphi}\\
\ast & r^2(K_{lm}Y_{lm}+G_{lm}W_{lm}) & r^2G_{lm}X_{lm} \\
\ast & \ast & r^2\sin^2\theta(K_{lm}Y_{lm}-G_{lm}W_{lm}) \\
\end{array}
\right) \nonumber \\
&&+\left(\begin{array}{ccc}
0 &  -C_{lm} \pa_{\varphi} Y_{lm}/\sin\theta
& C_{lm} \pa_{\theta}Y_{lm}\sin\theta  \\
\ast & r^2D_{lm}X_{lm}/\sin\theta
             & -r^2 D_{lm}W_{lm}\sin\theta  \\
\ast & \ast & -r^2 D_{lm}X_{lm}\sin\theta \\
\end{array}
\right). 
\eeqn
Here, $\ast$ denotes the symmetric components. The quantities 
$H_{2lm}$, $h_{1lm}$, $K_{lm}$, $G_{lm}$, $C_{lm}$, and 
$D_{lm}$ are functions of $r$ and $t$, and are calculated 
by performing integrals over a two-sphere of a given coordinate radius 
[see \cite{SN} for details]. 
$Y_{lm}$ is the spherical harmonic function, and 
$W_{lm}$ and $X_{lm}$ are defined as
\beqn
W_{lm} \equiv \Bigl[ (\pa_{\theta})^2-\cot\theta \pa_{\theta}
-{1 \over \sin^2\theta} (\pa_{\varphi})^2 \Bigl] Y_{lm},
\hskip 5mm 
X_{lm} \equiv 2 \pa_{\varphi} \Bigl[ \pa_{\theta}-\cot\theta \Bigr] Y_{lm}. 
\eeqn
The gauge-invariant variables of even and odd parities are defined by 
\beqn
&&R_{lm}^{\rm E}(t,r) \equiv 
\sqrt{2(l-2)! \over (l+2)!}
\Bigl\{ 4k_{2lm}+l(l+1)k_{1lm} \Bigr\}, \\
&&R_{lm}^{\rm O}(t,r) \equiv \sqrt{2(l+2)! \over (l-2)!}
\biggl({C_{lm} \over r}+r \pa_r D_{lm}\biggr),
\eeqn
where
\beqn
k_{1lm}&& \equiv K_{lm}+l(l+1)G_{lm}+2r \pa_r G_{lm}-2{h_{1lm} \over r},\\
k_{2lm}&& \equiv {H_{2lm} \over 2} - {1 \over 2}{\pa \over \pa r}
\Bigl[r\{ K_{lm}+l(l+1)G_{lm} \} \Bigr].
\eeqn
The cosine and sine components of the gauge-invariant 
variables, which are real quantities, are also defined as 
\beq
R_{lm+}^{\rm E}={R_{lm}^{\rm E}+R_{l~-m}^{\rm E} \over \sqrt{2}}~~~
{\rm and}~~~
R_{lm-}^{\rm E}={R_{lm}^{\rm E}-R_{l~-m}^{\rm E} \over \sqrt{2}}~~~
(m > 0). 
\eeq

Using the gauge-invariant variables, the energy luminosity and the angular
momentum flux of gravitational waves can be calculated as 
\beqn
&&{dE \over dt}={r^2 \over 32\pi}\sum_{l,m}\Bigl[
|\pa_t R_{lm}^{\rm E}|^2+|\pa_t R_{lm}^{\rm O}|^2 \Bigr],
\label{dedt} \\
&&{dJ \over dt}={r^2 \over 32\pi}\sum_{l,m}\Bigl[
 |m(\pa_t R_{lm}^{\rm E}) R_{lm}^{\rm E} |
+|m(\pa_t R_{lm}^{\rm O}) R_{lm}^{\rm O} | \Bigr]. 
\label{dJdt} 
\eeqn
The total radiated energy and angular momentum are defined as
\beq
\Delta E(t) = \int_0^t dt {dE \over dt}, \hskip 5mm
\Delta J(t) = \int_0^t dt {dJ \over dt}.
\eeq
We have computed modes with $l=2$, 3, and 4, and found that 
the even modes with $l=|m|=2$ are dominant, and 
the even mode with $l=2$ and $m=0$ is secondly dominant.
For merger of unequal-mass binaries, the amplitude of the
even modes with $l=|m|=3$ are as large as that of $l=2$ and $m=0$ .
Thus attention is paid to these three modes. 

To search for the dominant frequencies of gravitational waves,
the Fourier spectra are computed by 
\beq
\hat R_{lm\pm}(f)=\int^{t_f}_{t_i} e^{2\pi i f t} R_{lm\pm} dt.
\eeq
In the analysis, $t_f$ is chosen as the time at which the simulation is
stopped. Before $t < r_{\rm obs}$ where $r_{\rm obs}$ denotes
a radius at which gravitational waves are extracted, no waves 
propagate to $r_{\rm obs}$, so that we choose $t_i \approx r_{\rm obs}$. 

Using the Fourier spectrum, the energy spectrum which is
often referred to in literature (e.g., \cite{C,FR2}) can be written as
\beq
{dE \over df}={\pi \over 2}r^2 \sum_{l,m\geq 0}|\hat R_{lm}(f) f|^2, 
\eeq
where for $m\not=0$, we define 
\beq
\hat R_{lm}\equiv \sqrt{|\hat R_{lm+}(f)|^2 + |\hat R_{lm-}(f)|^2}. 
\eeq
To help the calculation of $dE/df$, $|\hat R_{lm}(f)f|r$ is presented as
the Fourier spectrum in the following.

\section{Numerical results}

\subsection{Setup for simulation}

Several quantities that characterize 
quasiequilibrium states of irrotational binary neutron stars 
used as initial conditions for the present simulations are
summarized in Table I. All quantities are appropriately scaled
with respect to $\kappa$ to be dimensionless. 

As the initial conditions,
we choose binaries of an orbital separation
which is slightly (by $\sim 10\%$) larger than that for an innermost
orbit. Here, the innermost orbit is defined as
a close orbit for which Lagrange points appear at the
inner edge of neutron stars \cite{USE,GBM}. 
Models M1414 and M1616 are equal-mass binaries, and others are
unequal-mass ones. Total baryon rest-masses for 
models M1616, M1517, and M1418 or for M1414 and M1315 are
almost identical, while the baryon rest-mass ratios are identical
for models M1517 and M159183 as 0.925. 

The frequency of gravitational waves for binaries in these 
quasiequilibria is given by 
\beq
f_{\rm QE}={\Omega \over \pi} \approx 960~{\rm Hz}
\biggl({2.8M_{\odot} \over M_0}\biggr)
\biggl({C_0 \over 0.12}\biggr)^{3/2},
\label{fqe}
\eeq
and, thus, the orbital period of the quasiequilibria, $P_{t=0}$, is
\beq
P_{t=0} \approx 2.08~{\rm msec}
\biggl({2.8M_{\odot} \over M_{0}}\biggr)^{-1}
\biggl({C_0 \over 0.12}\biggr)^{-3/2}, 
\eeq
where $C_0$ is a compactness parameter of an orbit defined by 
\beq
C_0 \equiv (M_{0}\Omega)^{2/3} \equiv {M_{0} \over a_0}. 
\eeq
Here, $a_0$ is defined as the initial orbital separation. 
For the initial conditions chosen in this paper, $a_0 > 8.5M_0$.

The simulations were performed using a fixed uniform grid 
and assuming reflection symmetry with respect to the
equatorial plane 
(here, the equatorial plane is chosen as the orbital plane). 
The typical grid size is (633, 633, 317) for $(x, y, z)$. 
The grid covers the region $-L \leq x \leq L$, $-L \leq y \leq L$, and 
$0 \leq z \leq L$ where $L$ is a constant. 
The grid spacing (which is $L/316$ in the typical case) is determined 
from the condition that the major diameter of each star is covered 
with about 40 grid points initially. 

Numerical results depend weakly on the grid resolution and location of 
the outer boundaries. In order to investigate this, 
additional test simulations were performed 
choosing the smaller grid sizes with a fixed value of 
grid spacing and the larger grid spacings with a fixed value of $L$ 
for selected models. The setting for the test simulations are
summarized in Table II and 
the numerical results are presented in Sec. IV E. 

With a (633, 633, 317) grid size, about 240 GBytes 
computational memory is required. For the case of neutron star formation, 
the simulations are performed for about 20000 time steps and then 
stopped artificially. The computational time for one model in
such a calculation is about 100 CPU hours 
using 32 processors on FACOM VPP5000 in the data processing center of
National Astronomical Observatory of Japan (NAOJ).
For the case of black hole formation, the simulations 
crash soon after the formation of apparent horizon because of
the so-called grid stretching around the black hole formation region.
In this case, the computational time is about 50 CPU hours for
about 10000 time steps. 

In the above setting, the wavelength of gravitational waves 
at $t=0$ (denoted by $\lambda_0$) is about twice that
of $L$ (cf. Table I). As found in a previous paper \cite{bina2}, 
gravitational waves and radiation reaction are taken 
into account with a fair accuracy (within $\sim 10 \%$ numerical error) 
in this setting.
Since the typical wavelength of gravitational waves becomes shorter and 
shorter in the late inspiral phase, 
the accuracy of the wave extraction is improved with the evolution
of the system. As a result, the magnitude of the 
error in the total radiated energy and 
angular momentum would be much smaller than 10\%. 
This point will be reconfirmed in Sec. IV D. 
The wavelength of quasiperiodic waves emitted from the formed 
neutron star is much shorter than $\lambda_0$ and $L$, so that 
the waveforms in the merger stage can be computed accurately
in the case of neutron star formation. 

As found in \cite{USE,GBM}, orbits for irrotational binaries 
of equal mass with $\Gamma < 2.5~(n > 2/3)$ are dynamically stable from 
the infinite separation to the innermost orbit. 
Therefore the merger in reality should be triggered by the 
radiation reaction of gravitational waves for $\Gamma=2$. 
In the previous implementation, however, 
the radiation reaction for the late inspiral stage
was not very accurately computed. Thus the simulations 
were initiated by setting a binary at the innermost orbit and 
reducing the angular momentum slightly to induce prompt merger
\cite{bina,bina2}. 
In the new implementation, on the other hand, the radiation reaction 
can be computed with a good accuracy (within $\sim 1$--2 \% error
throughout a simulation, see Sec. IV D). Thus, 
in the present work, we prepared binaries of orbits slightly
far away from the innermost orbits and started simulations 
without adding any perturbation. With this setting, 
a transition from the inspiral to the merger
is triggered by the radiation reaction. 
This point will be demonstrated in Sec. IV D. 

%%%%%%% Sec. IV B

\subsection{Characteristics of the merger}

\begin{figure}[t]
\begin{center}
\epsfxsize=2.4in
\leavevmode
\epsffile{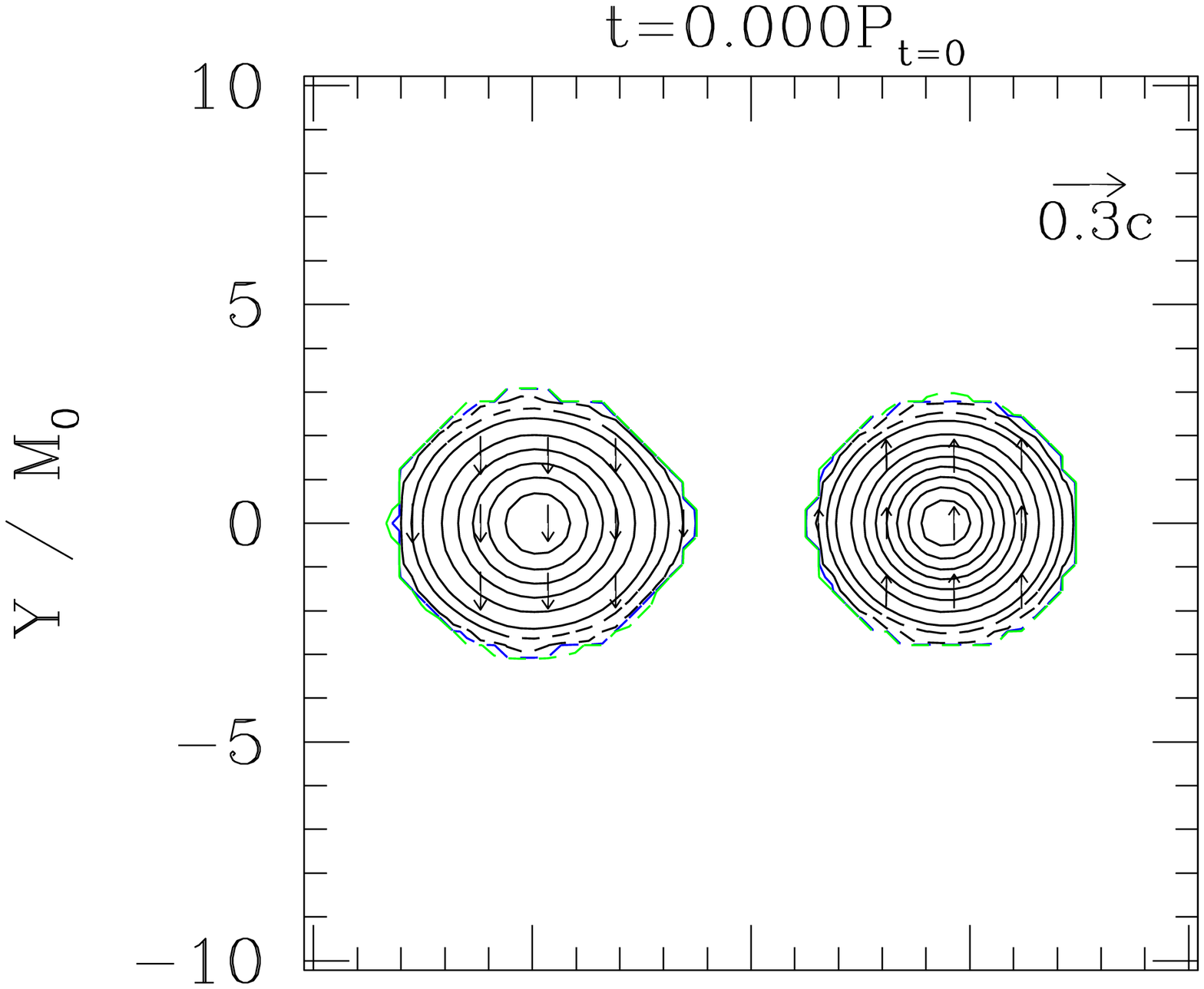}
\epsfxsize=2.4in
\leavevmode
\hspace{-1.2cm}\epsffile{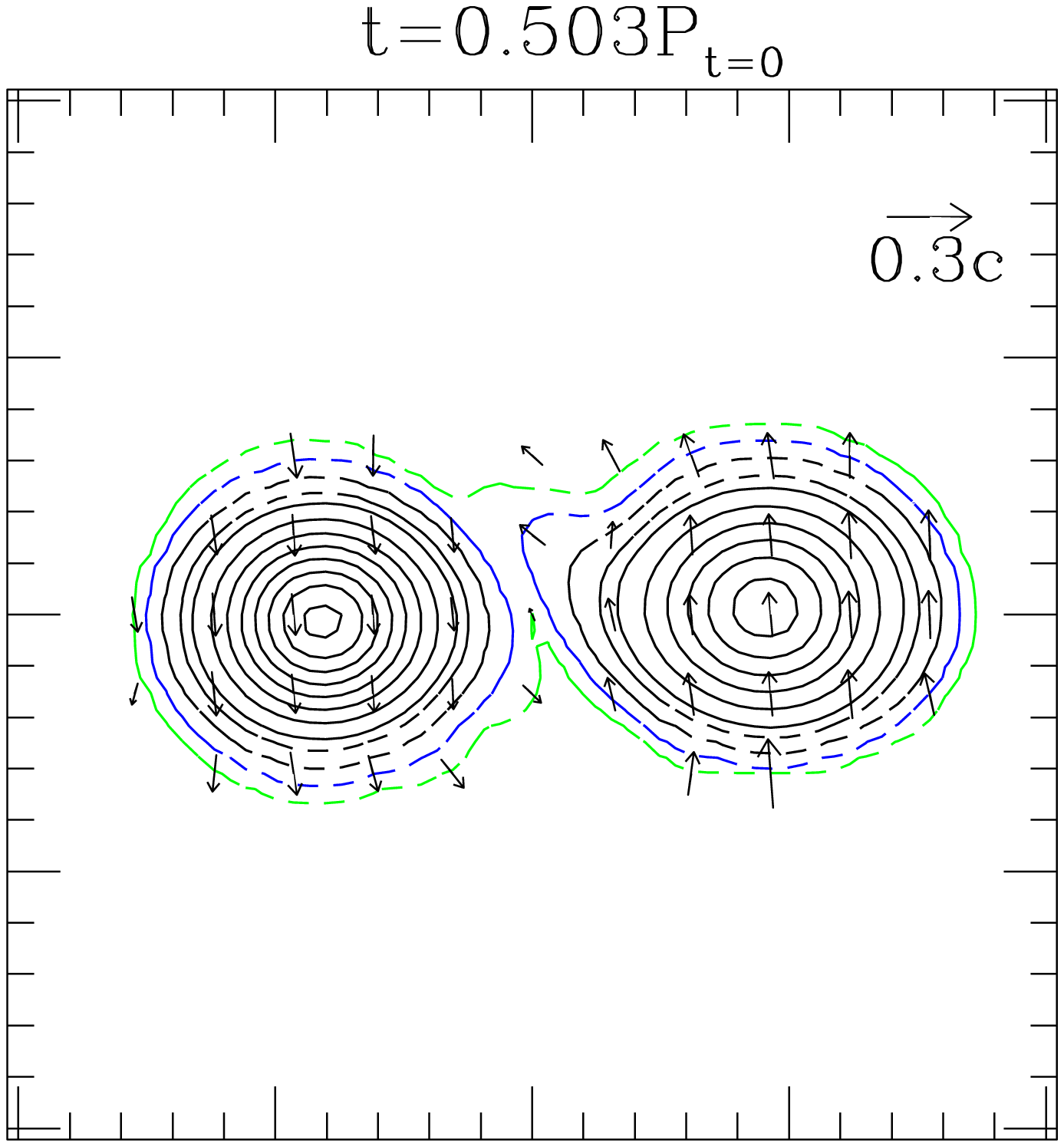} 
\epsfxsize=2.4in
\leavevmode
\hspace{-1.2cm}\epsffile{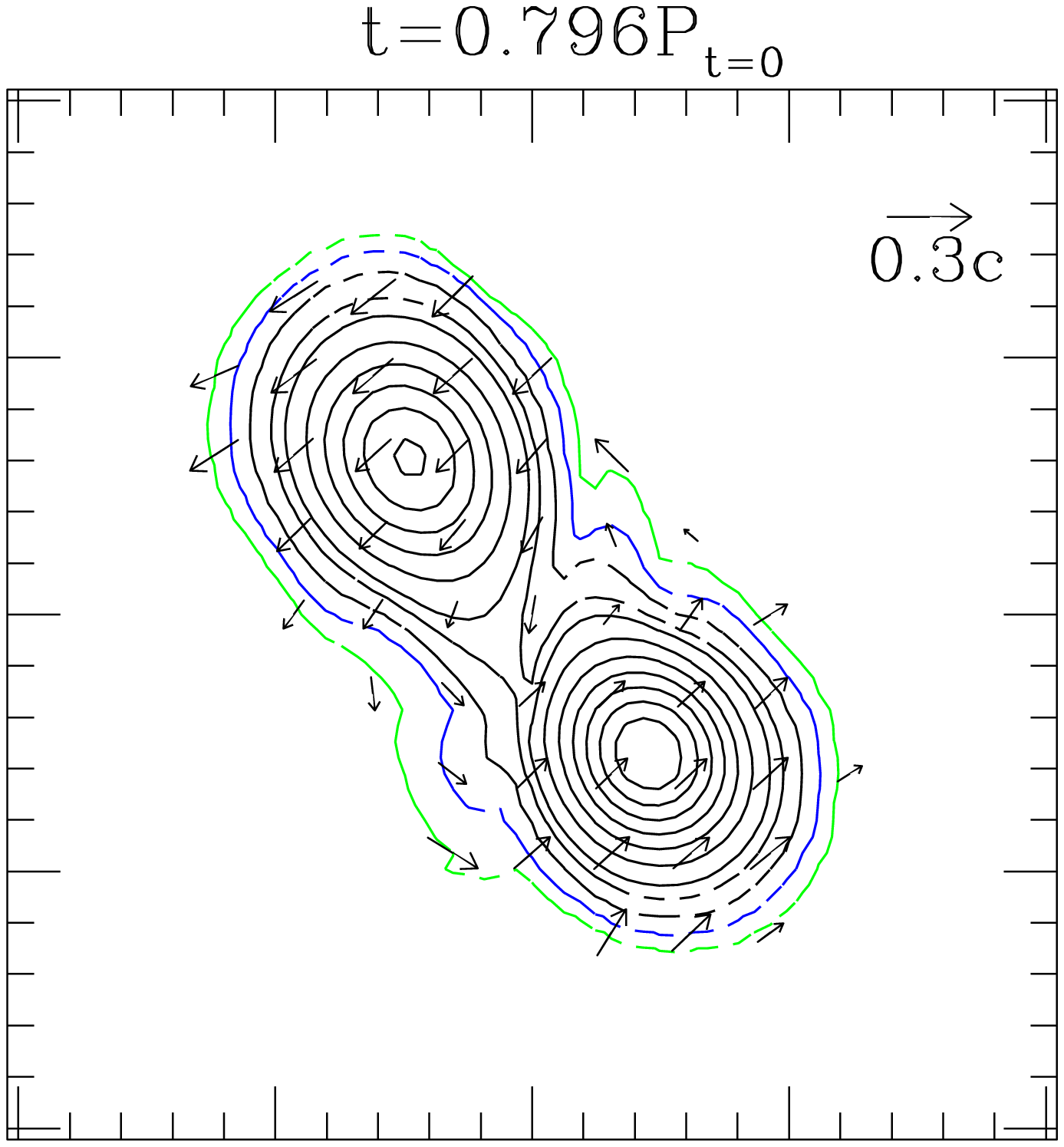} \\
\vspace{-1.2cm}
\epsfxsize=2.4in
\leavevmode
\epsffile{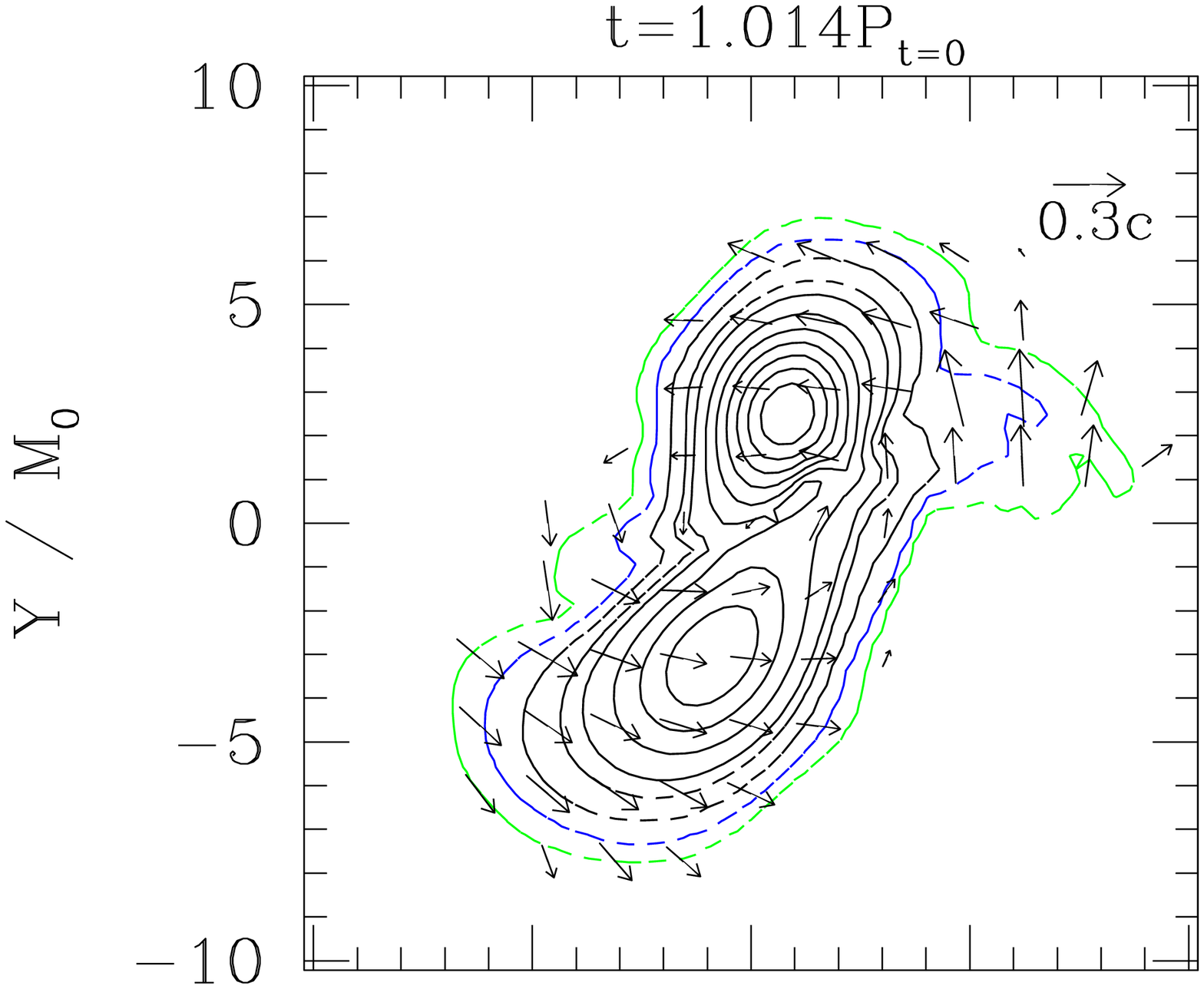} 
\epsfxsize=2.4in
\leavevmode
\hspace{-1.2cm}\epsffile{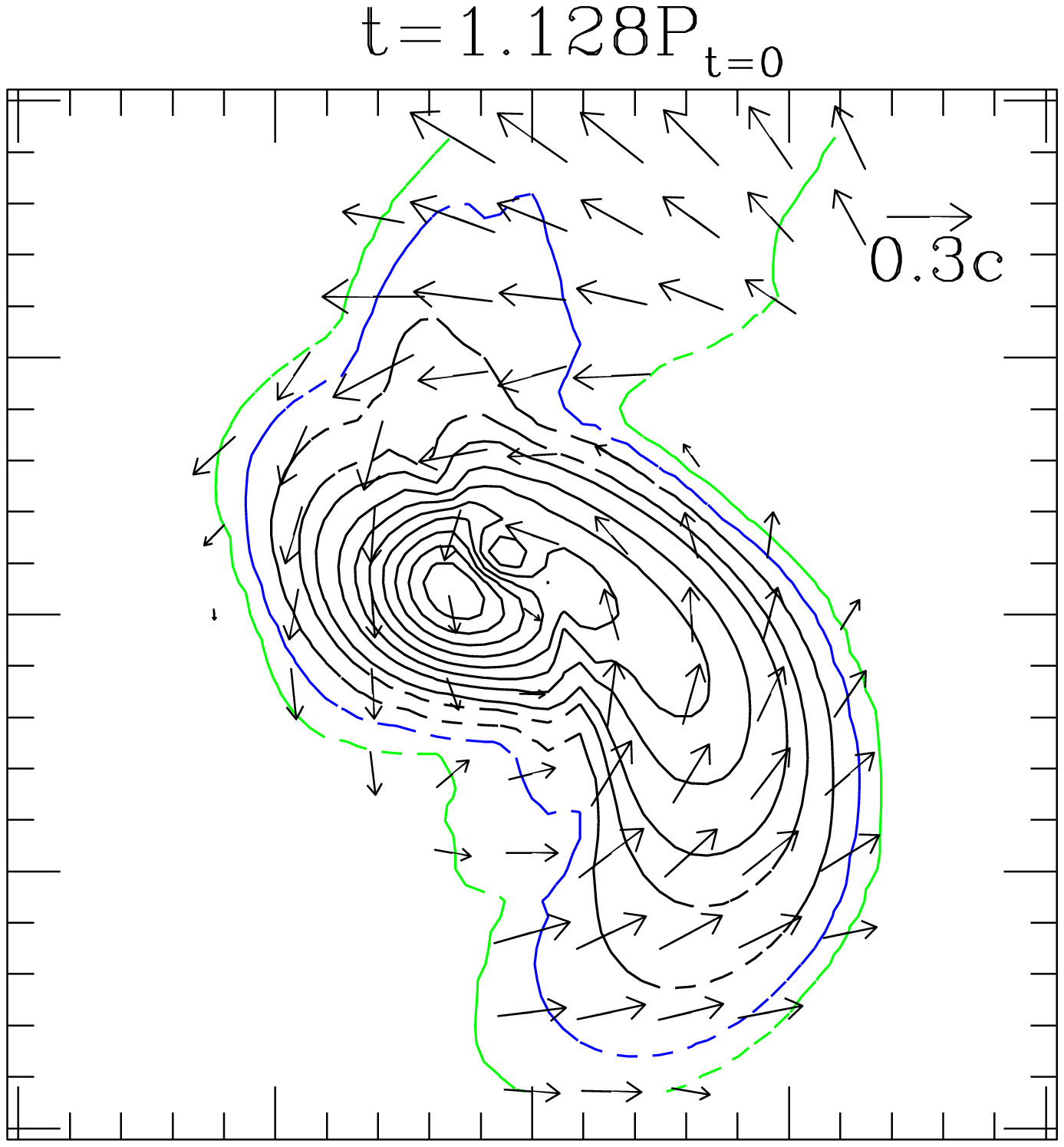}
\epsfxsize=2.4in
\leavevmode
\hspace{-1.2cm}\epsffile{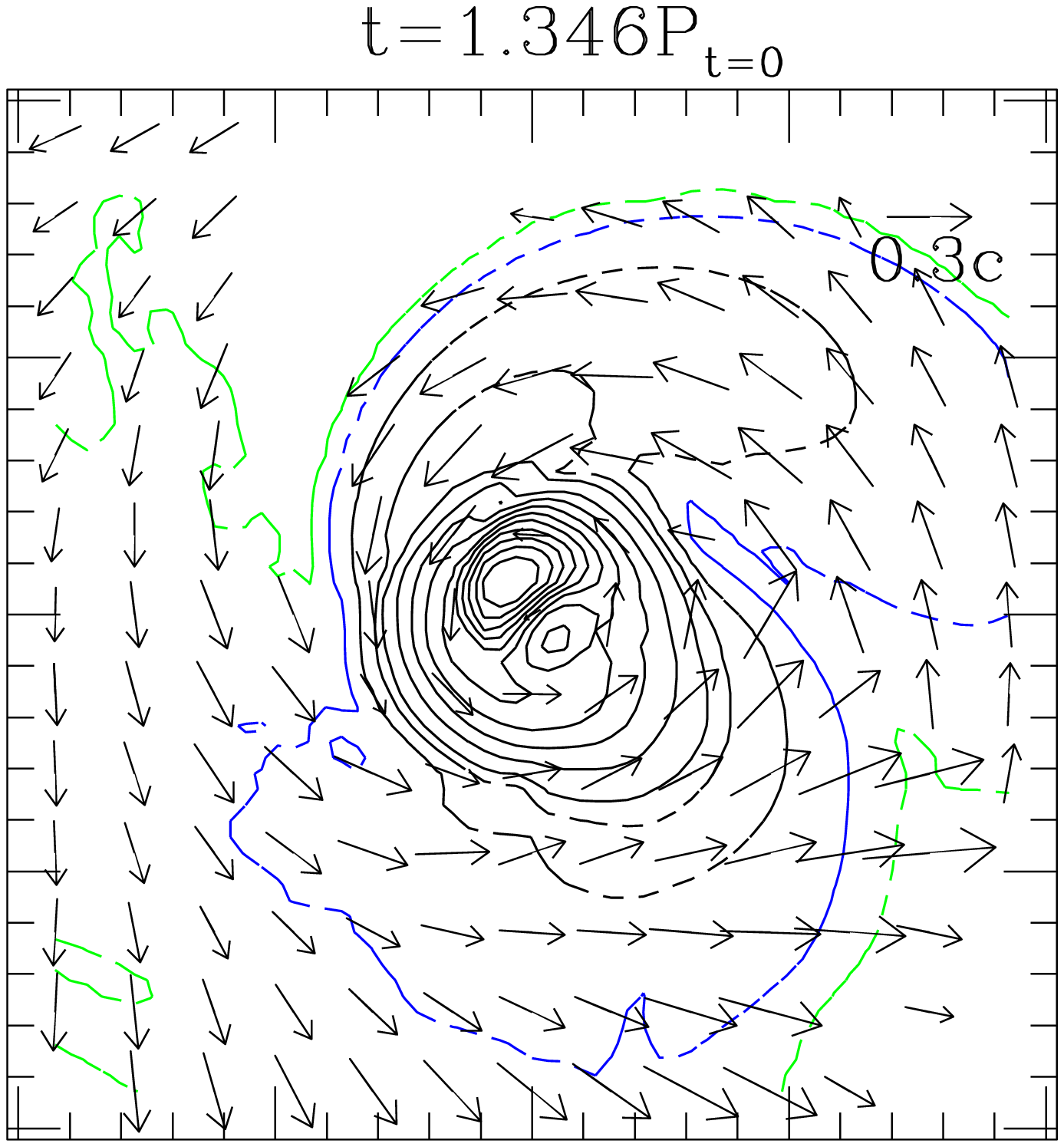}\\ 
\vspace{-1.2cm}
\epsfxsize=2.4in
\leavevmode
\epsffile{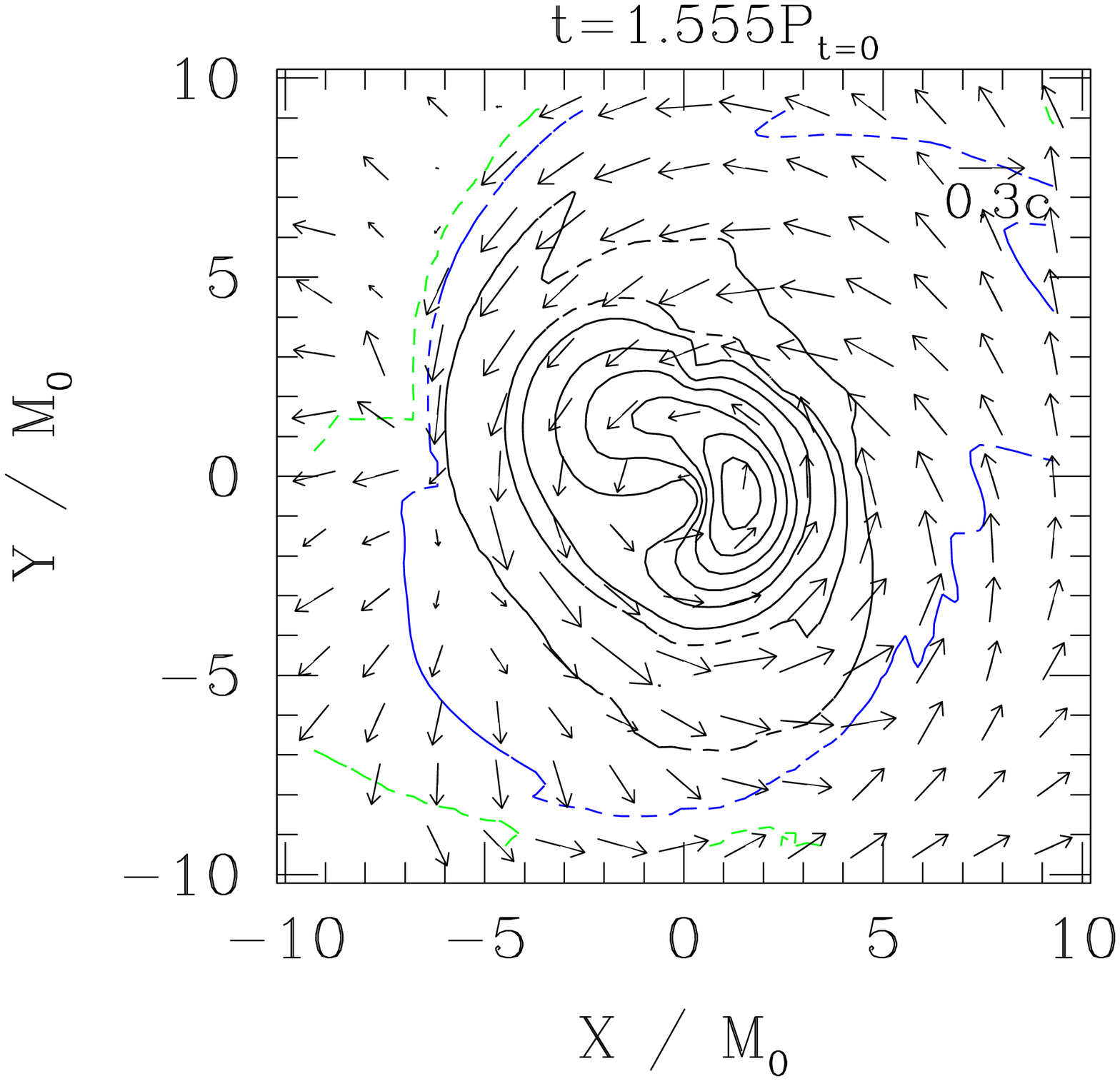} 
\epsfxsize=2.4in
\leavevmode
\hspace{-1.2cm}\epsffile{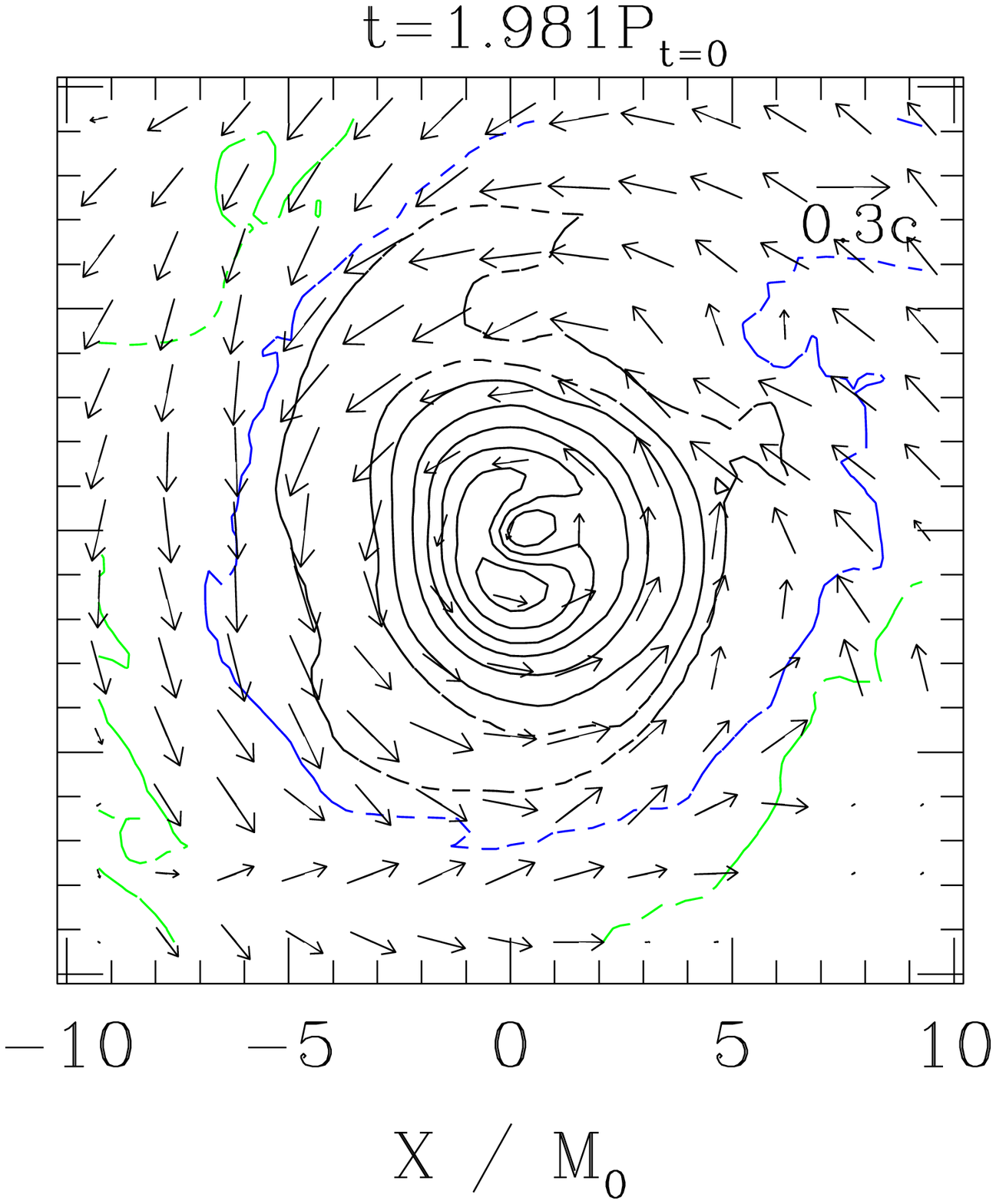}
\epsfxsize=2.4in
\leavevmode
\hspace{-1.2cm}\epsffile{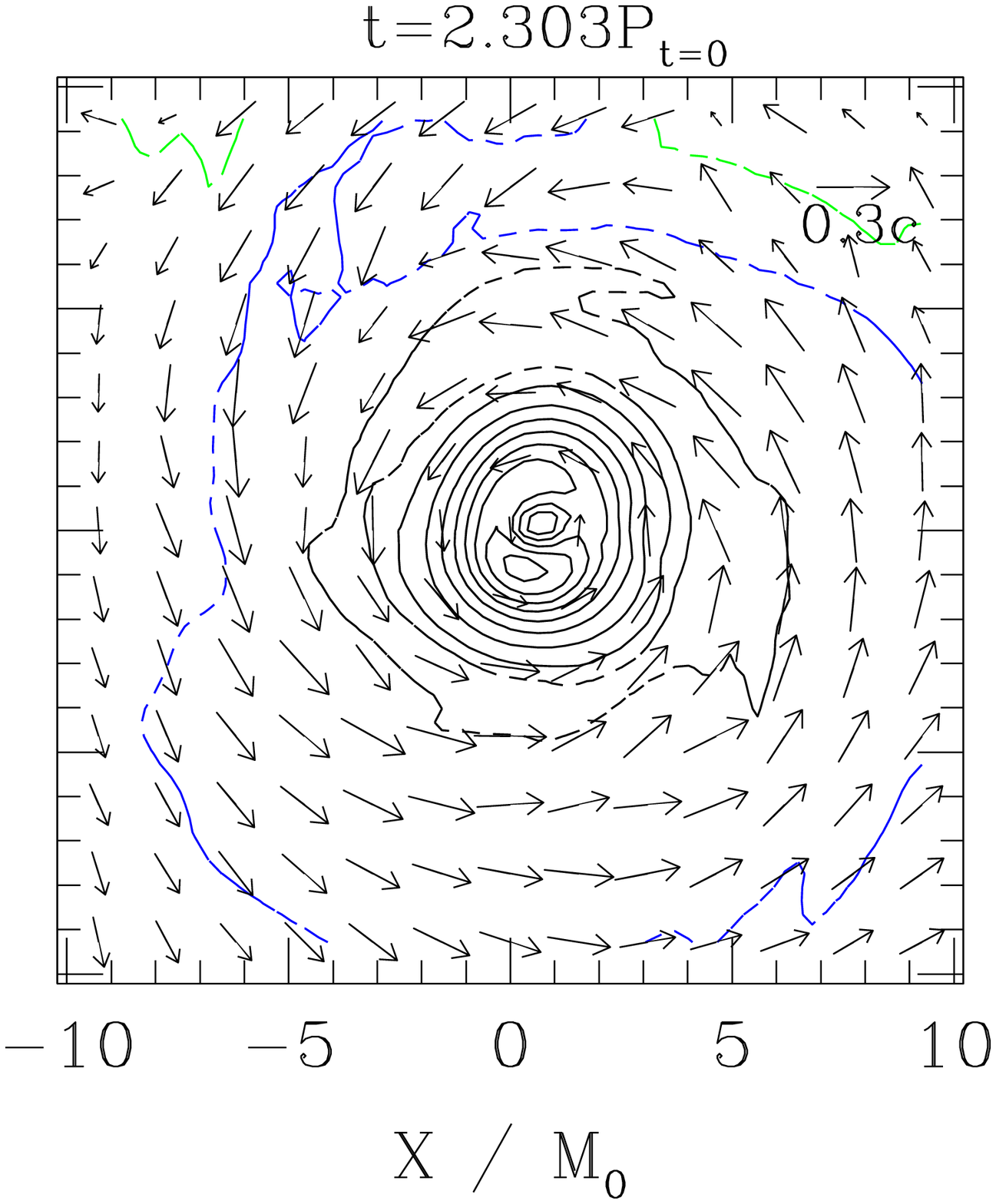} \\
\end{center}
\vspace{-5mm}
\caption{
Snapshots of the density contour curves for $\rho$ 
in the equatorial plane for model M1315. 
The solid contour curves are drawn for $\rho/0.15=1-0.1j$ 
for $j=0,1,2,\cdots,9$, and the dashed-solid curves are
for $\rho/0.15=0.05$, 0.01, $10^{-3}$ and $10^{-4}$.  
Vectors indicate the local velocity field $(v^x,v^y)$, and the scale 
is shown in the upper right-hand corner. $P_{\rm t=0}$ denotes the 
orbital period of the quasiequilibrium configuration given at $t=0$. 
The length scale is shown in units of $G M_{0}/c^2$, 
where $M_{0}$ is the gravitational mass computed at $t=0$. 
In the first panel, the primary neutron star is located at $x >0$. 
\label{FIG2} }
\end{figure}

\begin{figure}[t]
\begin{center}
\epsfxsize=2.4in
\leavevmode
\epsffile{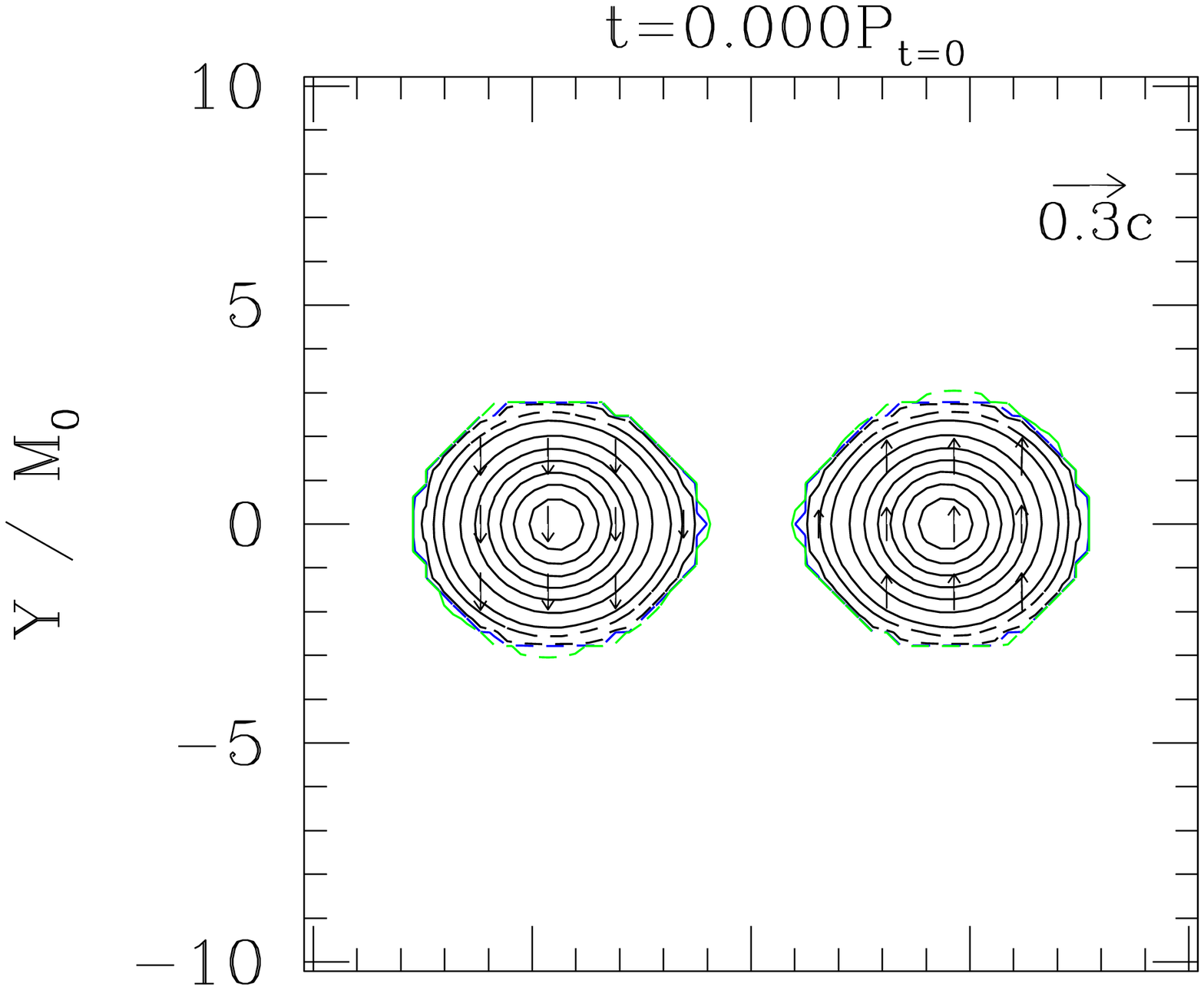}
\epsfxsize=2.4in
\leavevmode
\hspace{-1.2cm}\epsffile{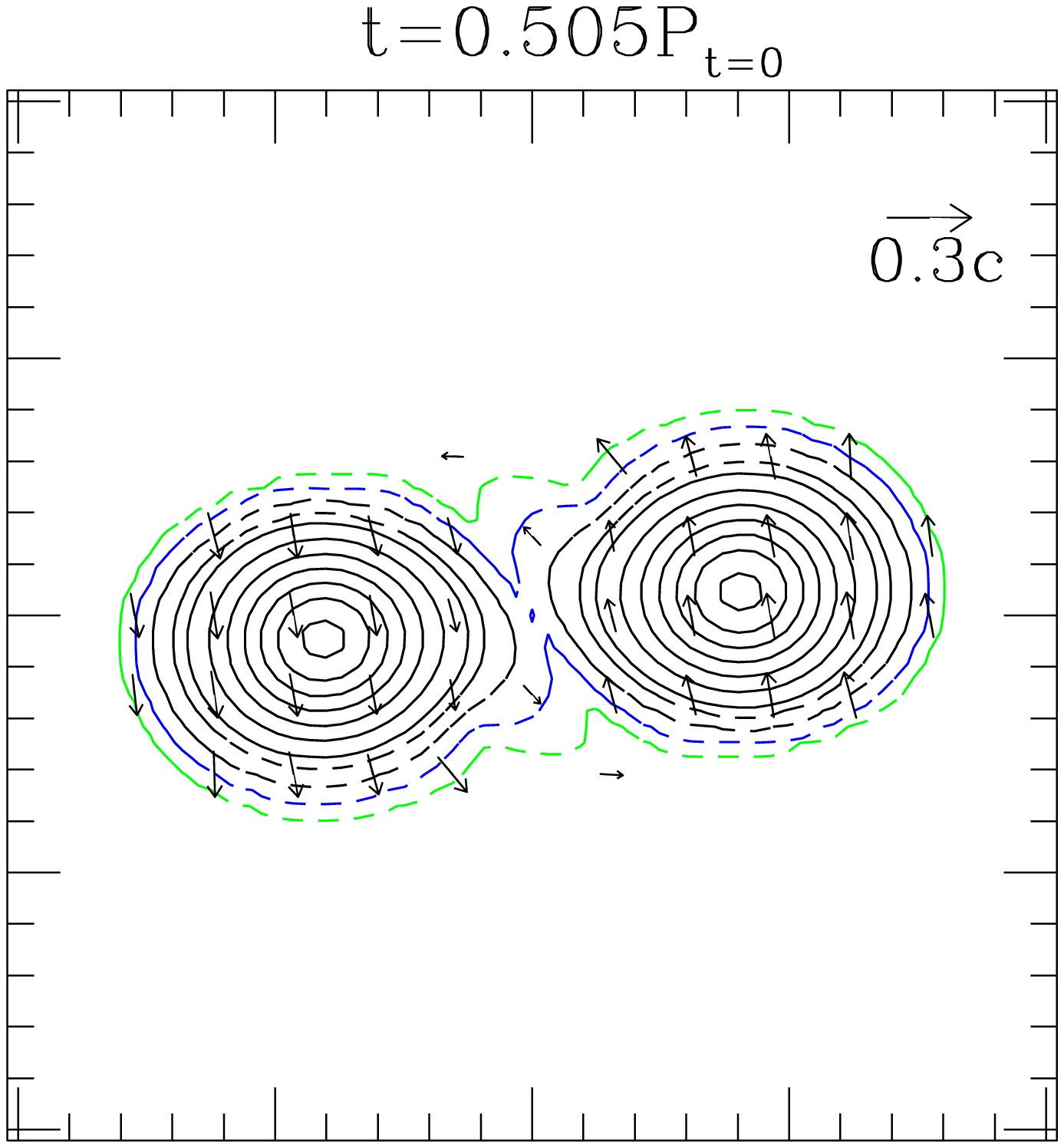} 
\epsfxsize=2.4in
\leavevmode
\hspace{-1.2cm}\epsffile{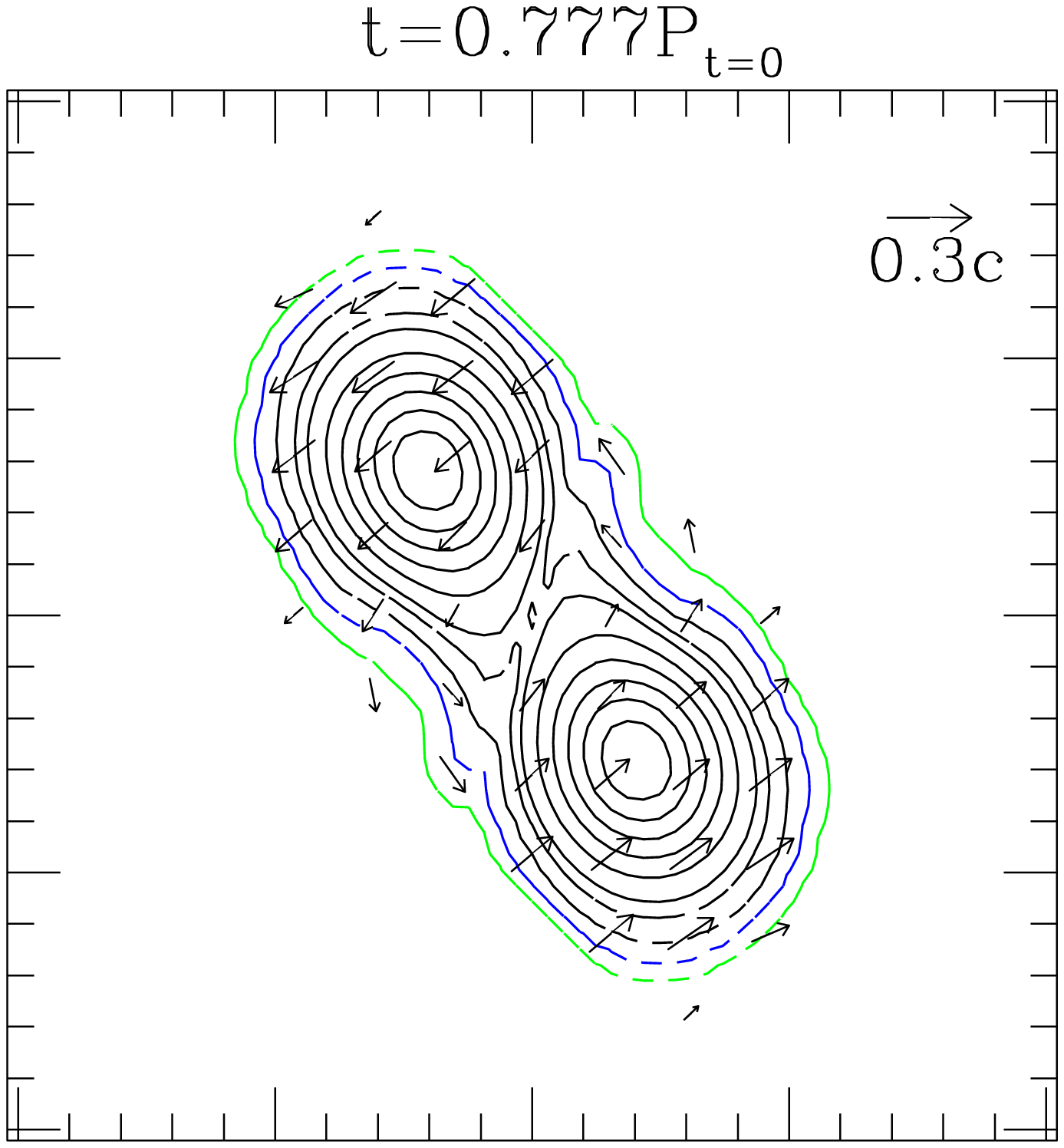} \\
\vspace{-1.2cm}
\epsfxsize=2.4in
\leavevmode
\epsffile{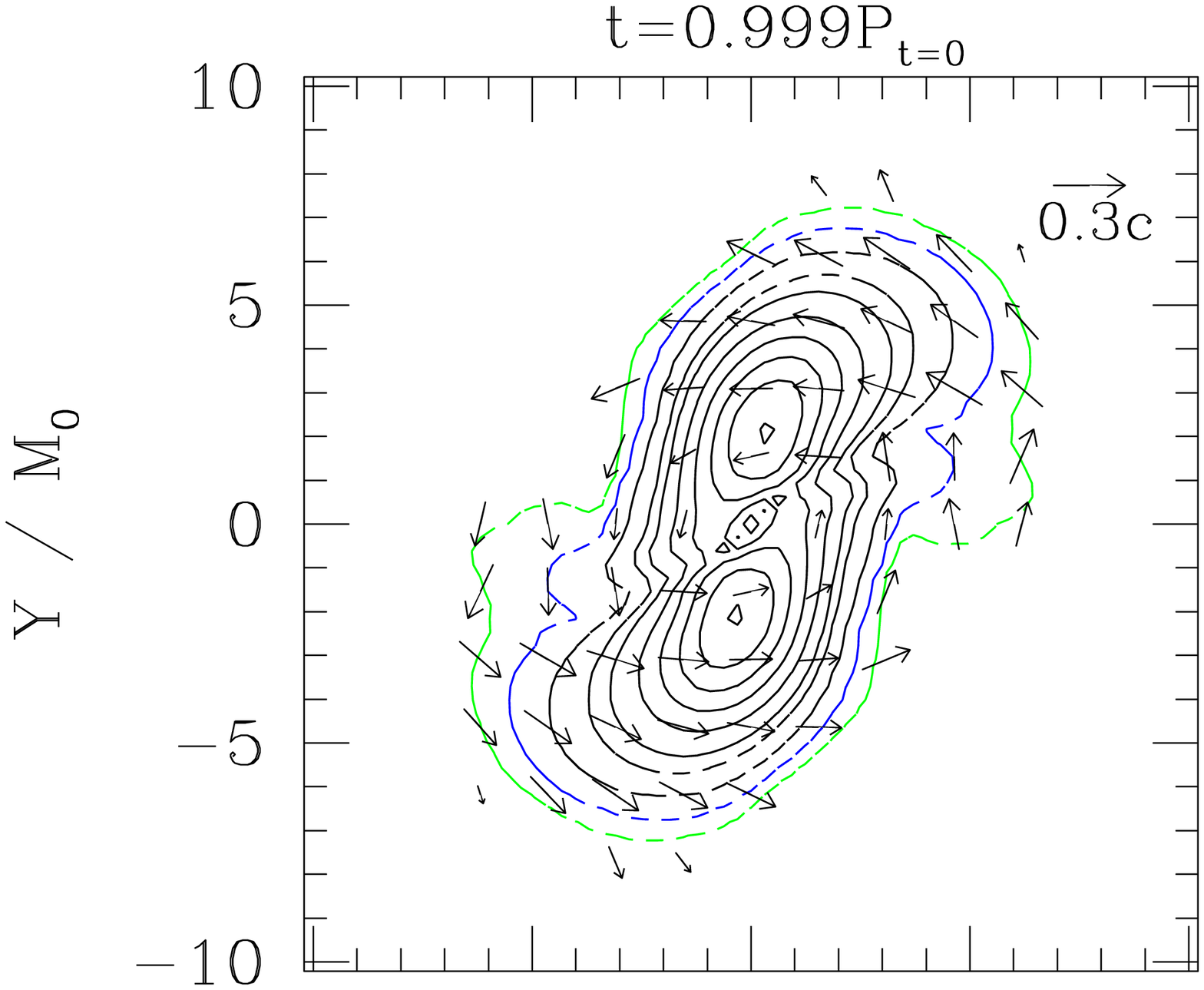} 
\epsfxsize=2.4in
\leavevmode
\hspace{-1.2cm}\epsffile{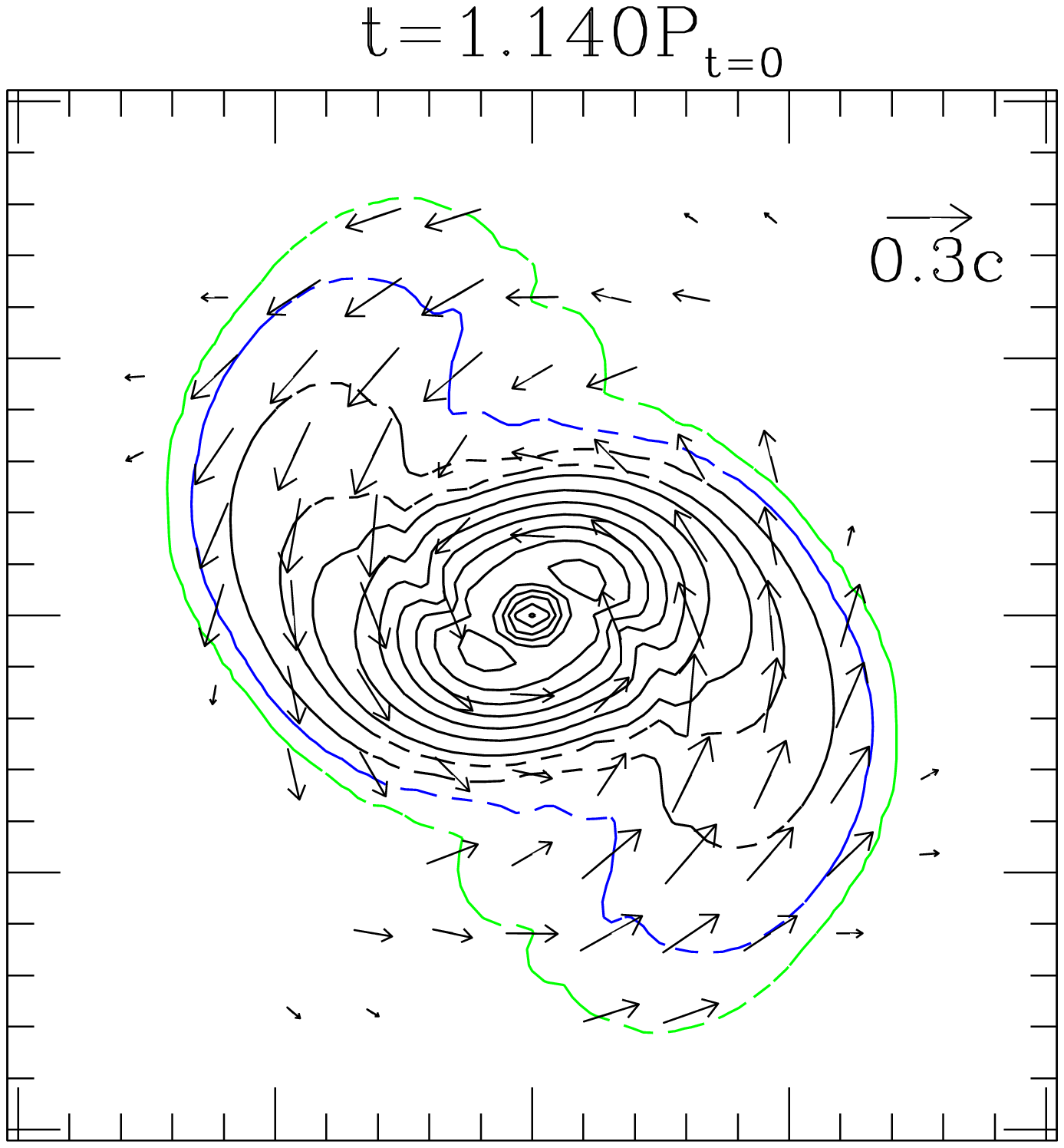}
\epsfxsize=2.4in
\leavevmode
\hspace{-1.2cm}\epsffile{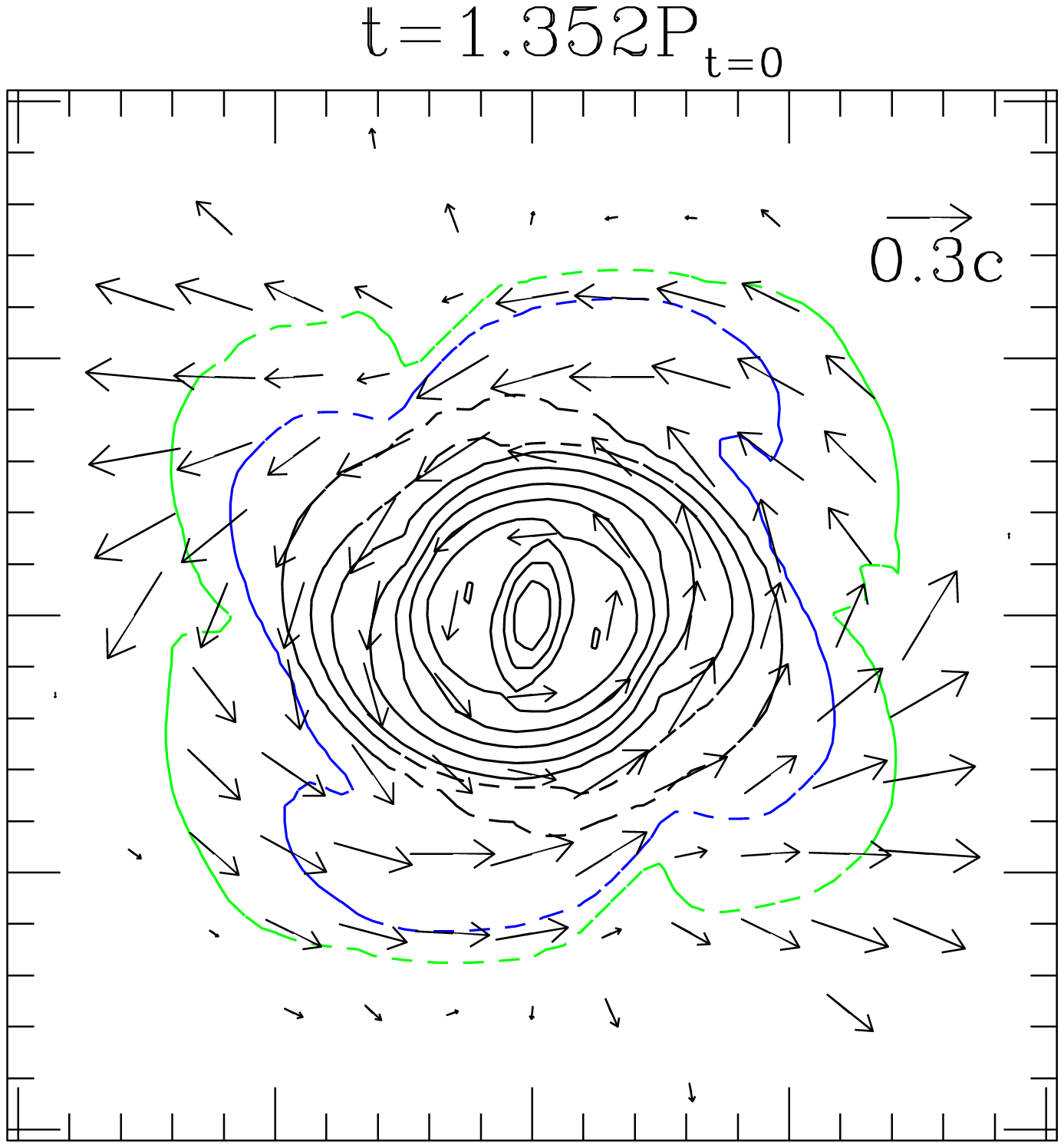}\\ 
\vspace{-1.2cm}
\epsfxsize=2.4in
\leavevmode
\epsffile{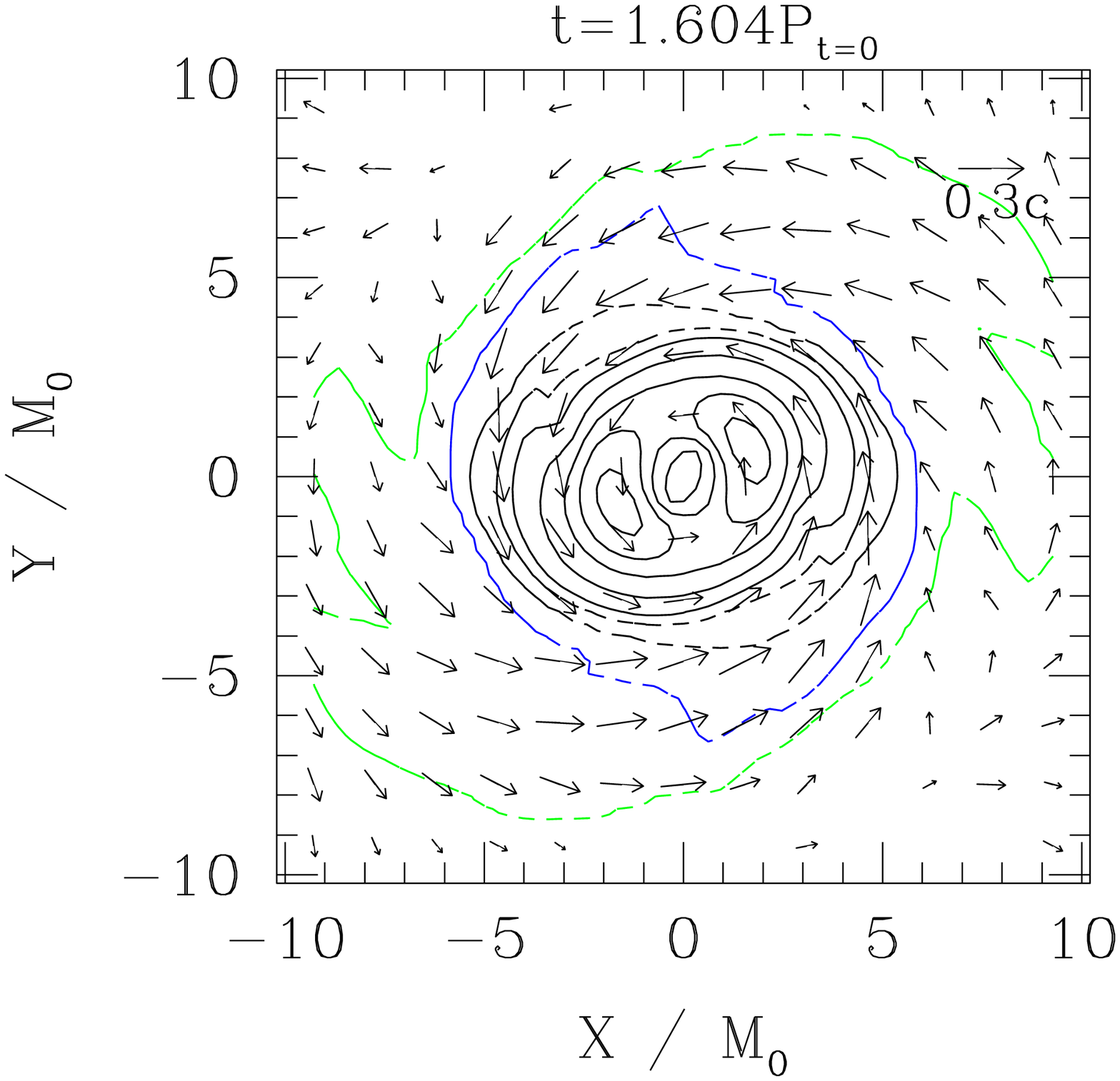} 
\epsfxsize=2.4in
\leavevmode
\hspace{-1.2cm}\epsffile{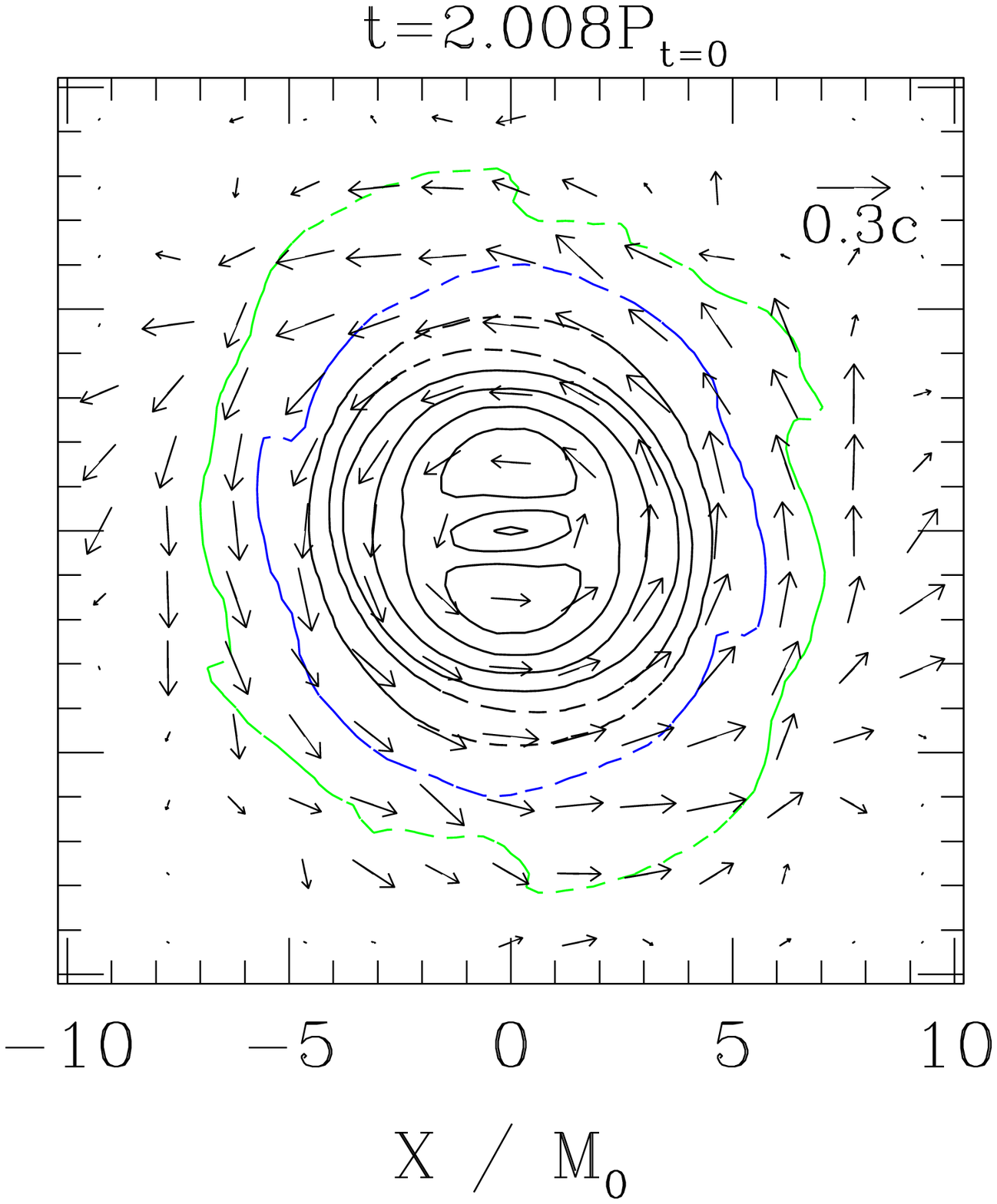}
\epsfxsize=2.4in
\leavevmode
\hspace{-1.2cm}\epsffile{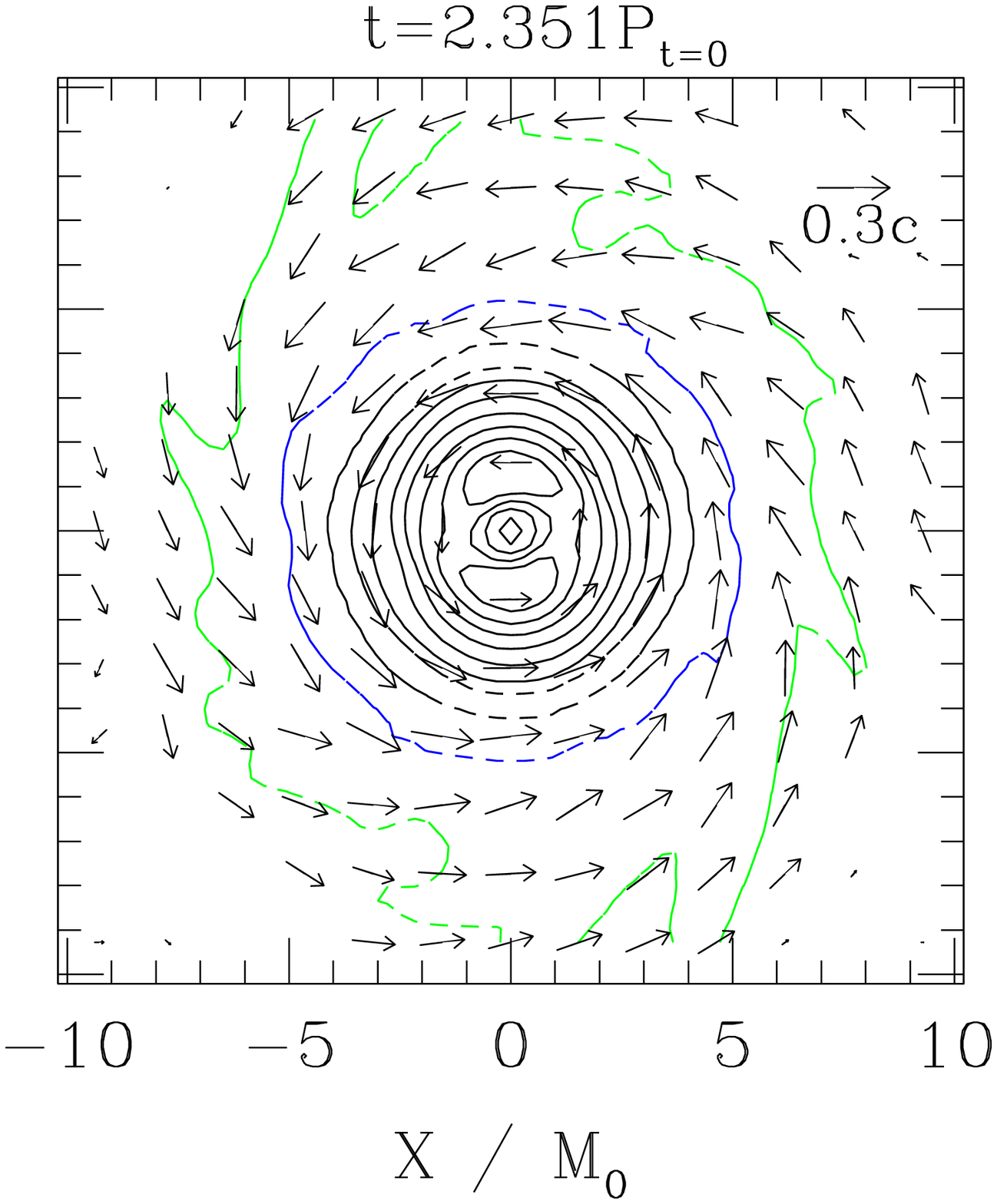} \\
\end{center}
\vspace{-5mm}
\caption{
The same as Fig. \ref{FIG2} but for model M1414. 
\label{FIG3} }
\end{figure}

In Figs. \ref{FIG2}--\ref{FIG5},
we display the snapshots of the density contour curves and
the velocity vectors at selected time steps for models 
M1315, M1414, M1418, and M1616 \cite{anim}.
Figure \ref{FIG6} shows the evolution of the maximum values of
$\rho$ and $\phi$, and the minimum value of $\alpha$ for
all the models adopted in this paper except for model M159183. 

The numerical results for models M1414 and M1616 
computed by an old implementation have been already presented
in \cite{bina,bina2}. 
With the present new implementation, however, the quality of the 
numerical results is significantly improved and, hence, 
the improved results are displayed as the updated ones. 

The simulations for models M1616, M1517, and M1418 
crashed soon after the formation of apparent horizons because
the black hole forming region was stretched significantly and 
the grid resolution became too poor to resolve such a region. 
On the other hand, we artificially stopped
the simulations for models M1414 and M1315 at $t \agt 3.5P_{t=0}$
to save the computational time. At the termination of these 
simulations, the averaged violation of the Hamiltonian 
constraint does not increase rapidly and remains of order 0.1. Therefore 
the simulations could be continued for a much longer time 
than $3.5P_{t=0}$ in the formation of the massive neutron stars. 

In every model, the merger is triggered by the radiation reaction:
For $t \alt P_{t=0}$, the orbital separation decreases gradually 
as a result of gravitational radiation reaction and each neutron star 
is elongated little by little.
The elongation is always larger for the smaller-mass star in
nonequal mass binaries. At a critical separation which is reached
at $t \sim 0.8$--0.9$P_{t=0}$, the orbit becomes unstable probably
against hydrodynamic instability to start the merger.
At this point, the lag angle which is defined to be the
angle in the equatorial plane between the major axis of
each star and the axis connecting the centers of mass of
two stars \cite{FR2} is $\sim 10$--15 degree.
In unequal-mass binaries, we always find a larger lag angle
for the smaller-mass star. This is in agreement with that in \cite{FR2}. 

In contrast to previous Newtonian \cite{RS,C},
post-Newtonian \cite{FR,FR2}, 
and approximately relativistic simulations \cite{ORT}, the 
formation of a black hole can be determined in fully general
relativistic simulations. In the black hole formation,
most of the fluid elements are swallowed into the black hole.
Therefore the evolution of the system  
is significantly different from that in the neutron star formation. 
From this reason, we describe the character of the merger
for the formation of neutron stars and black holes separately. 

\subsubsection{Formation of hypermassive neutron star}

For models M1315 and M1414, massive neutron stars are formed
(cf. Figs. \ref{FIG2} and \ref{FIG3}). 
In this case, the total baryon rest-mass of the system is 
about 1.62 times as large as $M_{*\rm max}^{\rm sph}$ for a 
polytropic equation of state with $n=1$. Thus the formed neutron star
is hypermassive in the sense that the mass is larger than the
maximum allowed value for rigidly rotating neutron stars with
$n=1$ \cite{BSS}. As indicated in the previous papers \cite{bina,bina2},
such a large mass is supported by a large centrifugal force 
due to rapid and differential rotation.

In \cite{bina,bina2}, we concluded that the 
merged object for model M1414 eventually collapses to a black hole 
in 2--3 $P_{t=0}$. However, in the present improved simulation, a 
hypermassive neutron star is formed instead of a black hole.
There are two plausible reasons for this discrepancy.
One is that shocks are calculated in a better accuracy with the new
implementation of a high-resolution shock-capturing scheme and, 
as a result, the thermal energy, which could play an important role for 
supporting the massive object, is increased in the 
new result. The other possible reason is 
an improvement on the treatment of the transport term for geometric
variables. This makes the angular momentum conservation 
more accurate, avoiding the spurious collapse. 

Besides the correction to the threshold for collapse of
merged objects, 
the qualitative properties during the merger of equal-mass
binaries are essentially the same
as those found in the previous simulations \cite{bina,bina2}: 
The merged object constitutes a double core structure soon after
the merger [cf. fifth panel of Fig. \ref{FIG3} and Fig. \ref{FIG7}(b)].
At the collision of two neutron stars, the radial infall velocity
is not so large that shock heating is not very important
around the mass center. In Fig. \ref{FIG8}, we show $\kappa'$ along the 
$x$ and $y$ axes, which denotes the efficiency of the shock heating.
In our notation, it is unity everywhere inside the neutron stars at $t=0$.
Thus the fluid elements for which the value of $\kappa'$
is larger than unity have 
experienced the shock heating. This figure shows that 
for $r \alt 4M_0$, $\kappa'$ is $\sim 1$--2, implying that the shocks 
do not play a very important role except for the outer envelops. 

In the outer region, small spiral arms are formed soon after the
merger sets in, but they do not 
spread outward widely because of insufficient angular momentum. 
The spiral arms subsequently wind around a central double core.
In the central region, the double core has not only a rotational
motion around the mass center but also has a quasiradial oscillation
which is originally excited by a radial plunge at a 
transition from the inspiral to the merger stage. Because of the 
quasiradial oscillation, weak shocks are formed in the outer envelops. 
As a result, the outer region is heated up and gains the 
kinetic energy to expand outward [cf. Fig. \ref{FIG8}(b)]. 
This process is repeated many times transferring 
the kinetic energy of the inner core to the outer region and, hence, 
the quasiradial oscillation of the core damps gradually
(see Fig. \ref{FIG6}).

For the merger of an unequal-mass binary (for model M1315),
the merger process is qualitatively different from that 
in the equal-mass case because tidal disruption of the smaller-mass star 
by the massive primary takes place (cf. fourth panel of Fig. \ref{FIG2}).
The tidally disrupted star subsequently forms a tidal tail.
During the formation of the tidal tail, the angular momentum
is efficiently transfered outward and, as a result, large 
spiral arms are formed. The spiral arms subsequently
wind around the central core to be accretion disks. 
Because of the angular momentum transfer at the tidal disruption
and at subsequent formation of spiral arms, 
the disk radius is much larger than that for model M1414
(cf. Fig. \ref{FIG9}). 

In the central region, a massive object with an asymmetric double core 
is formed [cf. the last panel of Fig. \ref{FIG2} and Fig. \ref{FIG7}(a)]. 
As in model M1414, the central core oscillates quasiradially 
(see Fig. \ref{FIG6}). 
This motion produces shocks around the outer part of the core and, 
as a result, the energy is transfered to the outer envelops 
[cf. Fig. \ref{FIG8}(a)]. Since 
this process is repeated, the quasiradial oscillation of the 
core damps gradually. The amplitude of the quasiradial oscillation 
for model M1315 is not as large as that for model M1414. 
This reflects the difference of the merger process between 
M1414 and M1315: For M1414, two neutron stars merge 
without tidal disruption and mass ejection outward.
Therefore all the mass elements in this system collide coherently. 
On the other hand, for M1315, the tidal disruption takes place. 
As a result, a fraction of mass elements in the smaller-mass
star do not have a plunging motion at the collision 
and, hence, the merger does not set in
as coherent as that for model M1414. Due to this reason, 
the amplitude of the quasiradial oscillation is suppressed. 

Figure \ref{FIG6}(a) shows that the maximum
density of the hypermassive neutron star
for model M1315 is larger than that for M1414 in spite of the fact
that the total baryon-rest mass is nearly identical.
This reflects the fact that the region around the mass center 
for model M1315 rotates less rapidly than that for M1414. This suggests 
that the hypermassive neutron stars formed from the merger of
the smaller rest-mass ratios are more compact. 

In Fig. \ref{FIG9}, we plot the evolution of the baryon
rest-mass fraction outside the spheres of radius
$3M_0$ (solid curve), $4.5M_0$ (dashed curve) and
$6M_0$ (dotted-dashed curve).
Here, $r=0$ is chosen as the center of the spheres. 
This shows the significance of the
angular momentum transfer for model M1315. For model M1414, the
baryon rest-mass outside the spheres of fixed radii 
simply oscillates with a mean value which is approximately constant 
with time evolution. The fraction of the rest-mass outside the
sphere of radius $6M_0$ is $\sim 1\%$ in this case.
On the other hand, for M1315, the 
baryon rest-mass outside spheres of fixed radii increases gradually 
with time. This result reflects an efficient angular momentum transfer. 
Figure \ref{FIG9} indicates that 
the fraction of the rest-mass outside the 
sphere of radius $6M_0$ is $\sim 5\%$, implying that disks
of $\agt 0.1M_{\odot}$ are formed around the hypermassive neutron star. 

A post-Newtonian simulation reports that the fraction of the disk mass 
for an equal-mass merger is $\sim 6\%$ \cite{FR2}. 
This value is much larger than 
the value obtained in this paper. A plausible reason for this discrepancy 
is that in the post-Newtonian approximation,
the gravity of the hypermassive neutron star is underestimated
and, hence, the mass captured by it is also underestimated. 

Because of the nonaxisymmetric and quasiradial oscillations of the
hypermassive neutron stars, quasiperiodic gravitational waves
of a few characteristic oscillation modes are simultaneously
excited for models M1315 and M1414 for a long duration after the
merger. This point will be discussed in Sec. IV C. 

\subsubsection{Formation of rotating black hole}

\begin{figure}[t]
\begin{center}
\epsfxsize=2.4in
\leavevmode
\epsffile{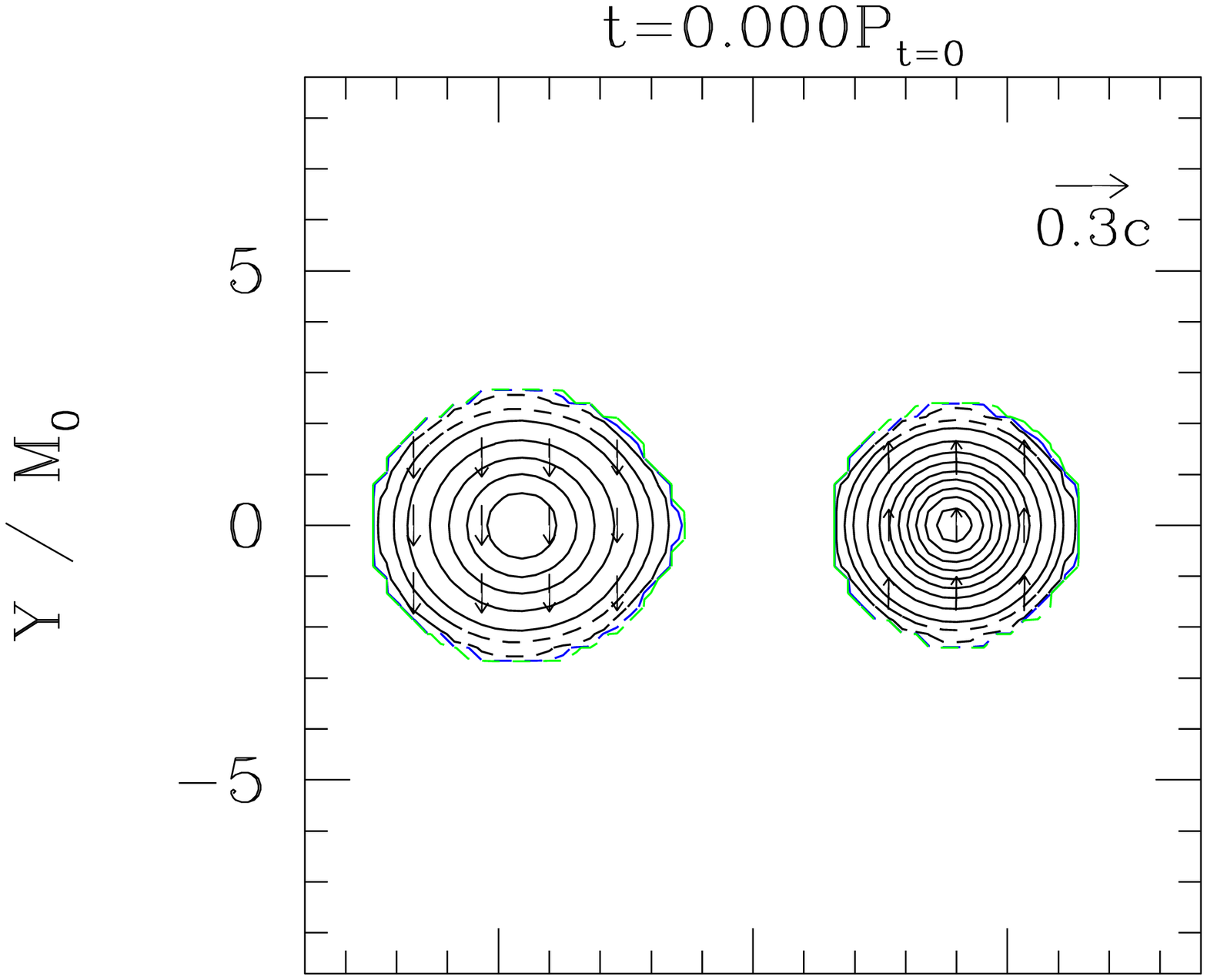}
\epsfxsize=2.4in
\leavevmode
\hspace{-1.2cm}\epsffile{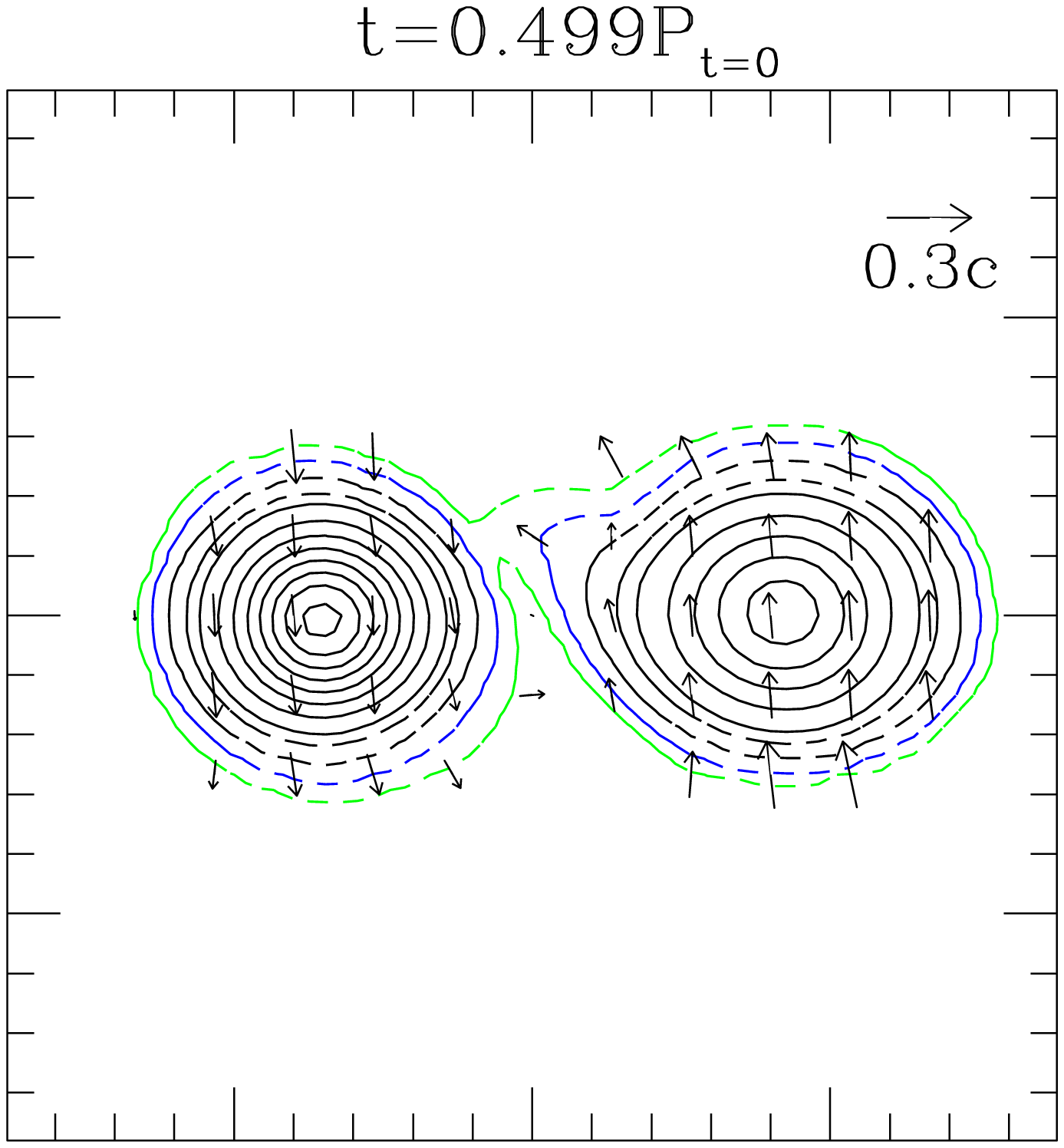} 
\epsfxsize=2.4in
\leavevmode
\hspace{-1.2cm}\epsffile{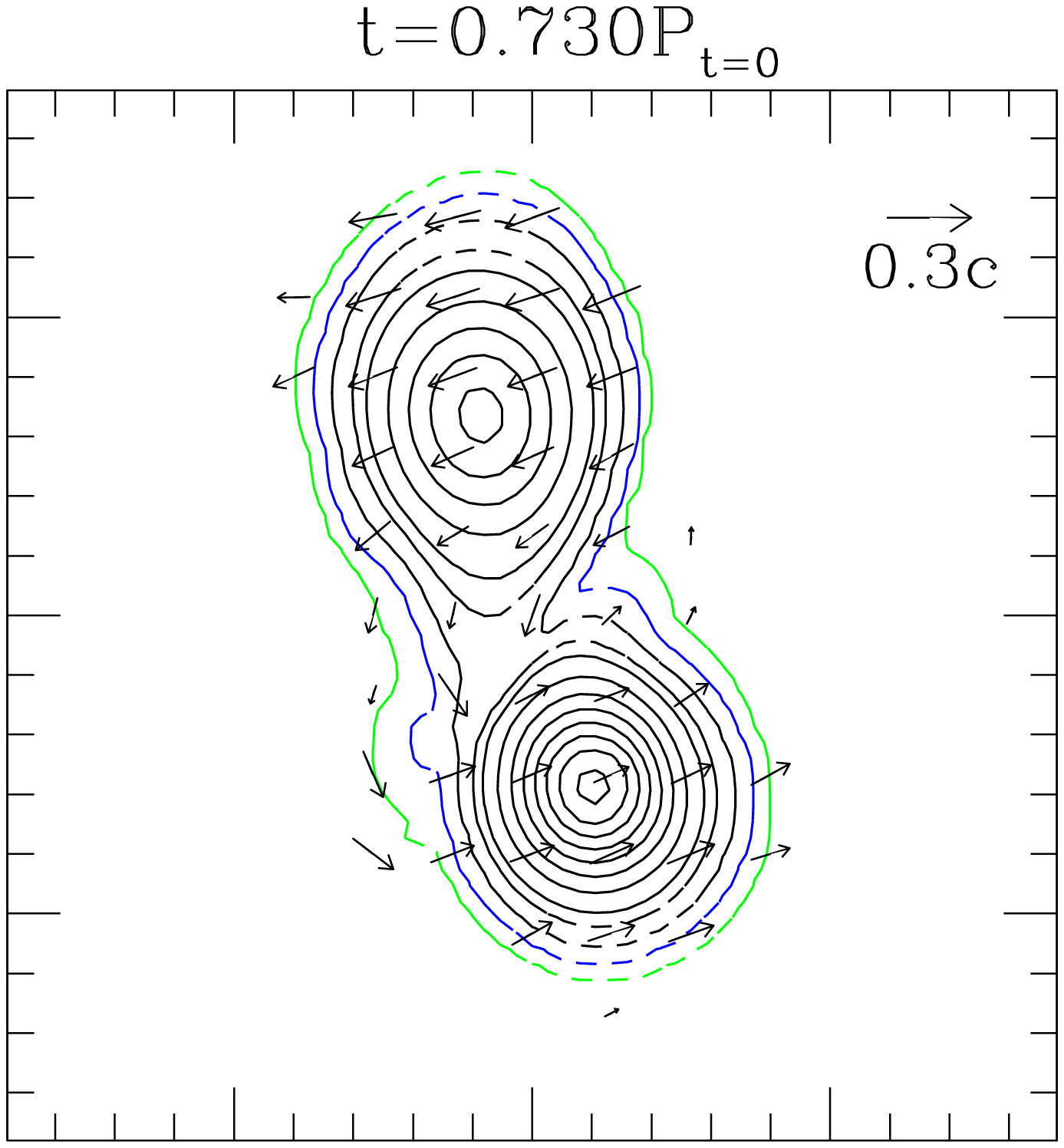} \\
\vspace{-1.2cm}
\epsfxsize=2.4in
\leavevmode
\epsffile{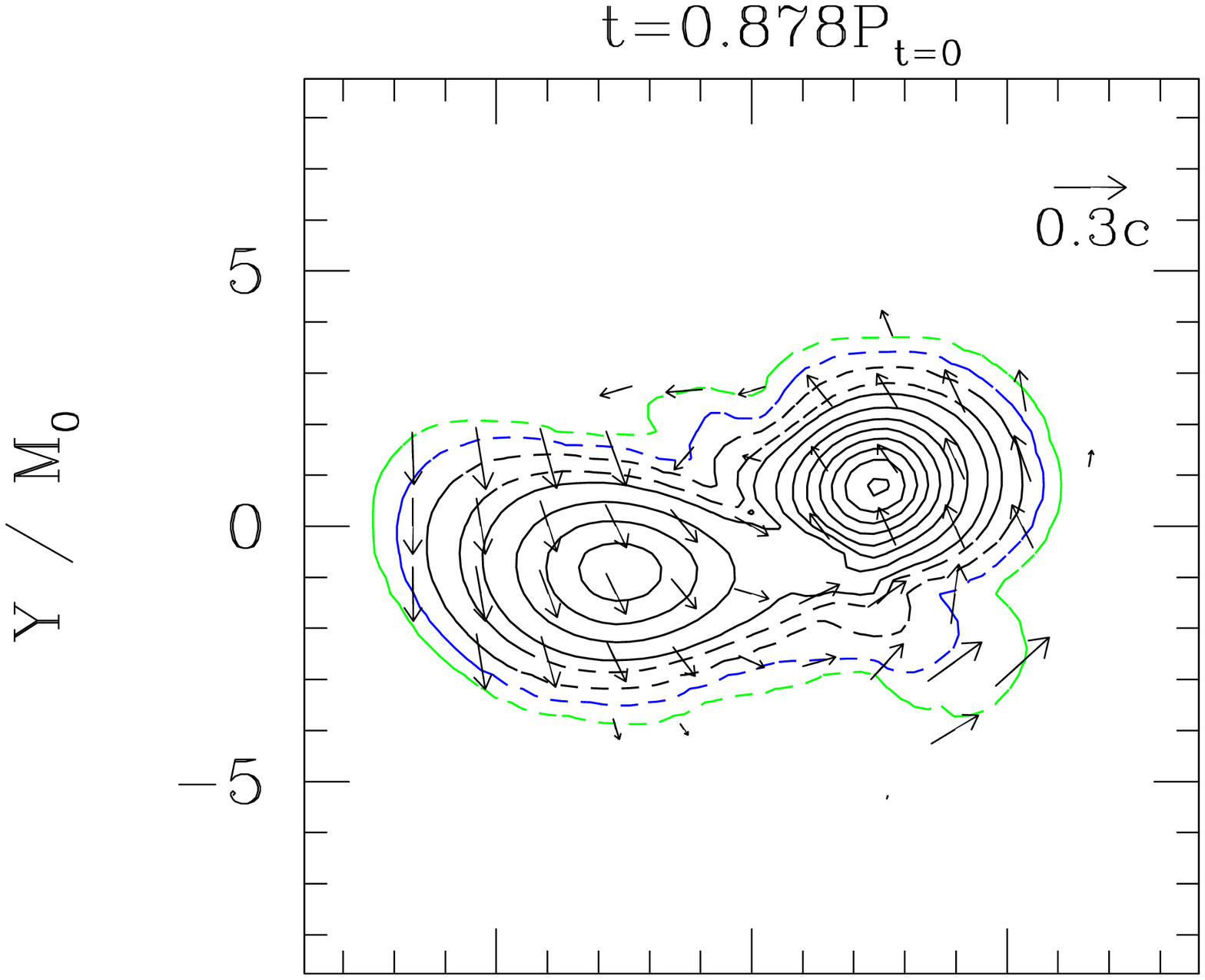} 
\epsfxsize=2.4in
\leavevmode
\hspace{-1.2cm}\epsffile{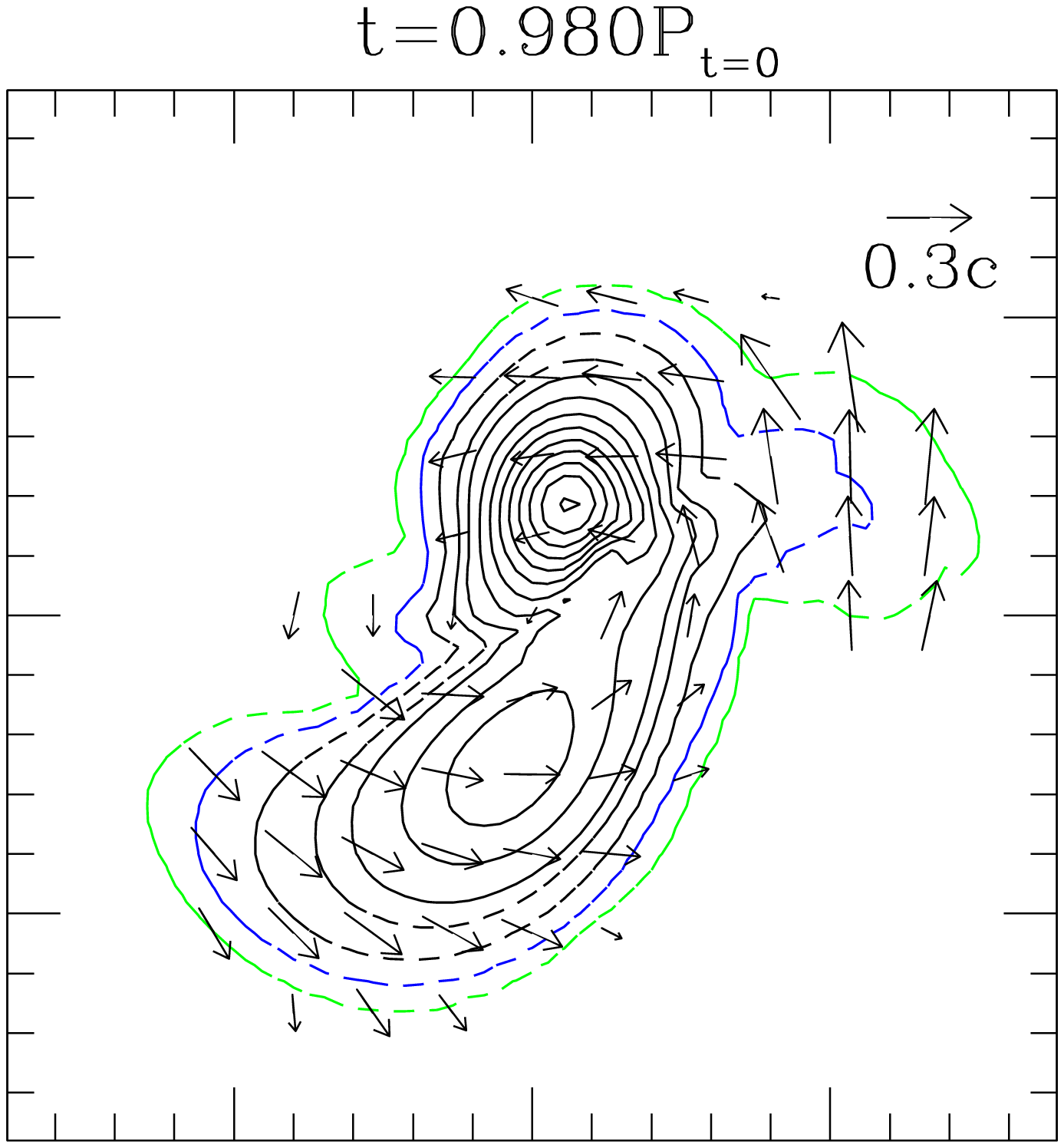}
\epsfxsize=2.4in
\leavevmode
\hspace{-1.2cm}\epsffile{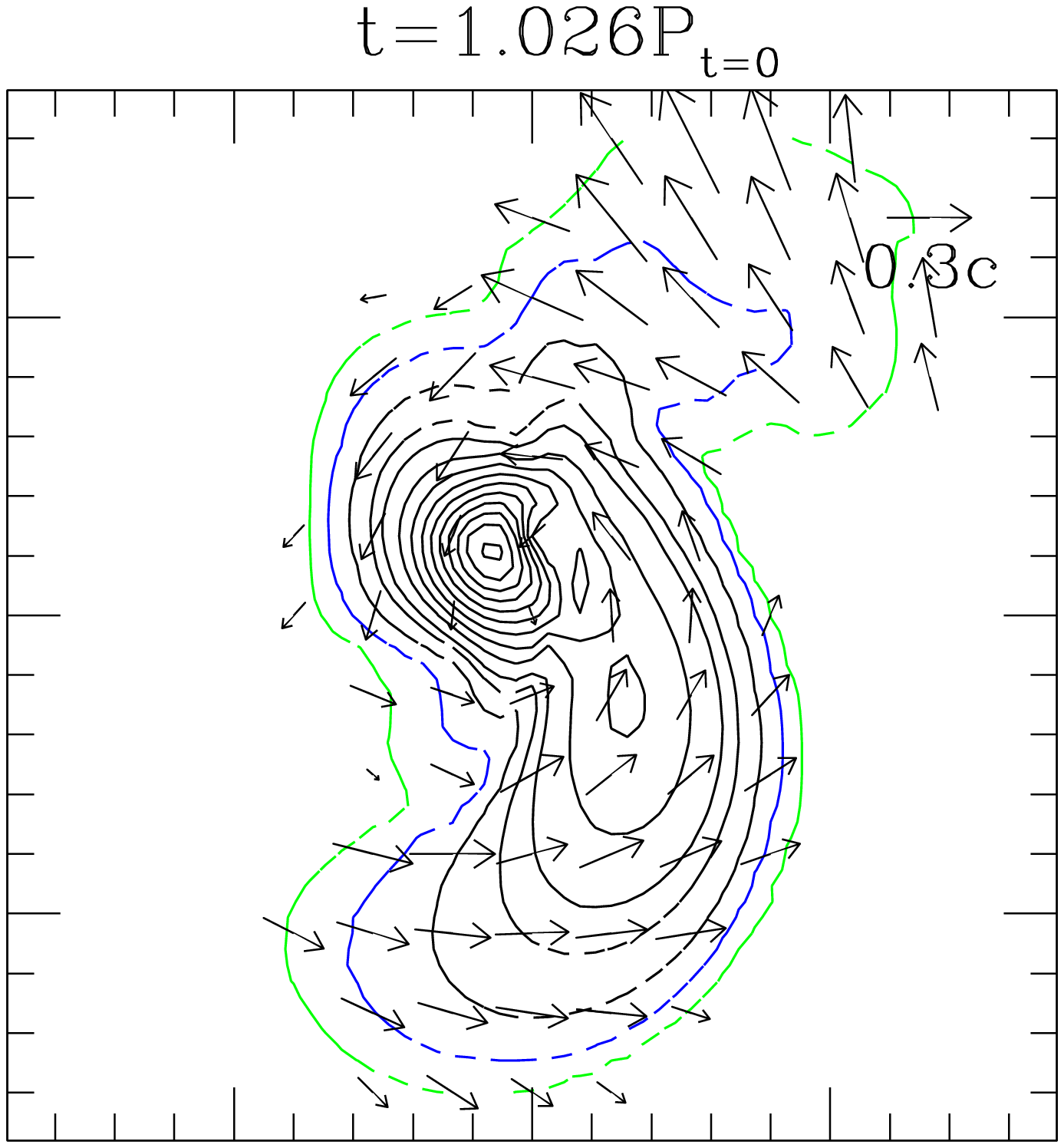}\\ 
\vspace{-1.2cm}
\epsfxsize=2.4in
\leavevmode
\epsffile{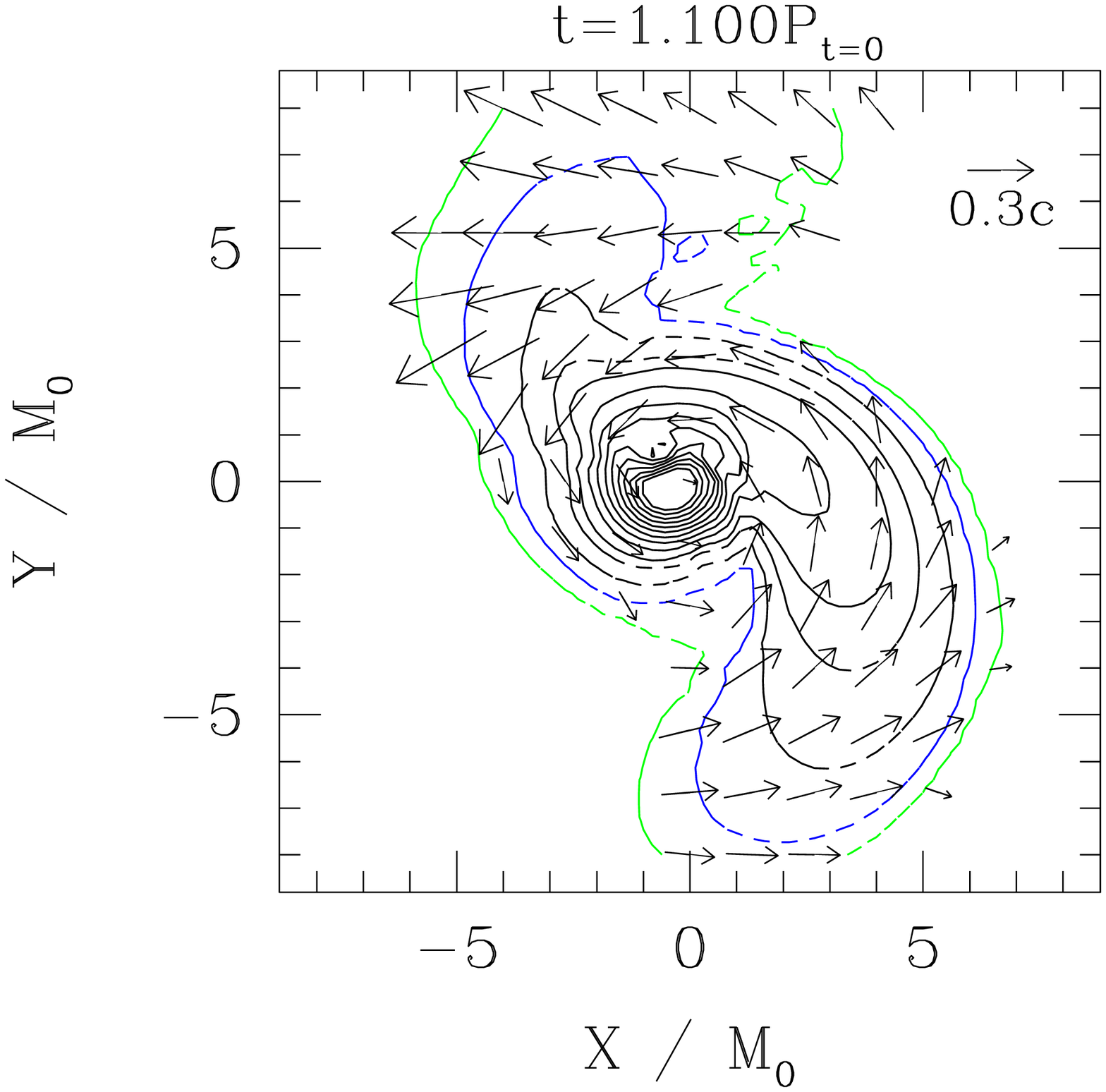} 
\epsfxsize=2.4in
\leavevmode
\hspace{-1.2cm}\epsffile{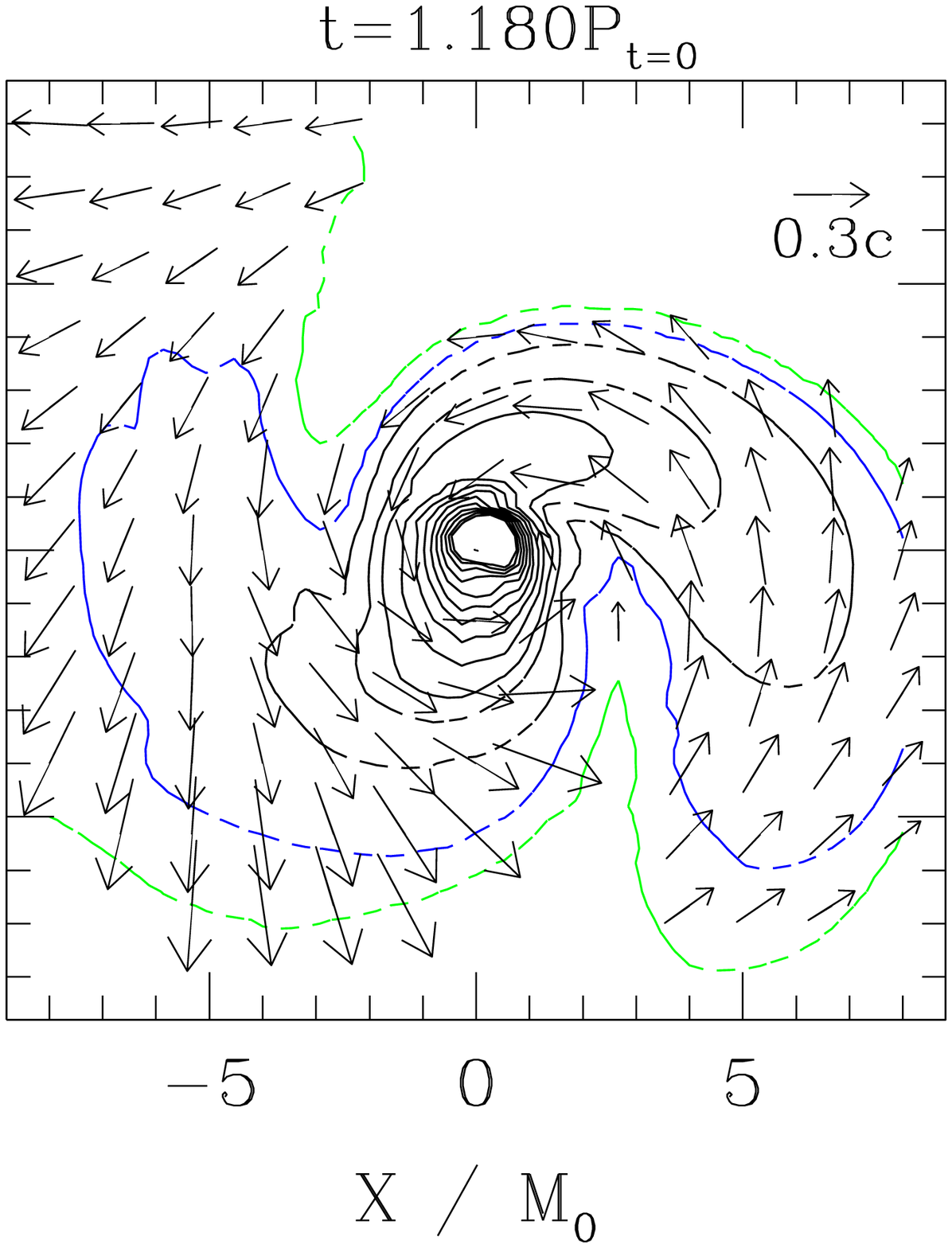}
\epsfxsize=2.4in
\leavevmode
\hspace{-1.2cm}\epsffile{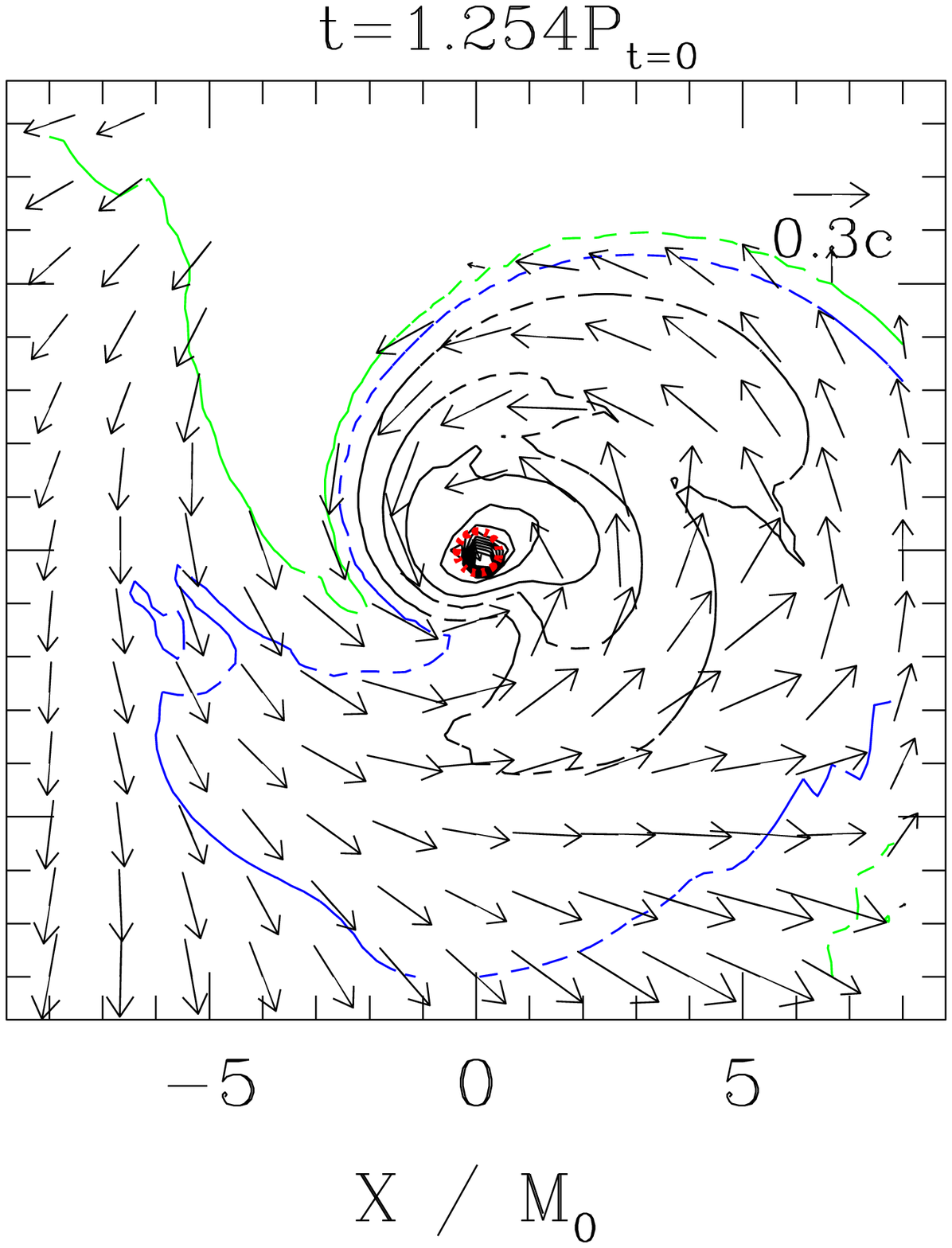} 
\end{center}
\vspace{-5mm}
\caption{
The same as Fig. 2 but for model M1418. 
Here, the solid contour curves are drawn for $\rho/0.20=1-0.1j$ 
for $j=0,1,2,\cdots,9$, and the dashed-solid curves are
for $\rho/0.20=0.05$, 0.01, $10^{-3}$ and $10^{-4}$.  
The thick dotted circle in the last panel of radius $r \sim 0.5M_0$
denotes the location of the apparent horizon. 
At $t=0$, the primary neutron star is located at $x >0$. 
\label{FIG4} }
\end{figure}

\begin{figure}[t]
\begin{center}
\epsfxsize=2.4in
\leavevmode
\epsffile{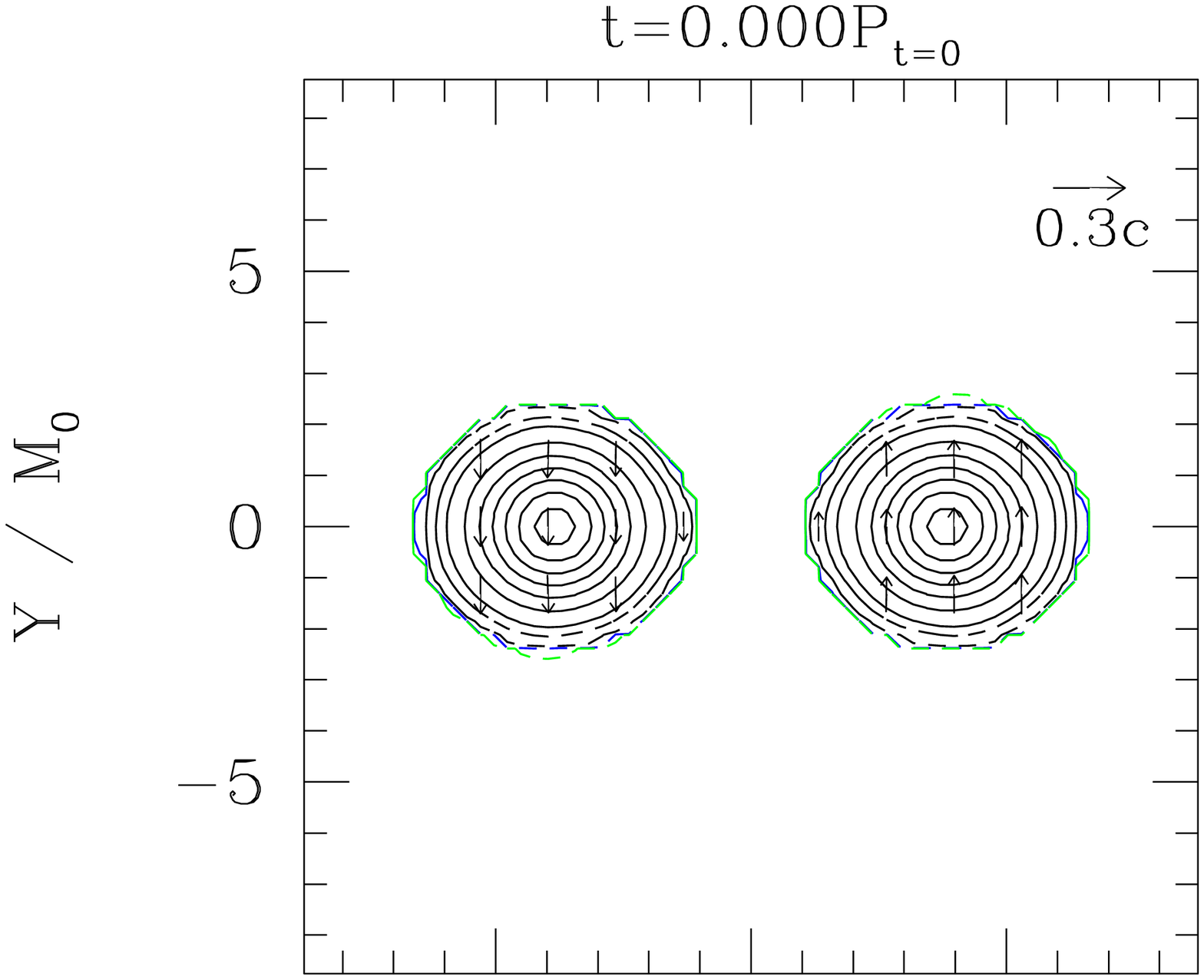}
\epsfxsize=2.4in
\leavevmode
\hspace{-1.2cm}\epsffile{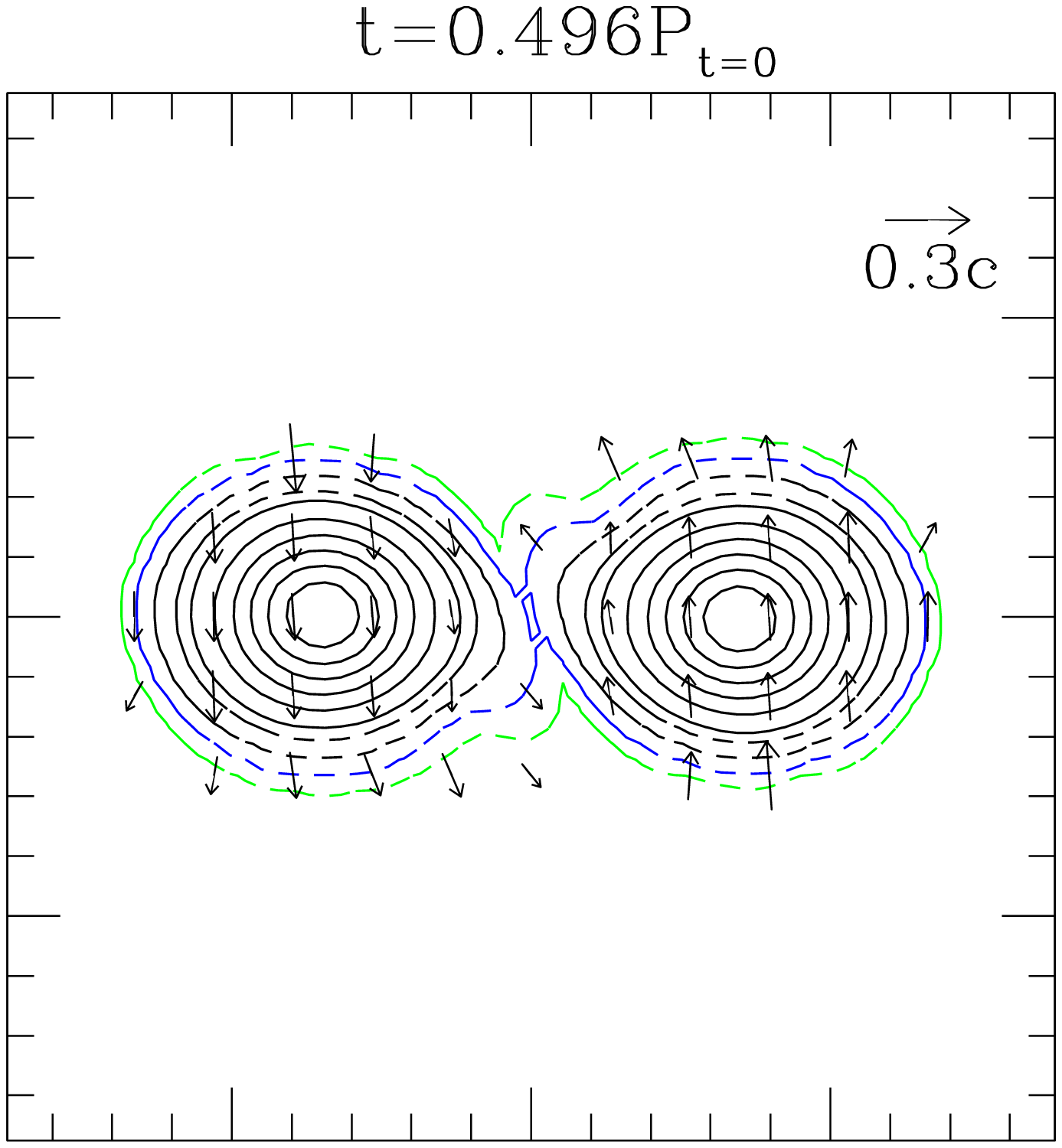} 
\epsfxsize=2.4in
\leavevmode
\hspace{-1.2cm}\epsffile{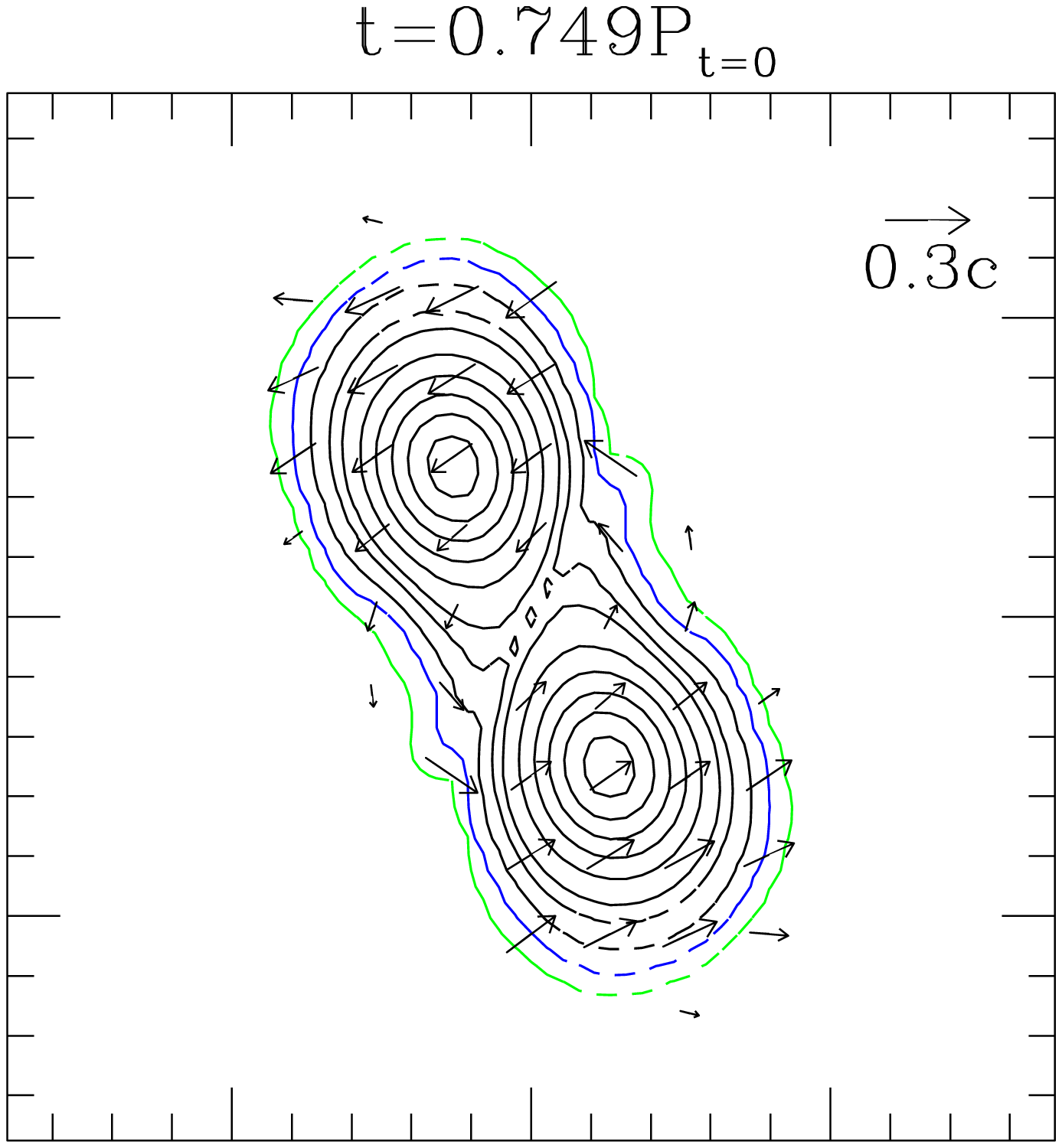} \\
\vspace{-1.2cm}
\epsfxsize=2.4in
\leavevmode
\epsffile{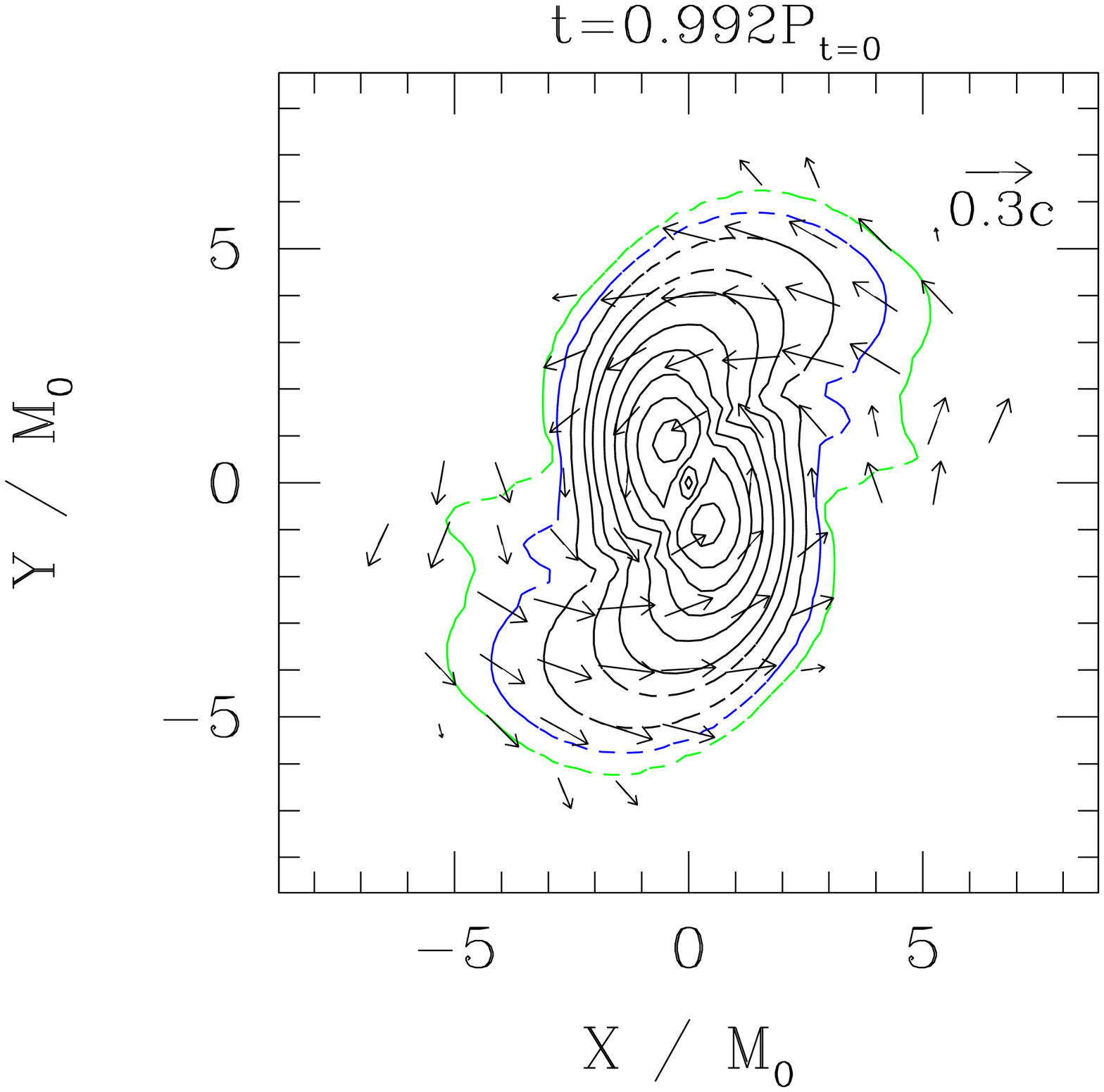} 
\epsfxsize=2.4in
\leavevmode
\hspace{-1.2cm}\epsffile{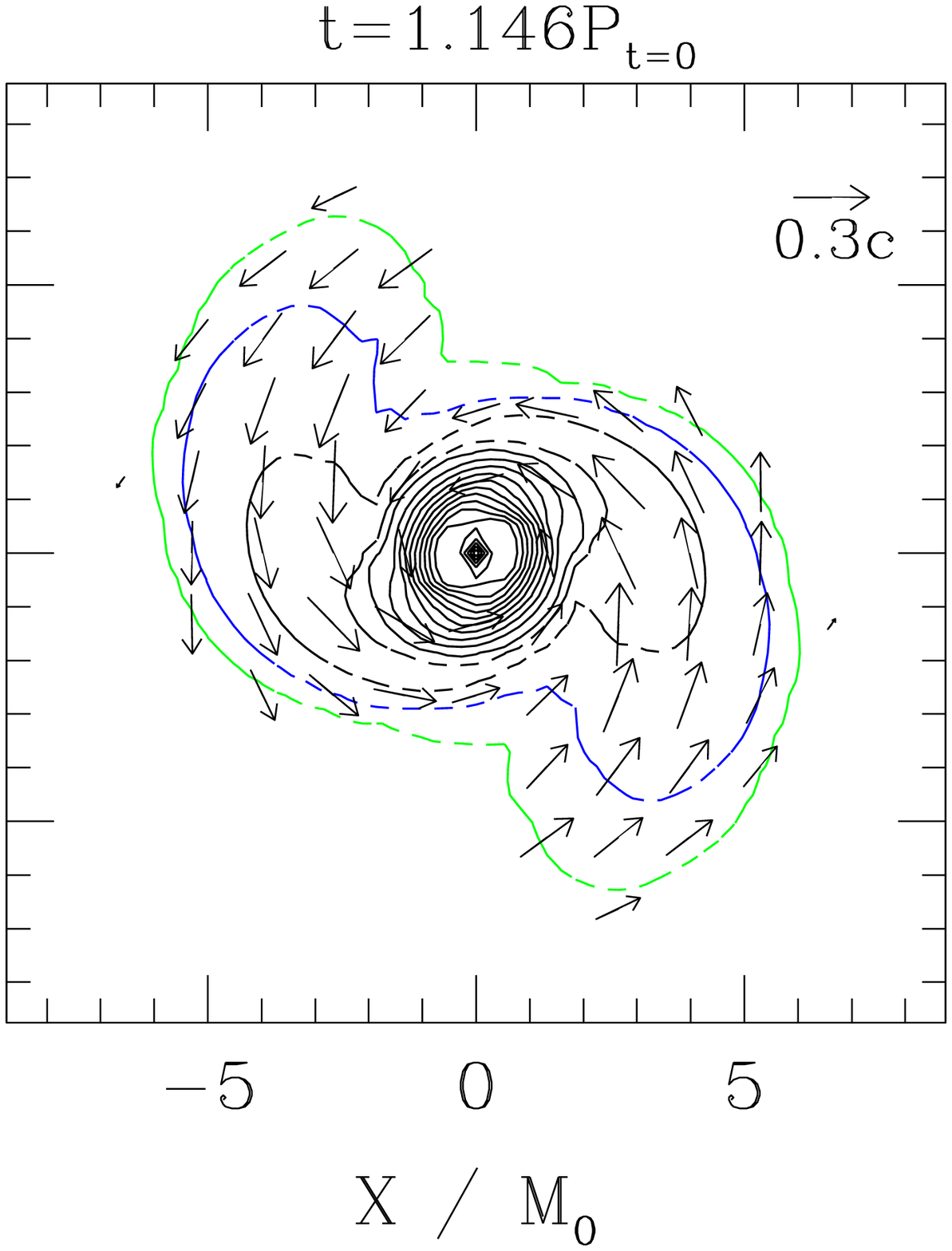}
\epsfxsize=2.4in
\leavevmode
\hspace{-1.2cm}\epsffile{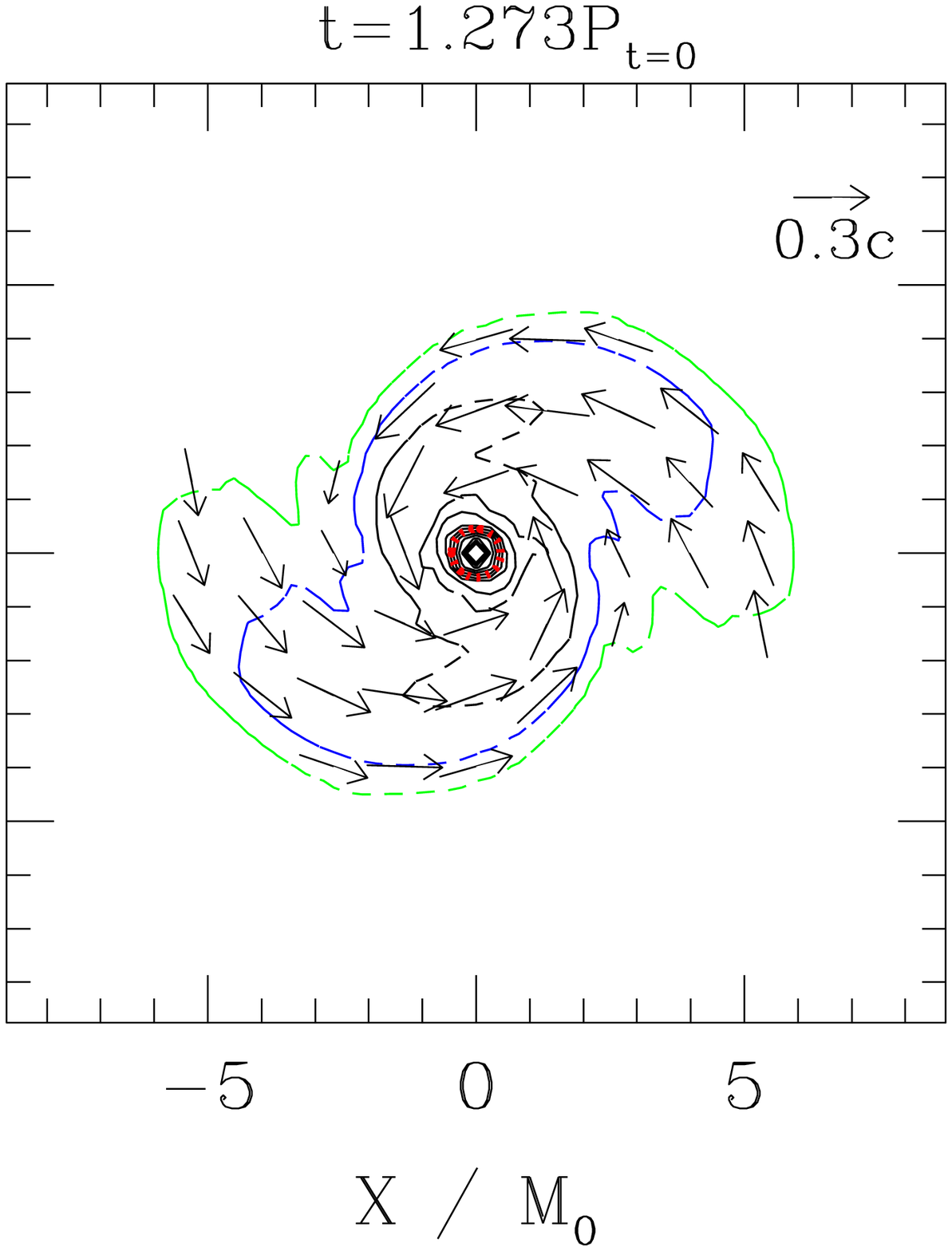}
\end{center}
\vspace{-5mm}
\caption{The same as Fig. \ref{FIG4} but for model M1616.
\label{FIG5} }
\end{figure}

\begin{figure}[t]
\begin{center}
\epsfxsize=2.8in
\leavevmode
(a)\epsffile{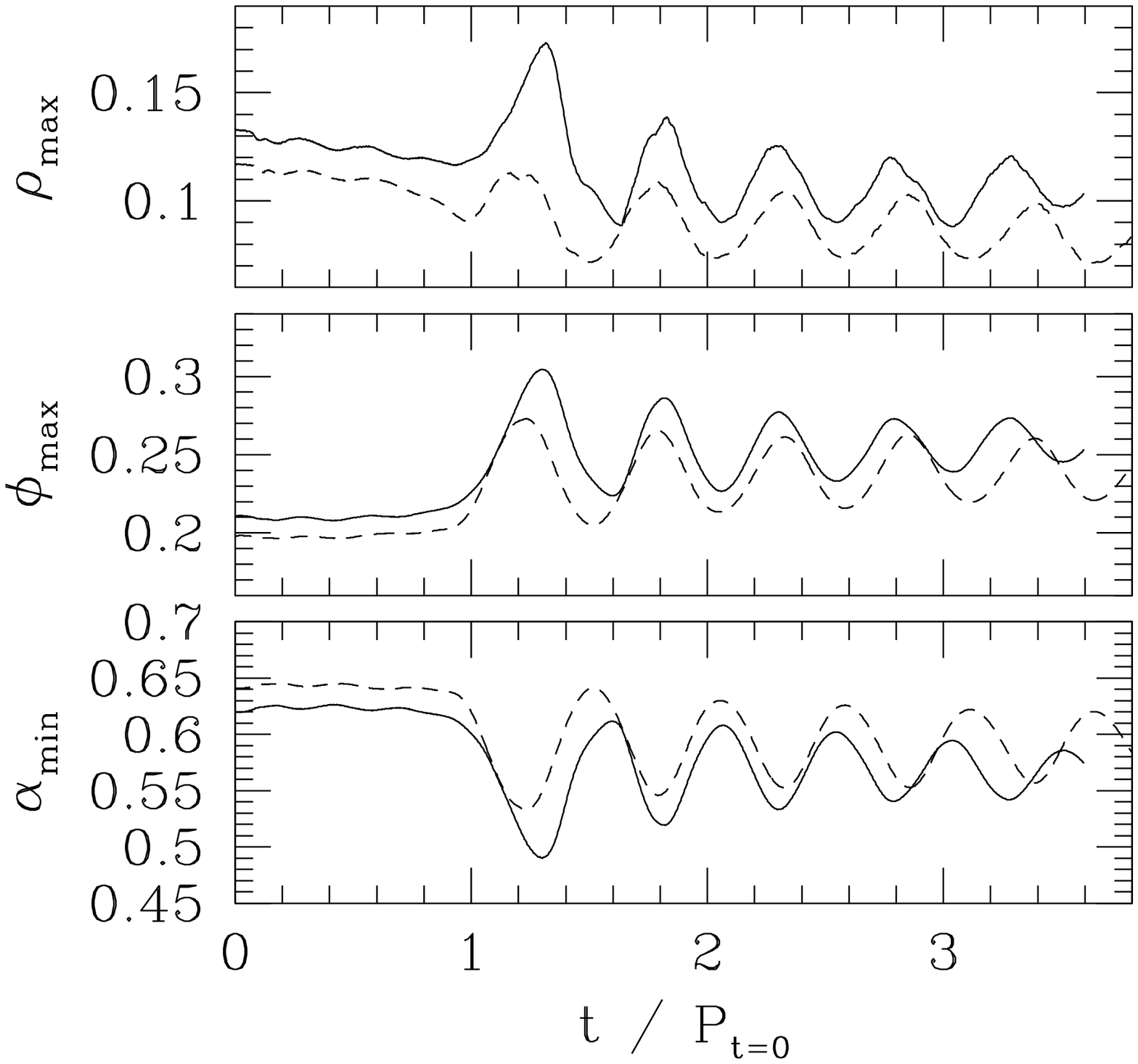}
\epsfxsize=2.8in
\leavevmode
~~~~(b)\epsffile{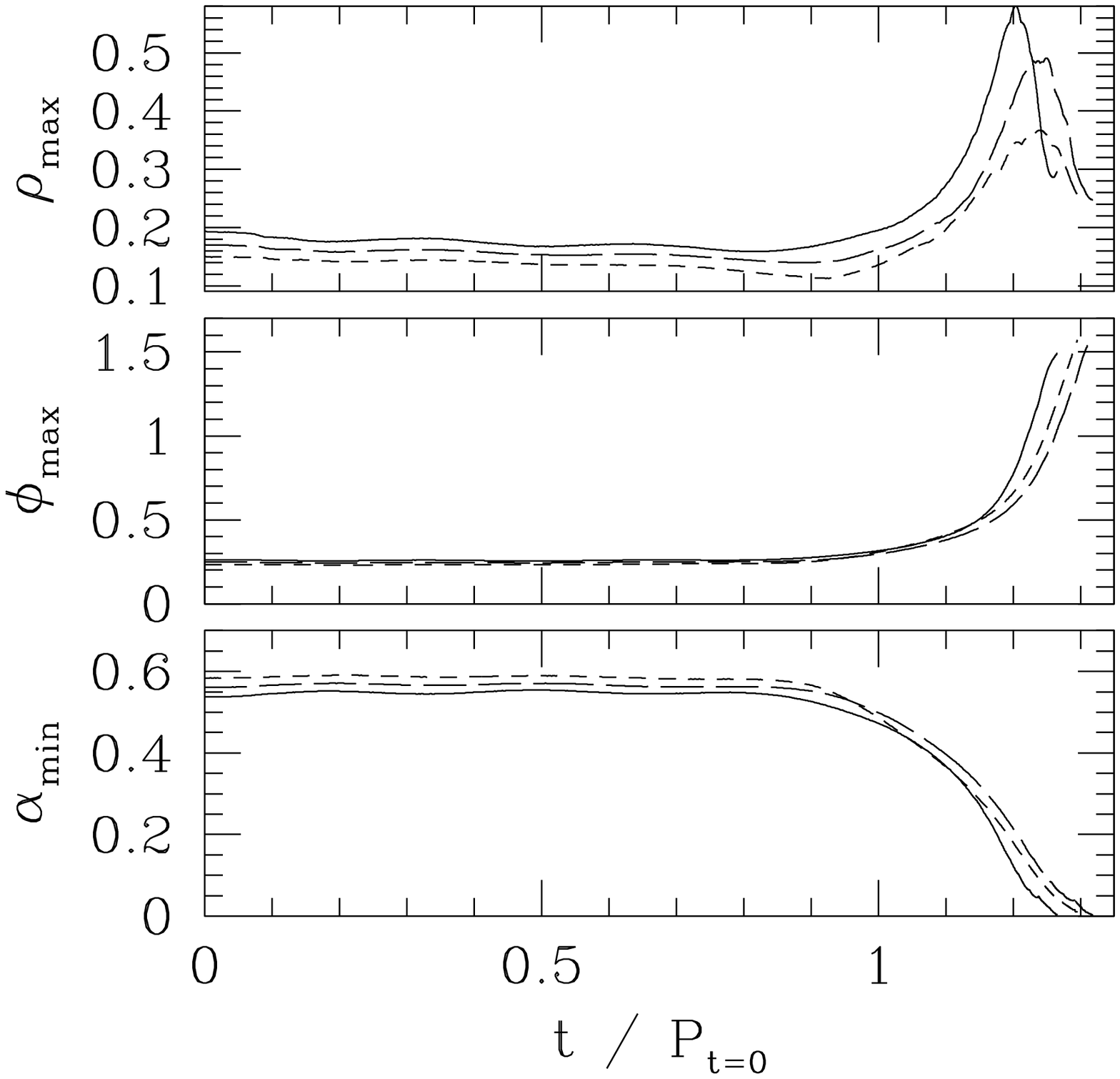}
\end{center}
\vspace{-2mm}
\caption{Evolution of the maximum values of $\rho$ and $\phi$, and
the minimum value of $\alpha$
(a) for models M1414 (dashed curves) and M1315 (solid curves), and 
(b) for models M1616 (dashed curves), M1517 (long-dashed curves),
and M1418 (solid curves).
Note that for the case of black hole formation [Fig. 6(b)],
the maximum density decreases in the final stage. 
The reason is as follows: We choose $\rho_*$ as a fundamental variable
to be evolved and compute $\rho$ from $\rho_*/w/e^{6\phi}$.
In the final stage, $\phi$ is very large ($> 1$) and, hence, a
small error in $\phi$ results in a large error in $\rho$. Note that 
the maximum value of $\rho_*$ increases monotonically
by many orders of magnitude. 
\label{FIG6}}
\end{figure}

\begin{figure}[htb]
\begin{center}
\epsfxsize=2.8in
\leavevmode
(a)\epsffile{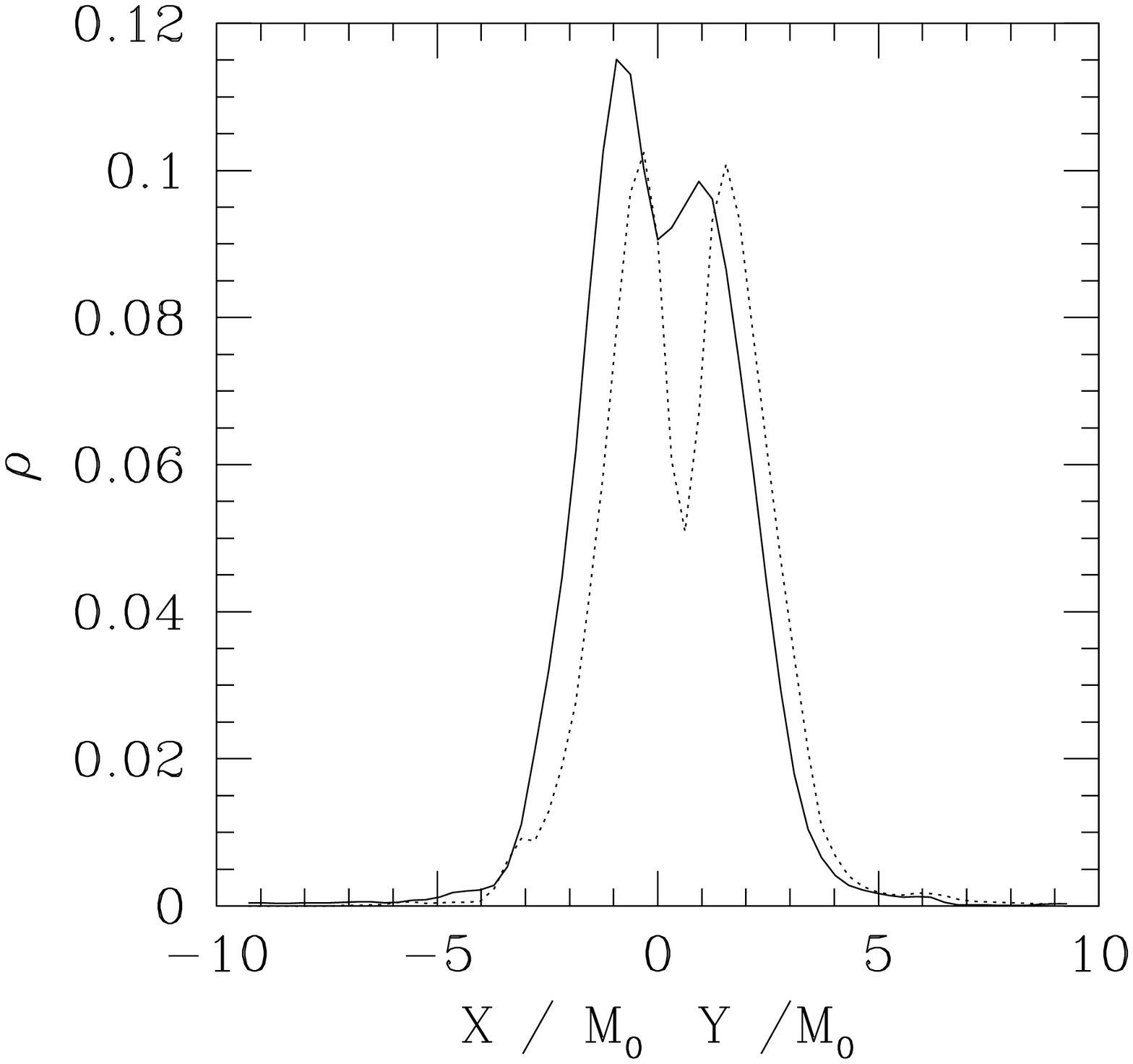}
\epsfxsize=2.8in
\leavevmode
~~~~(b)\epsffile{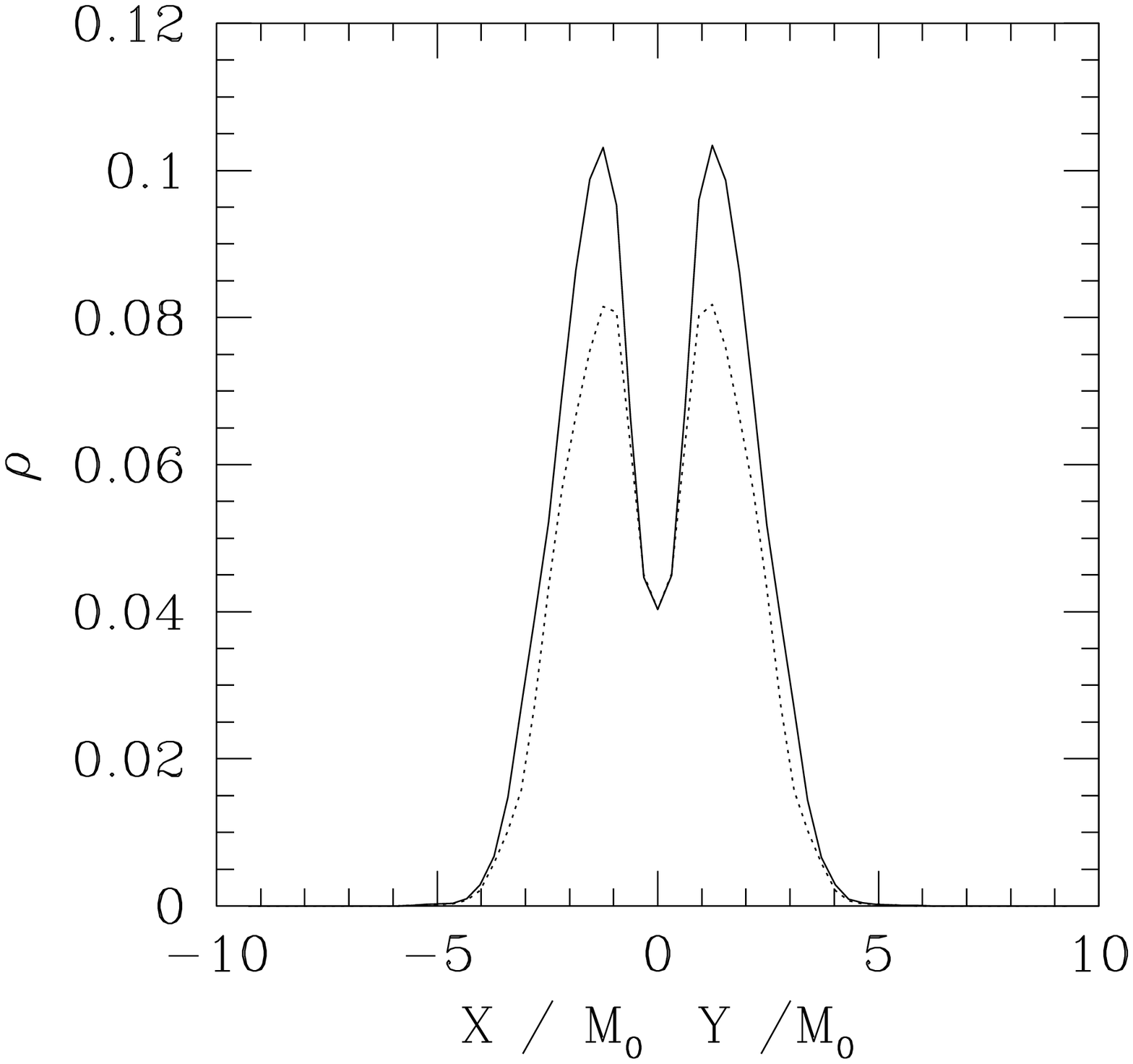}
\end{center}
\vspace{-2mm}
\caption{
The density profile along the $x$ (dotted curves) and
$y$ axes (solid curves)
(a) for model M1315 at $t=2.303P_{t=0}$ and
(b) for model M1414 at $t=2.351P_{t=0}$. 
\label{FIG7}}
\end{figure}

\begin{figure}[htb]
\begin{center}
\epsfxsize=2.8in
\leavevmode
(a)\epsffile{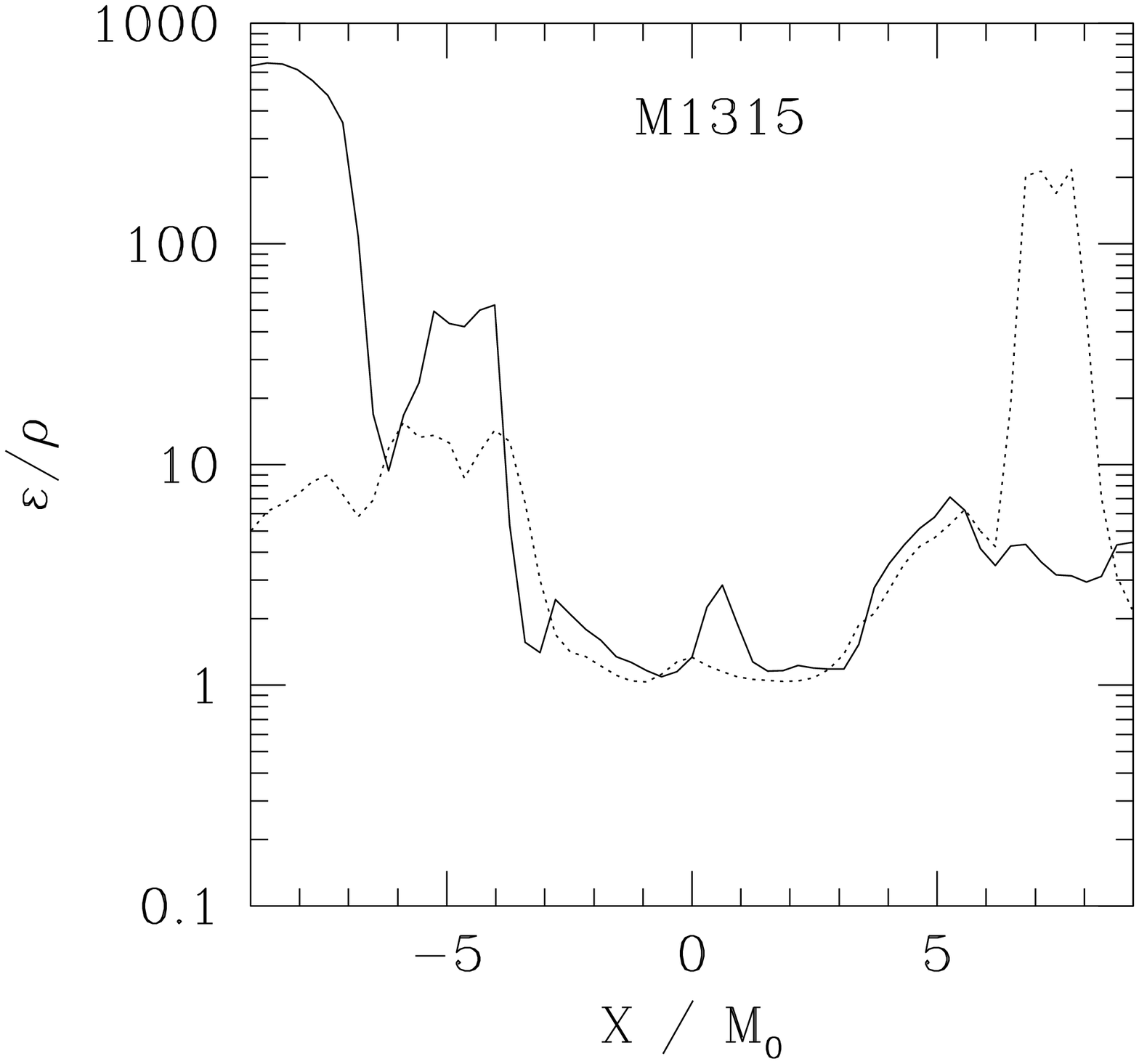}
\epsfxsize=2.8in
\leavevmode
~~~~(b)\epsffile{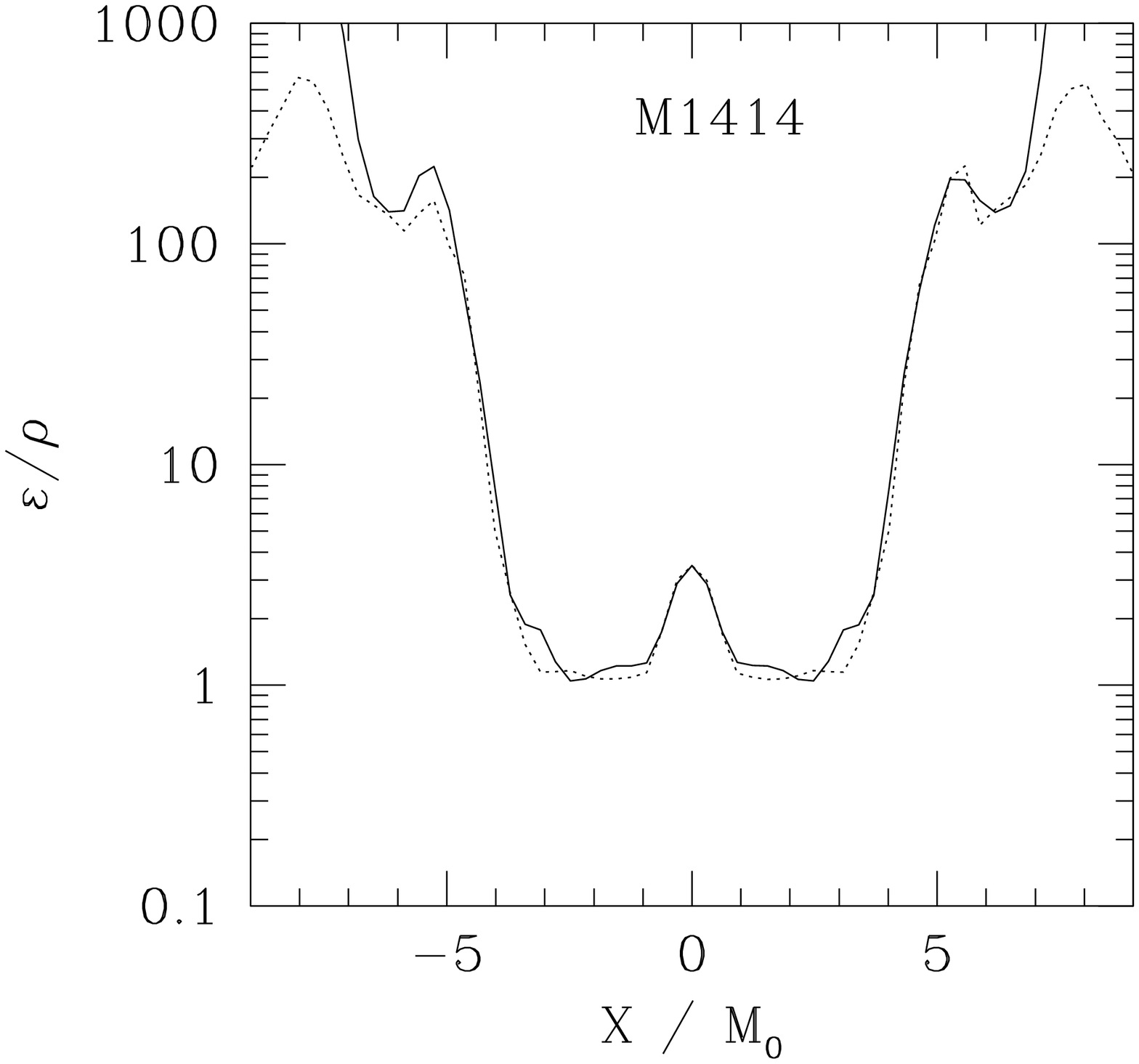}
\end{center}
\vspace{-2mm}
\caption{
$\kappa'(=\varepsilon/\rho)$
along the $x$ (solid curves) and $y$ axes (dotted curves)
(a) for model M1315 at $t=2.303P_{t=0}$ and
(b) for model M1414 at $t=2.351P_{t=0}$.
Note that at $t=0$, this quantity is unity everywhere inside neutron stars.
$\kappa'>1$ implies that shock heating is experienced. 
\label{FIG8}}
\end{figure}

\begin{figure}[htb]
\begin{center}
\epsfxsize=2.8in
\leavevmode
(a)\epsffile{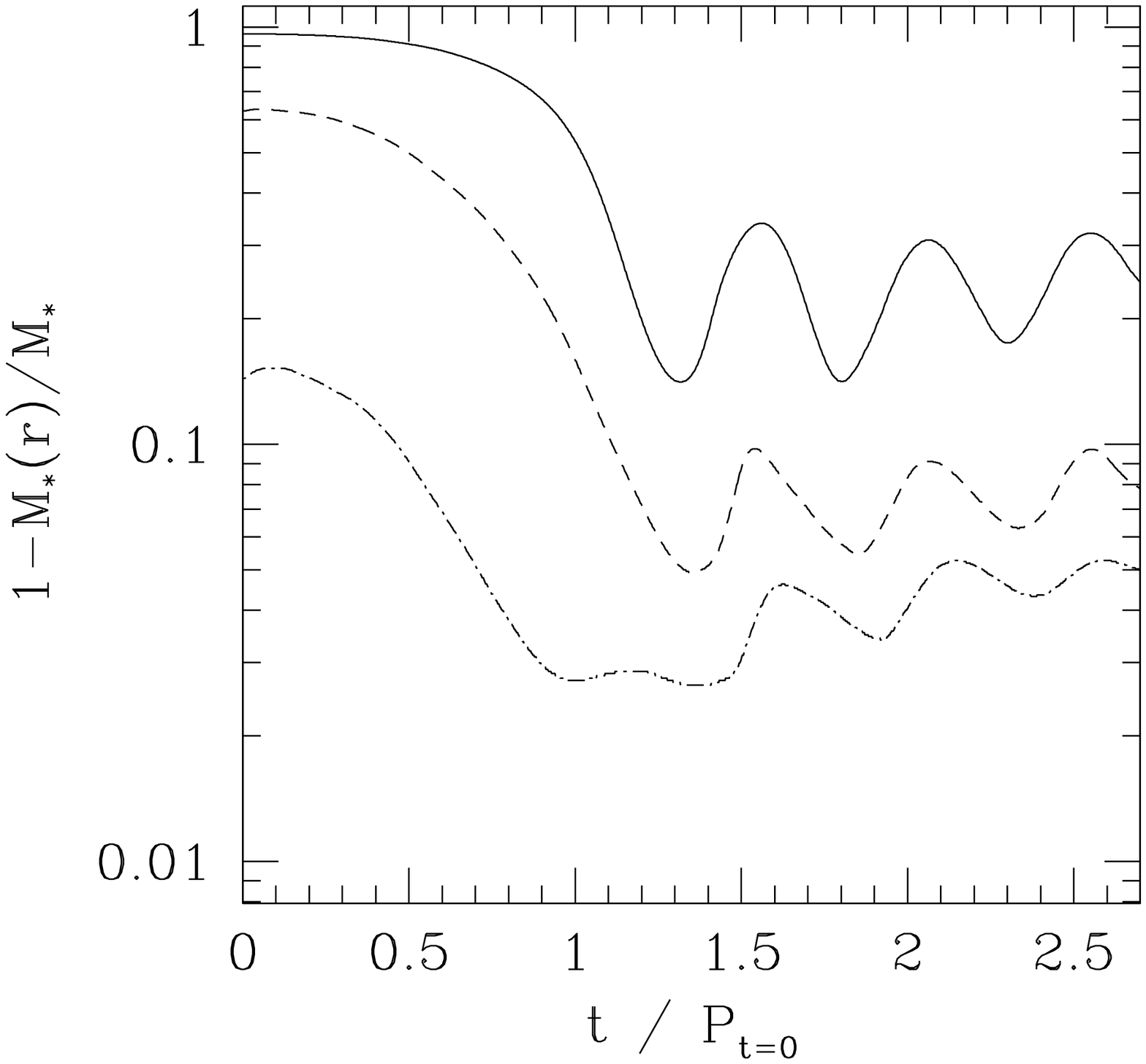}
\epsfxsize=2.8in
\leavevmode
~~~~(b)\epsffile{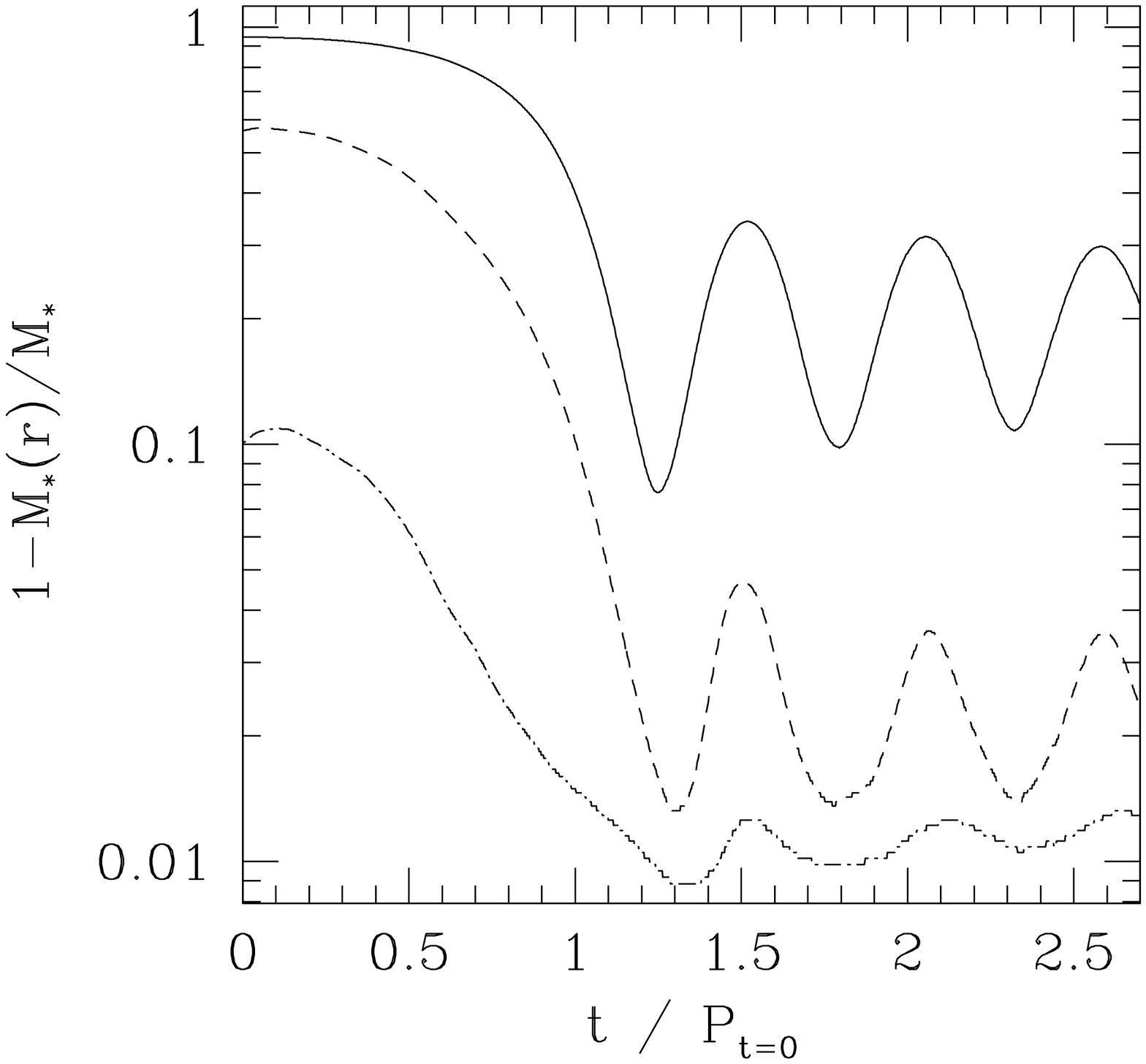}
\end{center}
\vspace{-2mm}
\caption{Evolution of the baryon
rest-mass fraction outside the spheres of radius
$3M_0$ (solid curve), $4.5M_0$ (dashed curve) and
$6M_0$ (dotted-dashed curve) (a) for model M1315 and (b) for
model M1414. 
\label{FIG9}}
\end{figure}

\begin{figure}[t]
\begin{center}
\epsfxsize=2.4in
\leavevmode
\epsffile{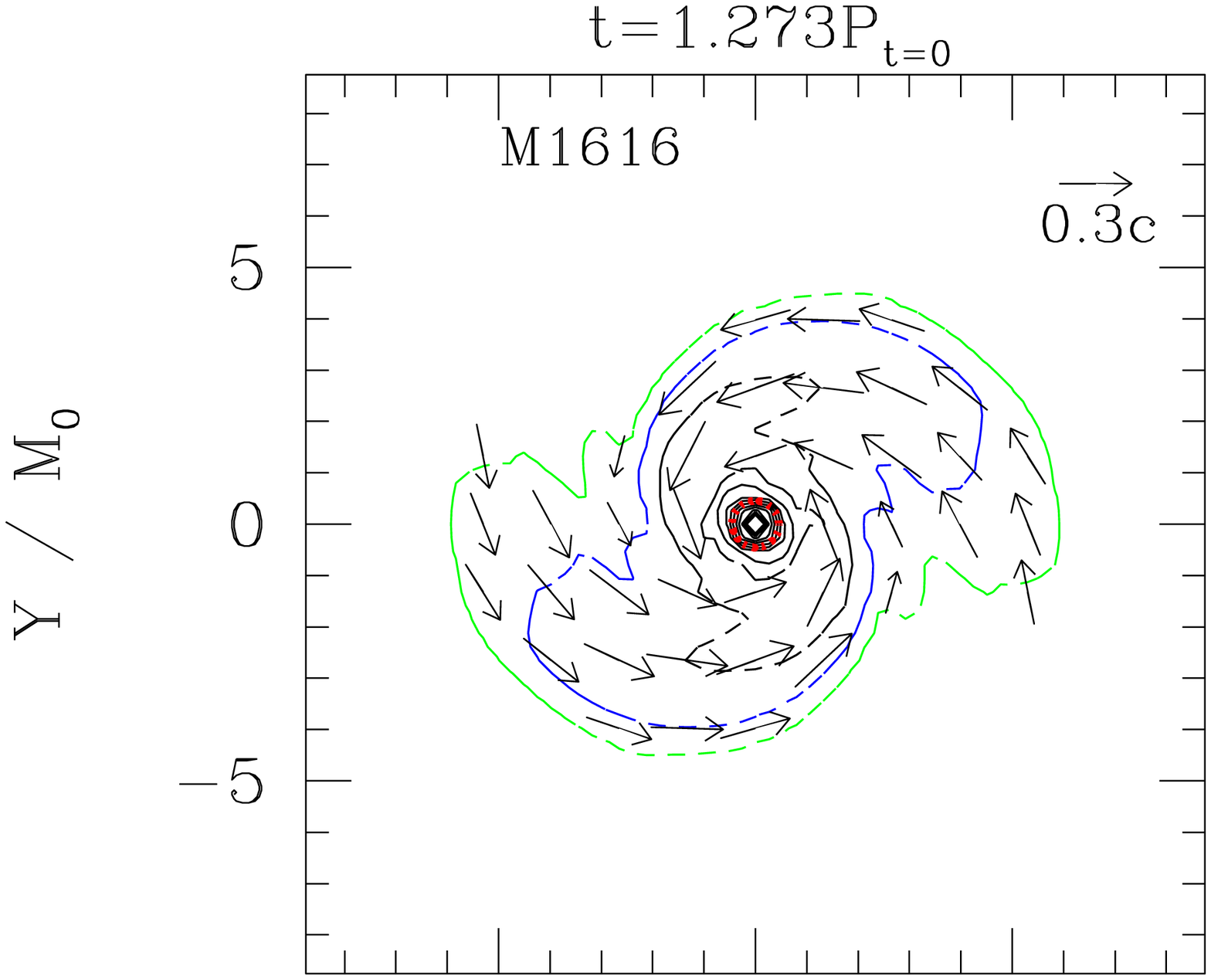}
\epsfxsize=2.4in
\leavevmode
\hspace{-1.2cm}\epsffile{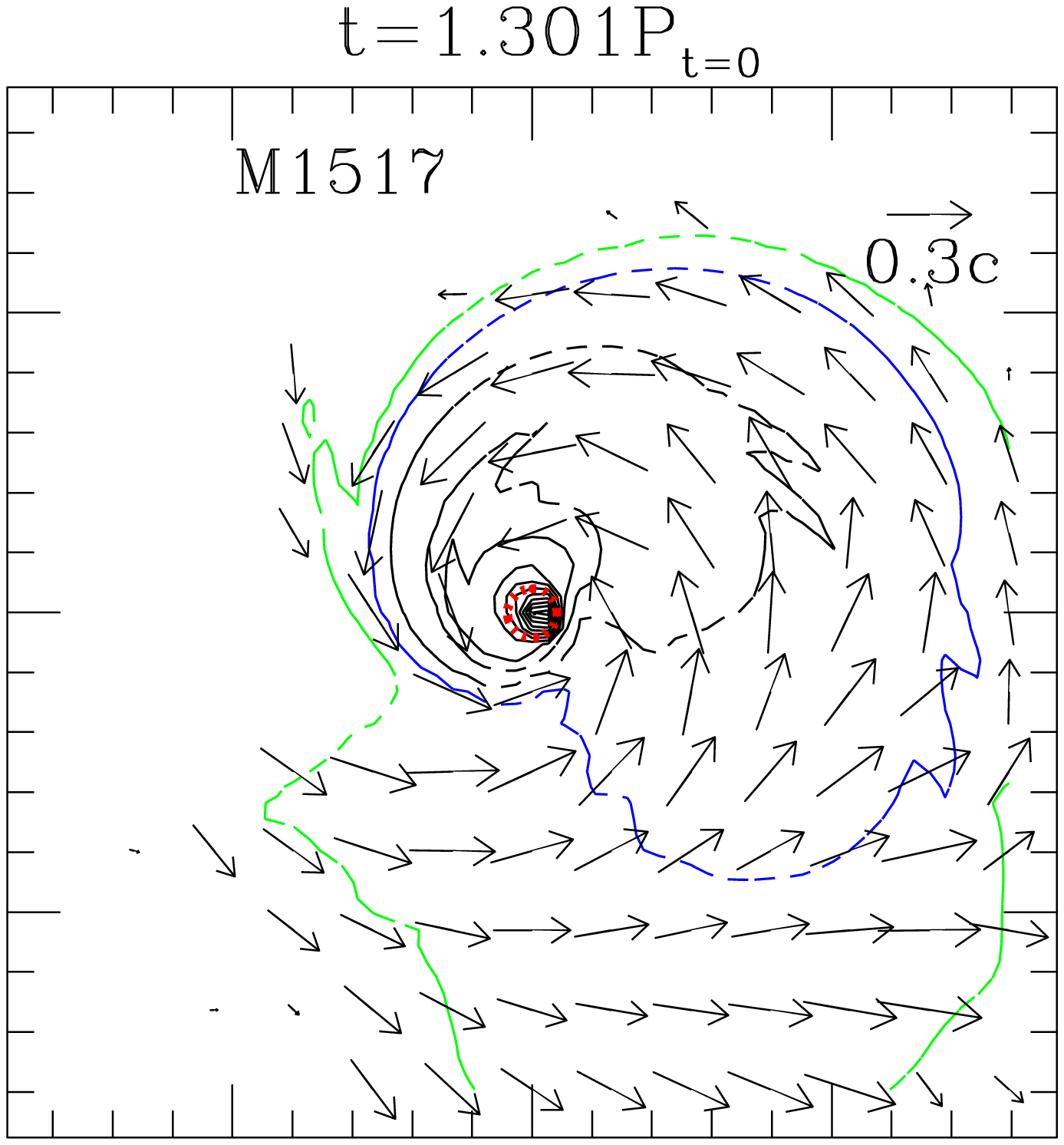} \\
\vspace{-1.2cm}
\epsfxsize=2.4in
\leavevmode
\epsffile{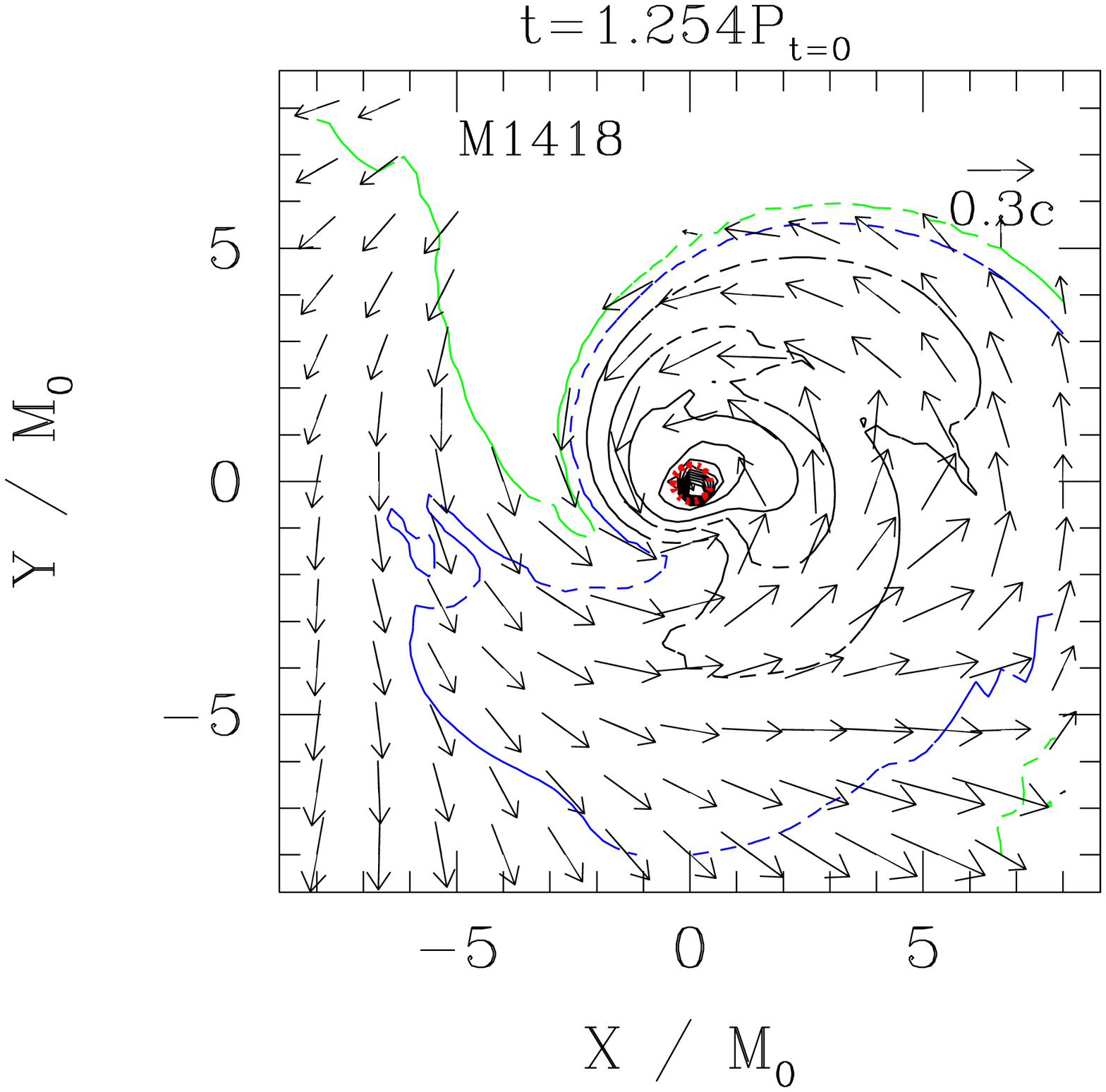} 
\epsfxsize=2.4in
\leavevmode
\hspace{-1.2cm}\epsffile{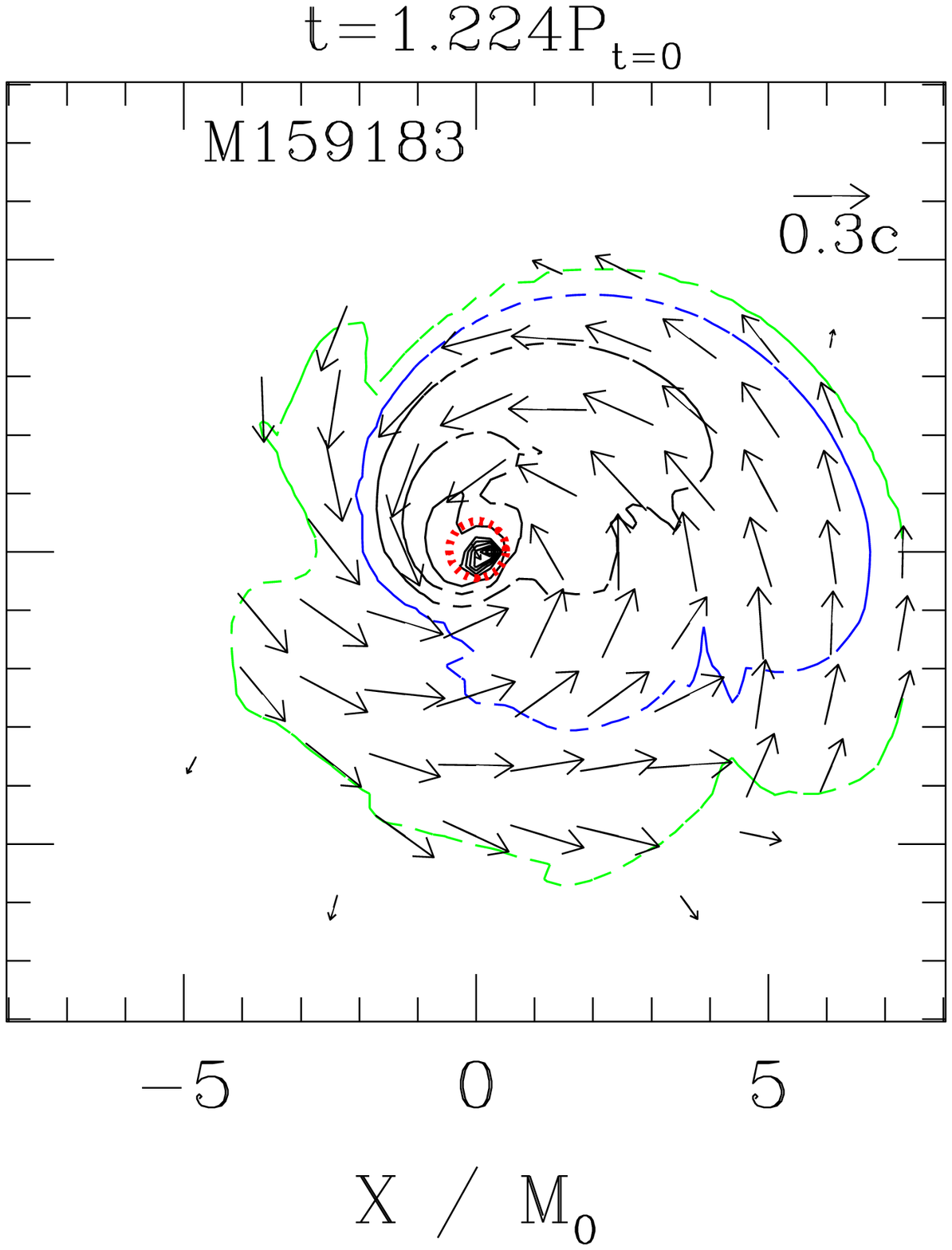} 
\vspace{-0.2cm}
\end{center}
\vspace{-5mm}
\caption{
Comparison of the density contour curves for $\rho$ 
in the equatorial plane near the end of the simulations for
models M1616, M1517, M1418, and M159183. 
The solid contour curves and the velocity vectors are 
drawn in the same manner as those for Fig. \ref{FIG4}. 
The thick dotted circle of radius $r \sim 0.5M_0$ 
denotes the location of the apparent horizon. 
\label{FIG10} }
\end{figure}

For models M1418, M1517, M1616, and M159183, black holes are formed
(cf. Figs. \ref{FIG4}, \ref{FIG5}, and \ref{FIG10})
in a dynamical time scale $\sim 1.3 P_{t=0}$ 
irrespective of the baryon rest-mass ratios. The formation of the black holes
is determined by finding the apparent horizons \cite{AH}. 
In all the cases, the total baryon rest-mass of the system is 
about 1.75 times as large as $M_{*\rm max}^{\rm sph}$ for a 
polytropic equation of state with $n=1$. 
Since the black holes are formed for $M_* \agt 1.75M_{*\rm max}^{\rm sph}$ 
while the hypermassive neutron stars are the outcomes for 
$M_* \alt 1.65M_{*\rm max}^{\rm sph}$, the threshold of the total
baryon rest-mass for the prompt black hole formation 
is between 1.65 and 1.75$M_{*\rm max}^{\rm sph}$ for $n=1$. 

The formation process of the black holes depends on the rest-mass ratios. 
For the merger of two equal-mass neutron stars, 
the merger results in a massive object of a double core
without tidal disruption and mass ejection outward.
The merged object is too massive to support
its self-gravity and, hence, collapses to a black hole promptly. 
Since the specific angular momentum of each fluid
element is too small \cite{bina2} and also since there is no efficient
transfer of angular momentum during the merger, 
the disk mass around the formed black hole is 
very small [cf. Figs. \ref{FIG10}(a) and (e)]. 

On the other hand, for the merger of two unequal-mass neutron stars, 
the black hole formation appears to be triggered by accretion 
to the primary star: First, 
the primary star tidally disrupts the smaller-mass companion 
at a critical separation. 
Subsequently, most of the tidal debris accrete
to the massive primary star and a small fraction of them form 
spiral arms. The accretion increases the mass
of the primary star rapidly, eventually, 
exceeding the critical value for formation of a black hole. 
During the merger, the angular momentum transfer works efficiently 
in the spiral arms, subsequently forming an accretion disk
around the formed black hole. 

\newpage

In Fig. \ref{FIG10}, we compare snapshots of density contour curves
soon after the formation of apparent horizons
for models M1616, M1517, M1418, and M159183.
Obviously, fractions of the disk mass
and the disk radius are larger for binaries of
the smaller rest-mass ratios. 
Comparing the figures for M1517 and M159183 for which
the rest-mass ratio is identical as 0.925, it is also found that
the disk mass and radius are smaller for the merger of 
the larger compactness. 

\begin{figure}[t]
\begin{center}
\epsfxsize=2.8in
\leavevmode
(a)\epsffile{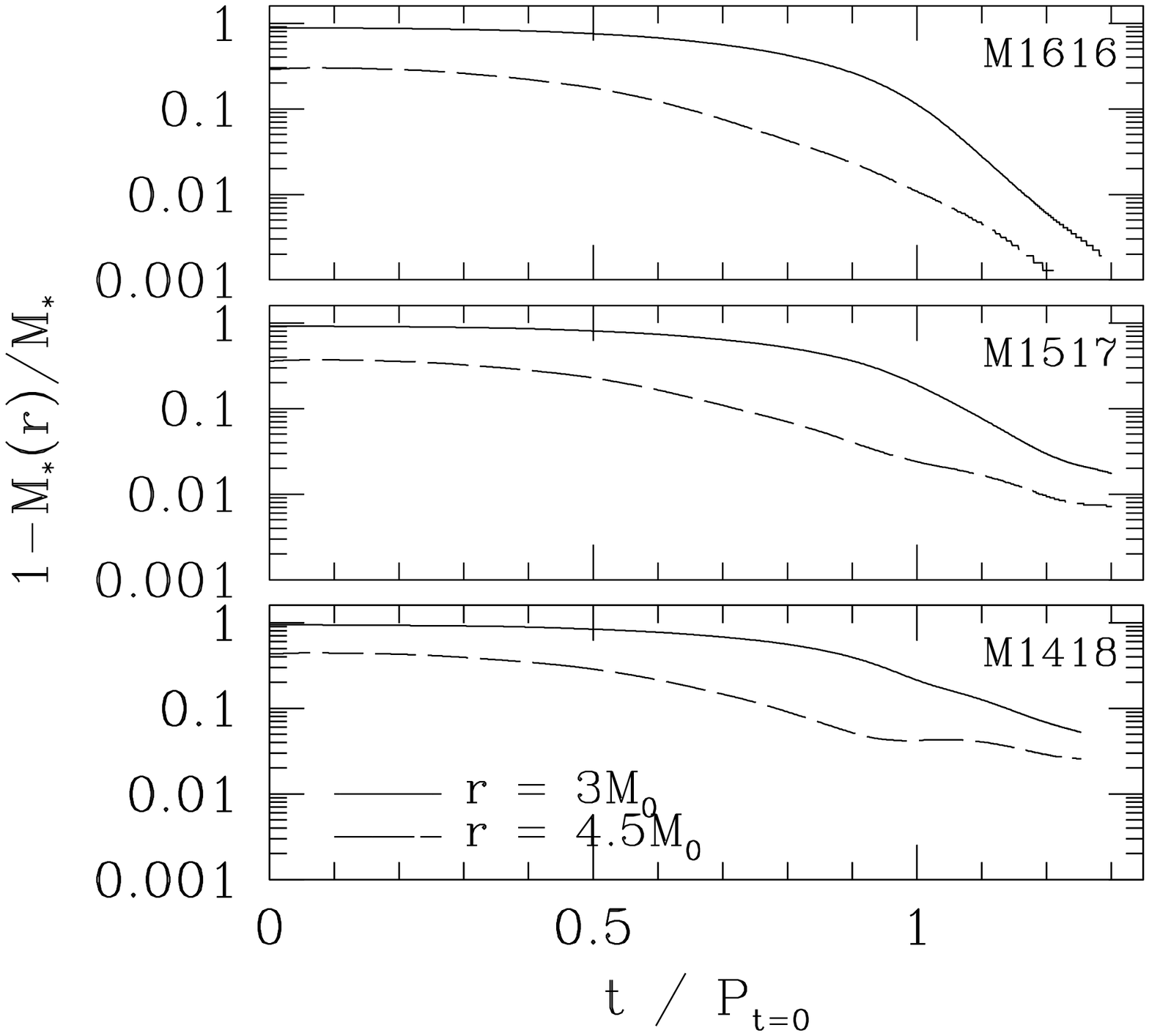}
\epsfxsize=2.8in
\leavevmode
~~(b)\epsffile{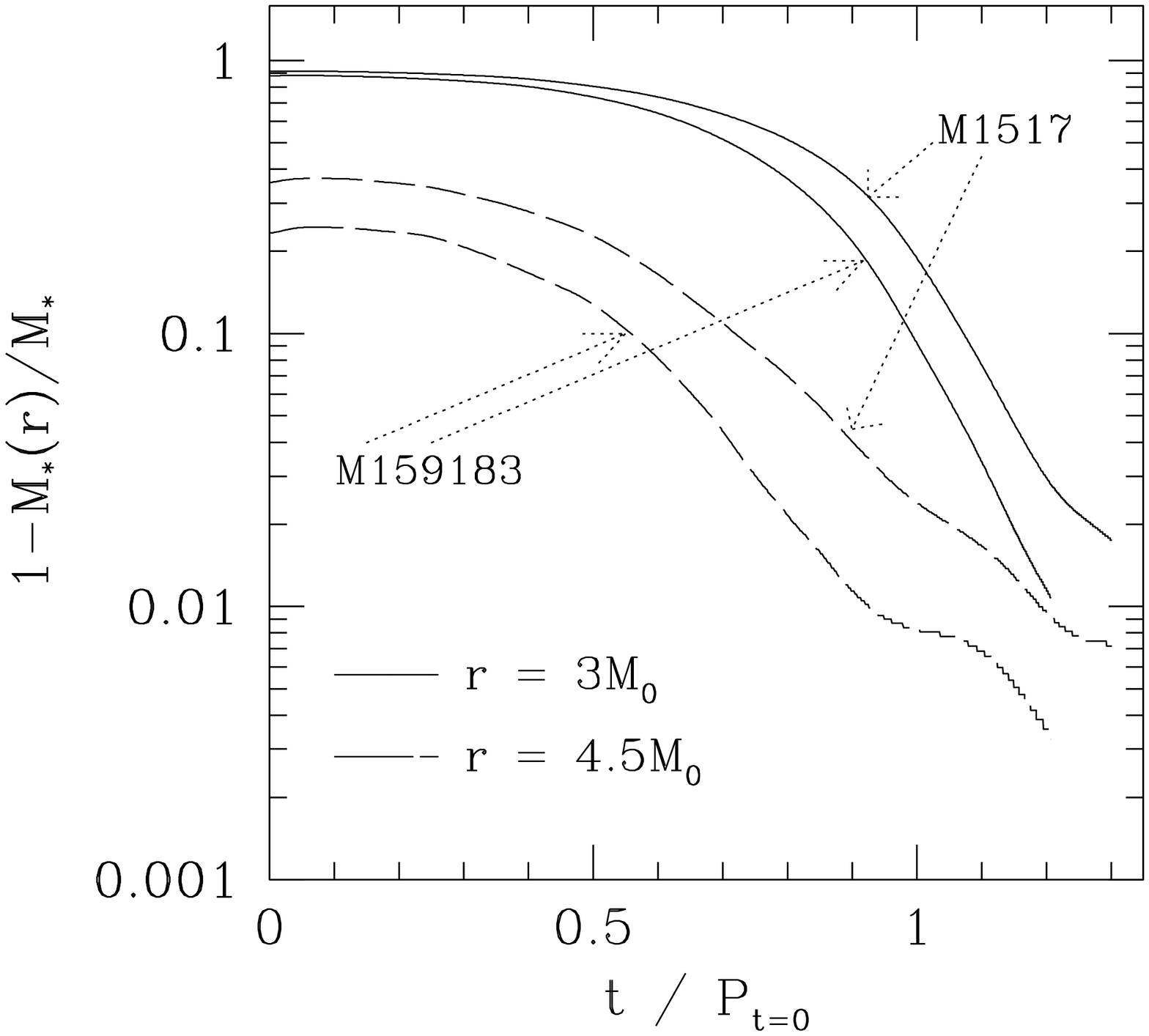}
\end{center}
\vspace{-2mm}
\caption{(a) $1-M_*(r)/M_*$ as a function of time for $r/M_0=3$ (solid curve)
and 4.5 (dashed curve) for models M1616, M1517, and M1418.
(b) The same as (a) but for models M1517 and M159183 for which
the rest-mass ratio is identical but the total baryon rest-mass is different. 
\label{FIG11}}
\end{figure}

Figure \ref{FIG11} shows the evolution of the baryon rest-mass 
fraction outside spheres of fixed coordinate radii 
for models M1616, M1517, M1418, and M159183. 
As the coordinate radii of the spheres, it is desirable to choose 
the radius of the innermost stable circular 
orbit around the formed black holes, but in practice 
it is difficult to determine it from numerical results exactly
for dynamical spacetimes.  
Thus we rather arbitrarily choose $3M_0$ and $4.5M_0$ which are 
not far away from the radius of the innermost stable circular orbit
for rotating black holes of nondimensional angular momentum parameter
$\sim 0.8$--0.9. For model M1616, the mass fraction outside these radii 
decreases monotonically, and at the termination of the simulation, 
less than 0.2\% of the total rest-mass of the system is 
outside the sphere of radius $3M_0$. 
On the other hand, the mass fraction for $r \geq 3M_0$ appears to approach 
$\sim 2$\% and $\sim 4\%$ for models M1517 and M1418, respectively. 
This indicates that for mergers of unequal-mass neutron stars, 
a certain fraction of the mass would form disks, and the 
mass fraction seems to increase in proportion to $1-Q_M$ 
for a given value of $M_*$. 

Comparing the results for models M1517 and M159183 for which
the rest-mass ratio is identical as 0.925, it is found that the
mass fraction of disks decreases with the increase of the compactness 
of neutron stars. The reason is simply that the gravity of the system 
is stronger for model M159183 and, as a result, a larger fraction of 
the mass is swallowed into the black hole. Obviously, smaller compactness of 
progenitor neutron stars is in favor of the formation of disks 
around a formed black hole. 

\begin{figure}[t]
\begin{center}
\epsfxsize=2.4in
\leavevmode
\epsffile{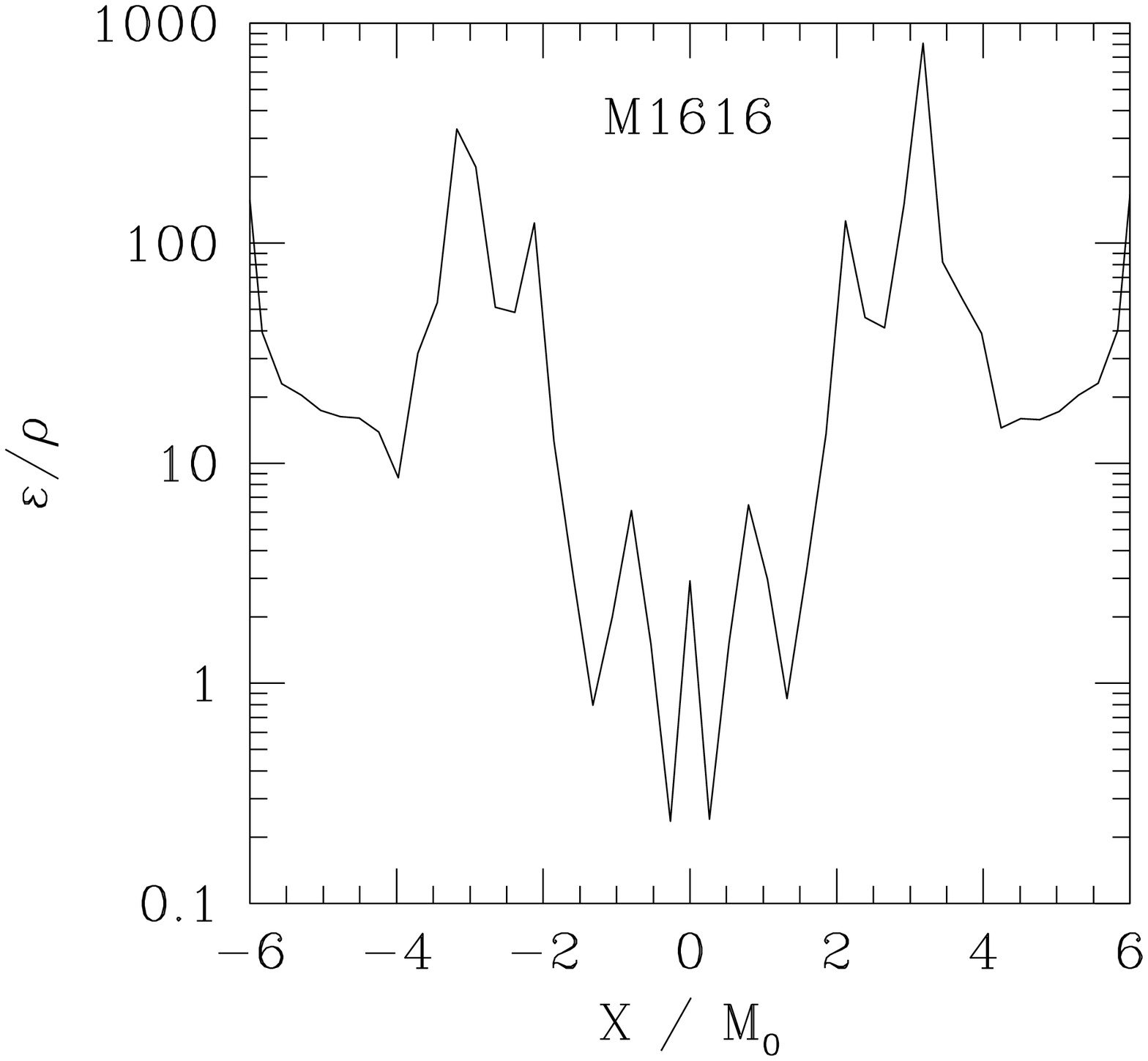}
\epsfxsize=2.4in
\leavevmode
\epsffile{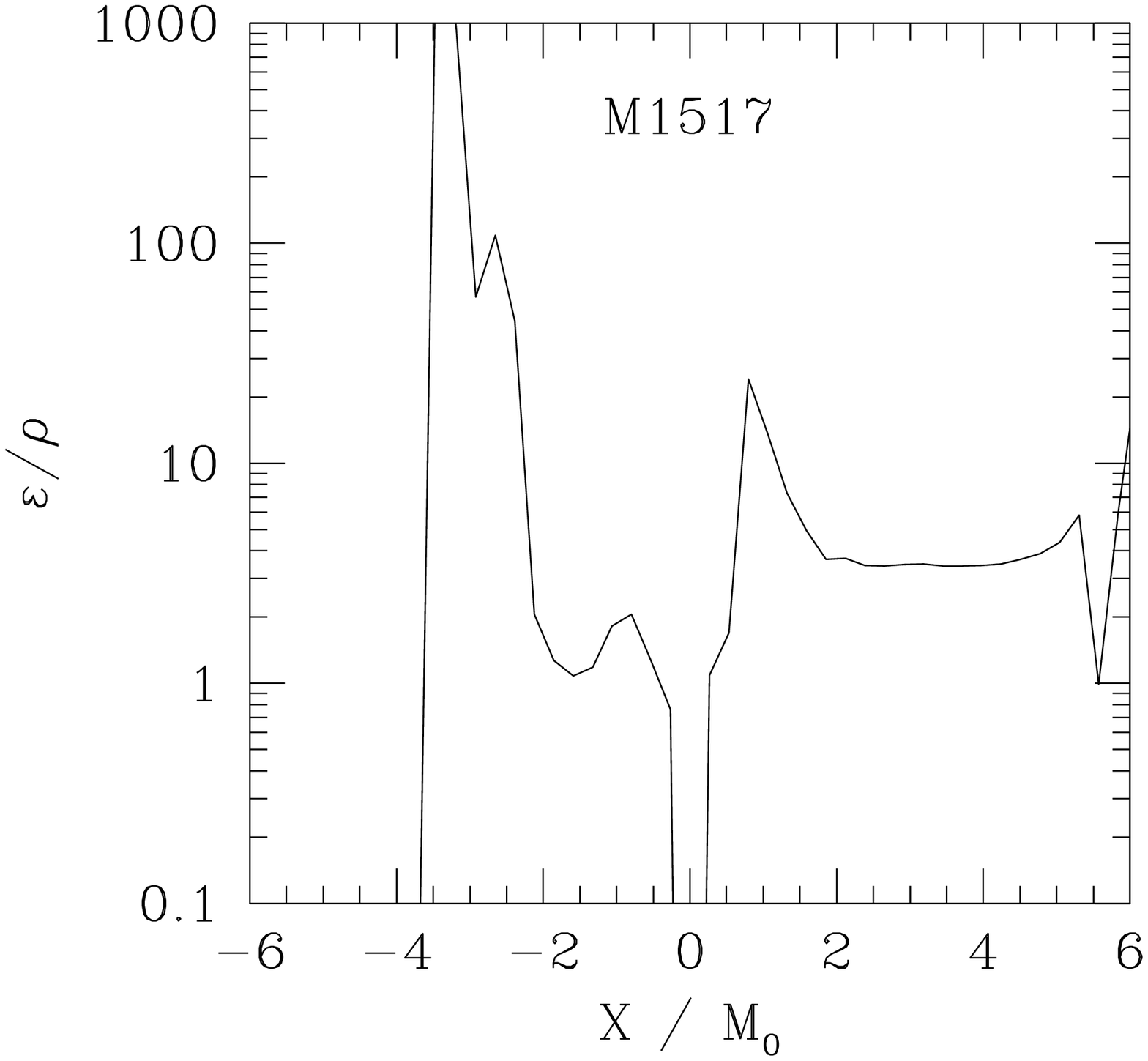} \\
\vspace{-0.4cm}
%\vspace{-1.2cm}
\epsfxsize=2.4in
\leavevmode
\epsffile{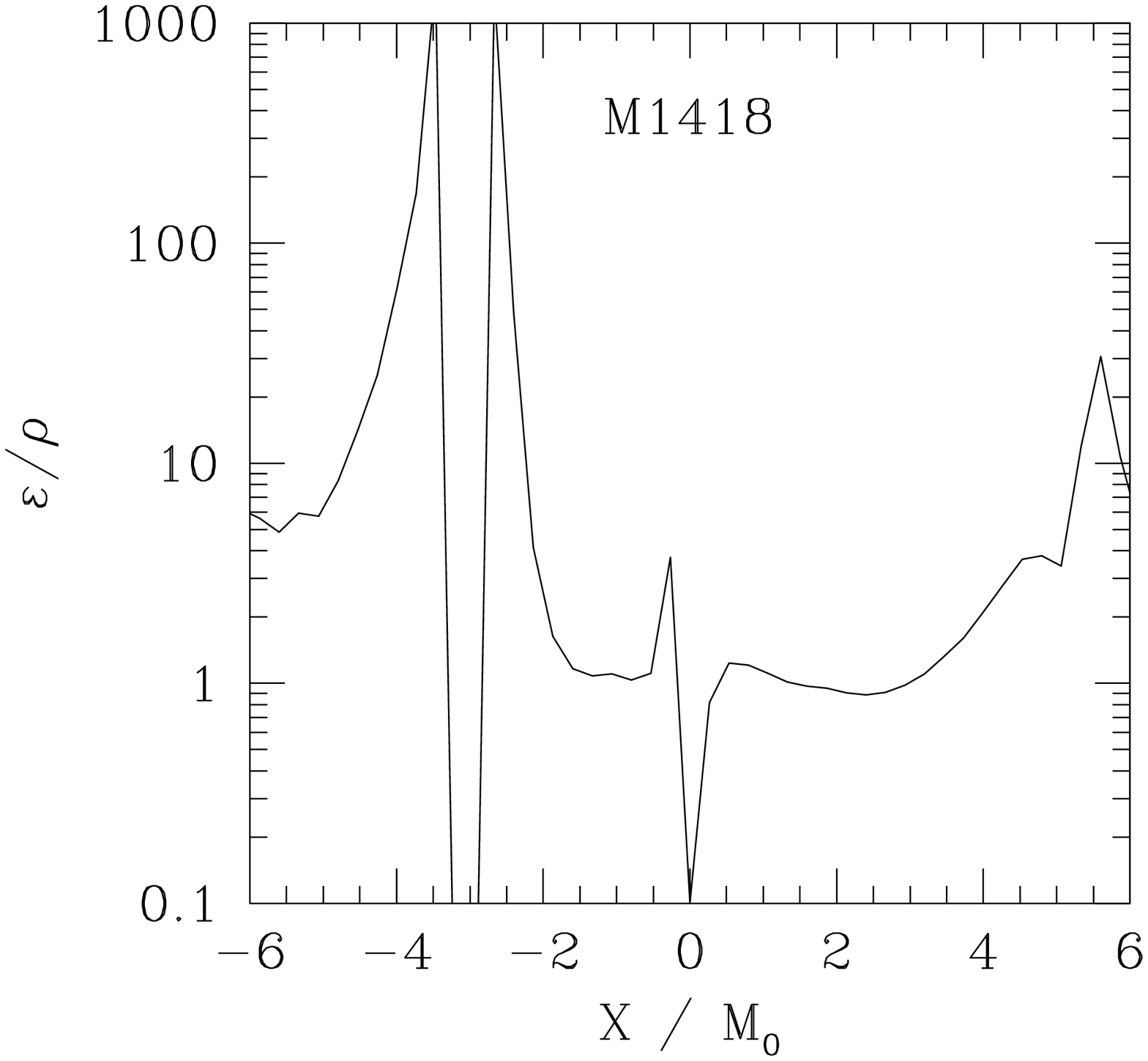} 
\epsfxsize=2.4in
\leavevmode
\epsffile{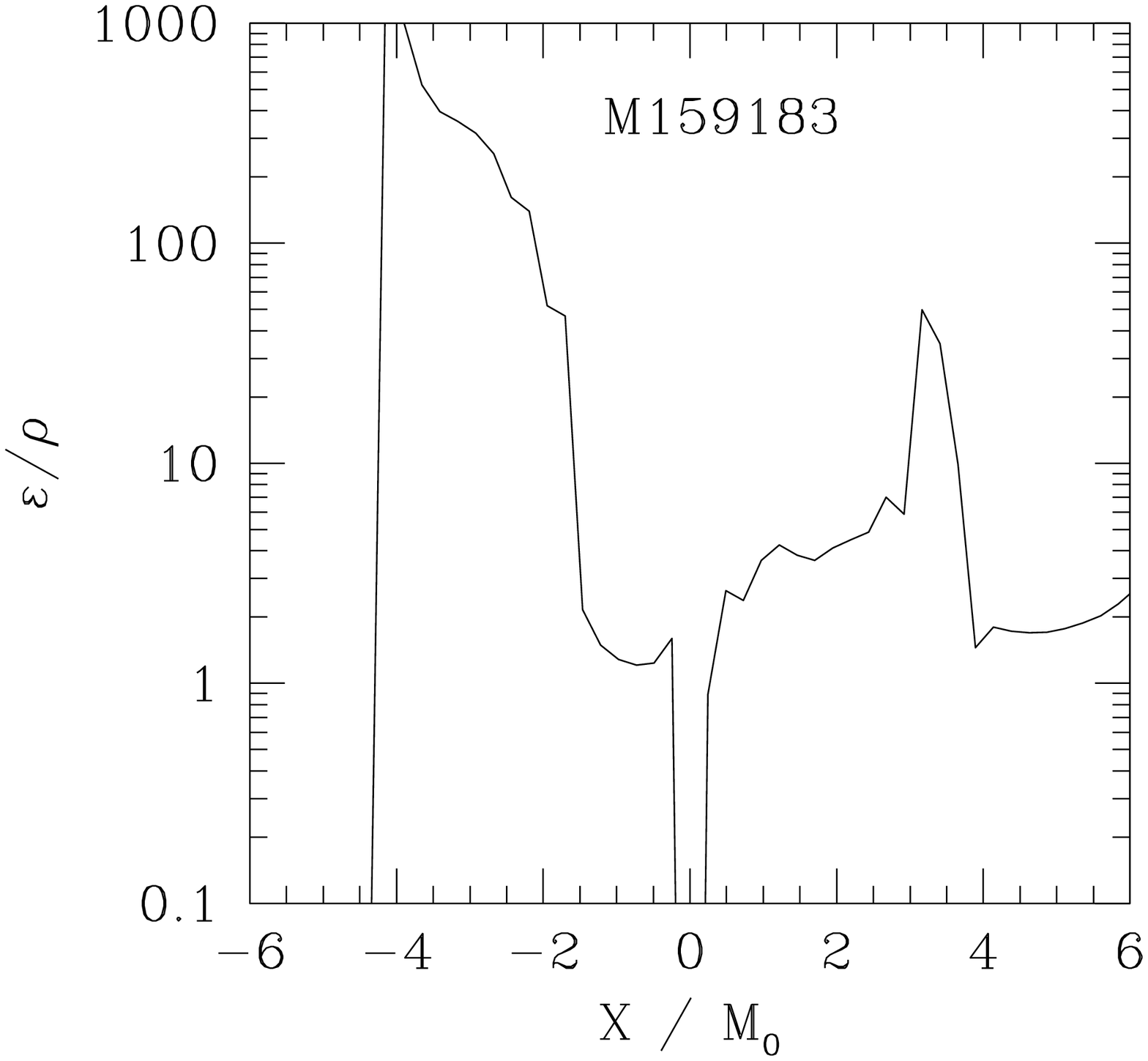} 
\vspace{-0.2cm}
\end{center}
\vspace{-5mm}
\caption{
$\kappa'(=\varepsilon/\rho)$ along the $x$ axis for 
models M1616, M1517, M1418, and M159183 at 
the same time steps as those of Fig.~\ref{FIG10}.
\label{FIG12} }
\end{figure}

Figures \ref{FIG12}(a)--(d) show $\kappa'$ along the $x$ axis 
for models M1616, M1517, M1418, and M159183. Figures indicate that
most of the fluid elements are heated up by shocks, except 
the inner region of the disk where the value of $\kappa'$ is 
less than 10.
This implies that the shock heating is not very important
for high density regions; i.e., at the collision of two neutron stars,
the shocks are not very strong.

\begin{figure}[t]
\begin{center}
\epsfxsize=3.0in
\leavevmode
\epsffile{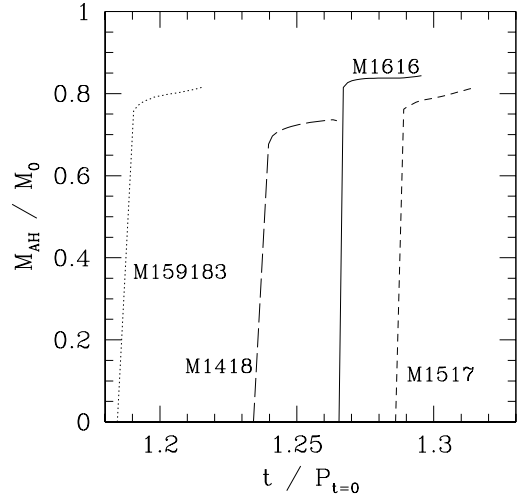}
\end{center}
\vspace{-2mm}
\caption{Mass of apparent horizon as a function of time 
for models M1616 (solid curve), M1517 (dashed curve), 
M1418 (long-dashed curve), and M159183 (dotted curve). 
\label{FIG13}}
\end{figure}

In Fig.~\ref{FIG13}, we display the time evolution of mass of
apparent horizons $M_{\rm AH}$ in units of $M_0$ for models M1616,
M1517, M1418, and M159183. $M_{\rm AH}$ is defined by 
\beqn
M_{\rm AH}=\sqrt{{S \over 16\pi}},\label{MAH}
\eeqn
where $S$ is the area of the apparent horizon. 
The figure indicates that $M_{\rm AH}/M_0$ appears to approach 
$\sim 0.85$ for models M1616, M1517, and M159183, 
and $\sim 0.75$ for model M1418. 
For models of the smaller rest-mass ratios, 
the value of $M_{\rm AH}/M_0$ is smaller; i.e., a larger fraction
of the mass element is not swallowed into a black hole. 

Since most of the mass elements fall into the black hole
and radiated energy of gravitational waves is less than $1\%$ 
of $M_0$ for model M1616 (see Sec. IV C), 
the black hole mass may be approximated by $M_0$ within $\sim 1\%$ error. 
Furthermore, recall that the area of a Kerr black hole of
mass $M$ and Kerr parameter $Mq$ ($q \equiv J/M^2$) is written by
\beq
S=8\pi M^2 \Bigl[1+(1-q^2)^{1/2}\Bigr]. 
\eeq
For model M1616, $M \sim M_0$ and $M_{\rm AH} \sim 0.85M_0$. 
This implies a value of $\sim 0.9$ for the nondimensional angular momentum 
$q$ of the formed black hole for model M1616. 
Since the initial value of the system, $q_0$, is about 0.913, and
$M$ and $J$ decrease by $\sim 0.5\%$ and
$\sim 7\%$ by gravitational radiation (see Sec. IV), respectively, the
expected final value of $q$ is $\sim 0.85$, which agrees with the
numerical value within $\sim 5\%$ error. (The disagreement is due to
a numerical error associated with insufficient grid resolution.
This point is confirmed by the convergence test presented in Sec. IV E.) 
The result presented here indicates that the location
and the area of the apparent horizon are determined within
$\sim 5\%$ error with the current grid resolution. 

%%%%%%%%%%%%%%%%%%%%%%%%%%%%%%%%%%%%%%%%%%%%%%%%%%
\subsection{Gravitational waves}

In Figs. \ref{FIG14}--\ref{FIG16}, we present the gravitational waveforms
(gauge-invariant quantities)
and the accumulated energy and angular momentum loss
by gravitational radiation as a function of the retarded time
$(t - r_{\rm obs})/P_{\rm t=0}$. 
(In the following, the retarded time is always normalized by $P_{t=0}$.)
It should be noted that $R_{22\pm}r/M_0=1$ implies the values of 
$r h_{+}$ and $r h_{\times}$ along the $z$ axis due to 
$l=m=2$ modes are $\approx 1.85$ km, where 
\beq
h_+ \equiv {1 \over 2r^{2}}
\biggl( \gamma_{\theta\theta}
-{\gamma_{\varphi\varphi} \over \sin^{2}\theta}\biggr), ~~~~~
h_{\times} \equiv {\gamma_{\theta\varphi} \over r^2 \sin\theta}. 
\eeq

Irrespective of the mass and the mass ratio of binaries, 
the inspiral waveforms are dominant for $t-r_{\rm obs} \alt P_{t=0}$,
and subsequently, the merger waveforms are excited. 
For hypermassive neutron star formation, 
quasiperiodic waves are excited because of its quasiradial and 
nonaxisymmetric oscillations. 
For the black hole formation case, the computation crashed 
soon after the formation of apparent horizon. 
As a result, we were not able to compute complete gravitational
waveforms for $t - r_{\rm obs} \agt P_{\rm t=0}$, for which 
gravitational waves would be dominated by quasinormal mode 
ringings \cite{Leaver}. 
A straightforward approach to compute such gravitational waves is 
to develop a black hole excision technique \cite{U}, 
by which it may be possible to continue the simulation for a long time 
duration even after the formation of the black hole.
An alternative approach is to extract gravitational waves 
from a restricted spacetime data set using the so-called 
Lazarus technique \cite{BCL}. 
Leaving the development of the two methods for future implementations, 
we focus our discussions below mainly on the character of gravitational 
waves for neutron star formation than for black hole formation.

\subsubsection{Hypermassive neutron star formation case}

\begin{figure}[t]
\begin{center}
\epsfxsize=2.75in
\leavevmode
(a)\epsffile{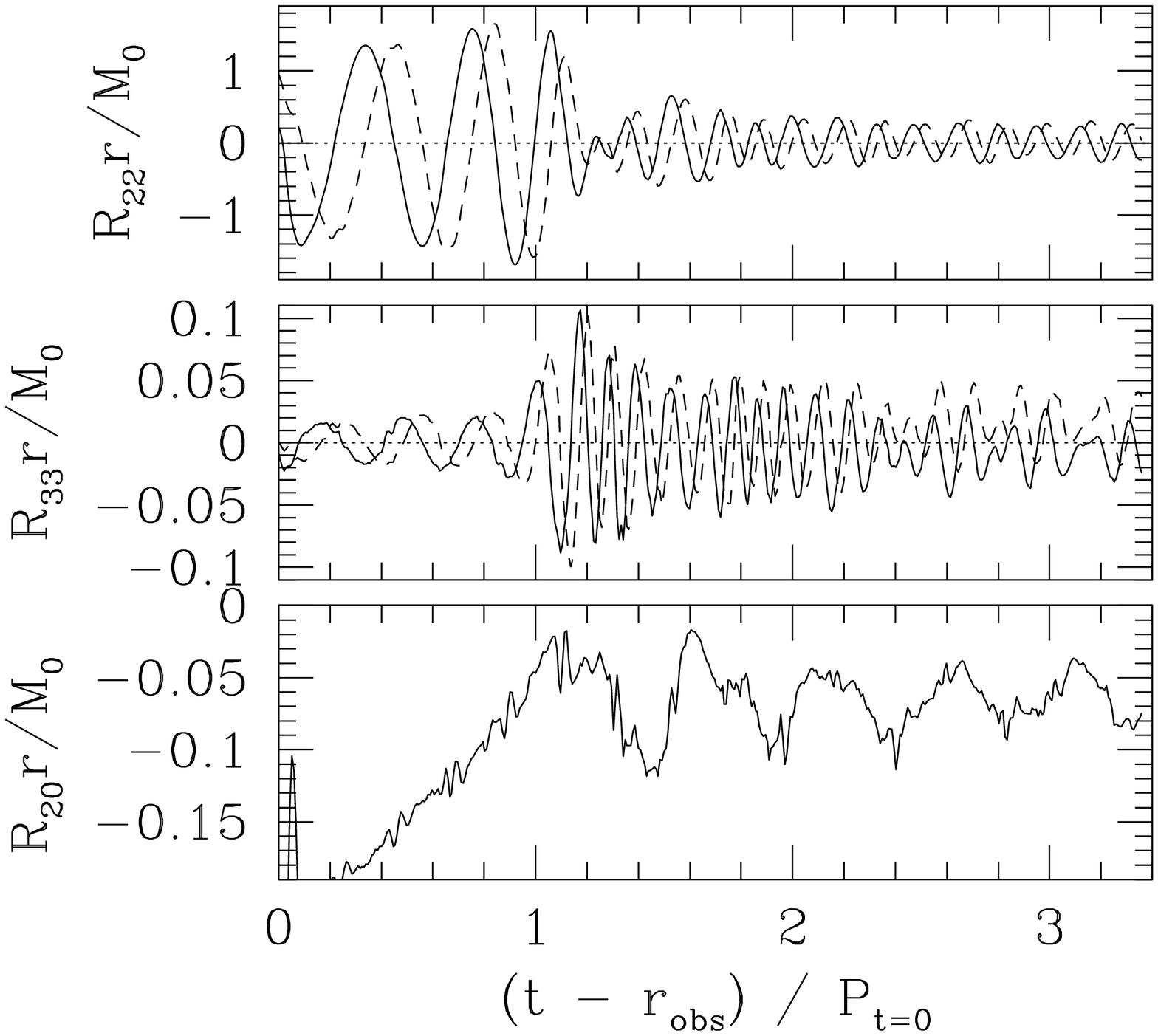}
\epsfxsize=2.75in
\leavevmode
~~~~~(b)\epsffile{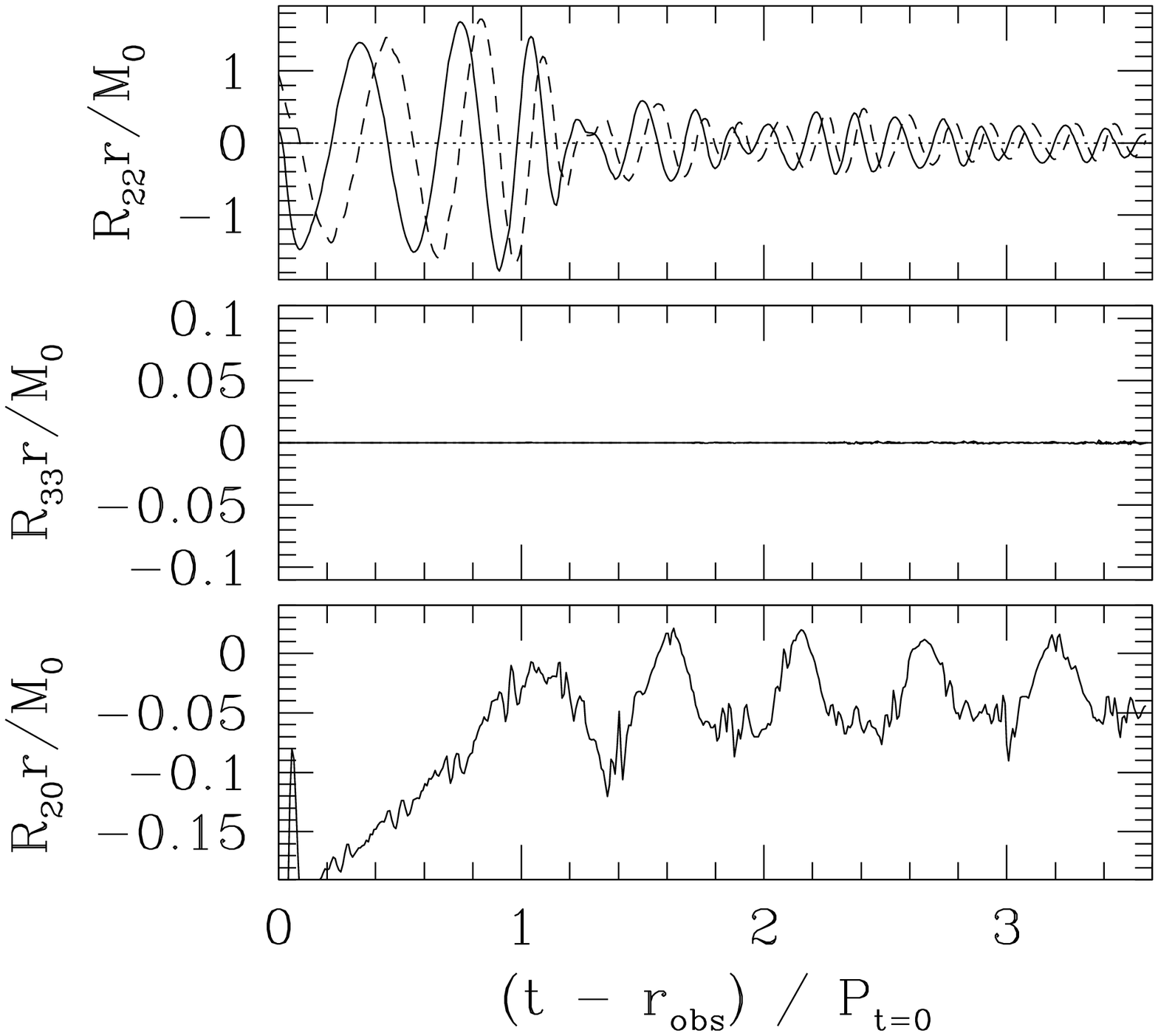}
\end{center}
\vspace{-2mm}
\caption{
$R_{22\pm}r$, $R_{33\pm}r$, and $R_{20}r$
as a function of the retarded time
(a) for model M1315 and (b) for model M1414.
The solid and dashed curves for $R_{22}$ and $R_{33}$
denote $R_{lm+}$ and $R_{lm-}$. 
\label{FIG14}}
\end{figure}

\begin{figure}[t]
\begin{center}
\epsfxsize=2.75in
\leavevmode
\epsffile{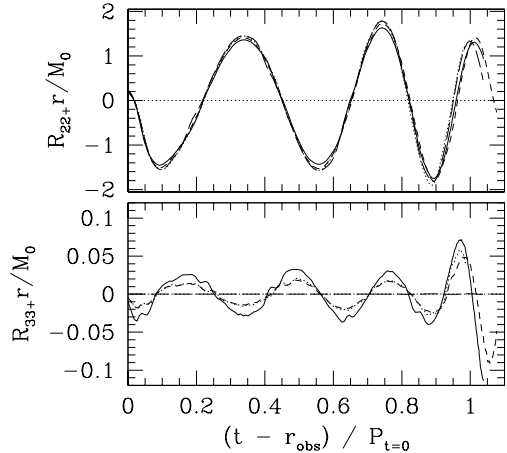}
\end{center}
\vspace{-2mm}
\caption{
$R_{22+}r$ and $R_{33+}r$ as a function of the retarded time
for models M1616 (long-dashed curves), M1517 (dashed curves), 
M1418 (solid curves), and M159183 (dotted curves). 
\label{FIG15}}
\end{figure}

In Fig. \ref{FIG14}, the gauge-invariant quantities 
for $(l,m)=(2,2)$, (3,3), and (2,0) for models M1315 and M1414 are 
displayed. In both cases, the inspiral waveforms are dominant for 
$t-r_{\rm obs} \alt P_{t=0}$ and quasiperiodic waves excited 
by nonaxisymmetric quasiperiodic oscillations are emitted 
after the merged objects are formed for $t-r_{\rm obs} \agt P_{t=0}$. 
It is found that the waveforms of the (2,2) mode for two models are 
similar but a slight difference can be seen in the quasiperiodic 
oscillation for $t - r_{\rm obs} \agt P_{\rm t=0}$. 
For model M1315, quasiperiodic waves appear to be mainly 
composed of a single oscillation mode. On the other hand, 
a non-negligible modulation can be observed in the waveforms 
for model M1414. This implies that they are composed of 
more than two dominant modes. 

Figure \ref{FIG14} shows that the gauge invariant variable for the
(2,0) mode does not oscillate around zero. This is due to the fact that
gravitational waves are extracted at a finite radius and, as a result,
this variable contains nonwave 
components associated with a stationary quadrupole moment. 
To calculate the Fourier spectrum of gravitational waves,
we first subtracted the stationary component
from $\hat R_{20}$ and, then, performed the Fourier transformation. 
In Fig. \ref{FIG17}, the Fourier spectra of the gauge-invariant
variables for models M1315 and M1414 are displayed. 
Here, $|\hat R_{lm}(f)f|r$ is plotted.
When taking a look at this figure, the following should be also kept in mind: 
(i) the spectra presented for the (2,2) mode are 
not realistic for $f < f_{\rm QE}$ because 
the spectra of inspiraling waveforms should be dominant in reality as
$|\hat R_{22}(f) f| \propto f^{-1/6}$ for $f \alt f_{\rm QE}$; 
(ii) the amplitude of the peaks found at $f \sim 2f_{\rm QE}$ 
and $3f_{\rm QE}$ for the (2,2) mode and at $f_{\rm QE}$ for the (2,0)
mode is underestimated because we stopped the simulations
during the oscillation of the formed hypermassive neutron stars 
to save computational time (see discussion below). 

For model M1315, a single peak is found at a frequency
$\approx 3.2f_{\rm QE}$ in the Fourier spectrum,
while for model M1414, two peaks of $\approx 2.0f_{\rm QE}$
and $2.95f_{\rm QE}$ are found for the (2,2) mode. 
The difference in the number of the peaks 
reflects the difference of the merger process. 
For model M1414, the merged object constitutes a nonaxisymmetric 
hypermassive neutron star of a double core structure, 
which quasiradially oscillates with a large amplitude. 
Therefore at least two modes (nonaxisymmetric 
and quasiradial oscillation modes) are contained. 
The peak at $f \sim 3f_{\rm QE}$ is associated with the nonaxisymmetric
bar-mode oscillation, while that at $f \sim 2f_{\rm QE}$ is 
produced by a modulation due to coupling between
the nonaxisymmetric and quasiradial oscillations 
because the difference of their frequencies is approximately
equal to $f_{\rm QE}$ which corresponds to the frequency of the
quasiradial oscillation. These two peaks in the Fourier spectra have
been also found in Newtonian \cite{C} and
post-Newtonian simulations \cite{FR,FR2}. 
On the other hand, for model M1315,
the merged object forms a hypermassive neutron star of
an asymmetric double core structure. In this object, the amplitude of the
quasiradial motion is not as large as that for model M1414. 
As a result, the peak at $f \sim 2f_{\rm QE}$ associated with
the modulation between the nonaxisymmetric and quasiradial 
oscillations is not as remarkable as that for model M1414. 
This feature has been also found in the post-Newtonian study \cite{FR2}. 

The frequency of the peaks for the (2,2) mode 
around $f \sim 3f_{\rm QE}$ for model M1315 is slightly larger than that
for M1414. This results from the fact that the maximum density
(or the compactness) of the formed hypermassive neutron stars
is larger for M1315 (see Fig. \ref{FIG6}). This fact indicates that
for the smaller rest-mass ratio, the gravitational wave frequency
associated with the nonaxisymmetric oscillation is higher
for a fixed total rest-mass of the system. 

Assuming that the total mass of the system is $2.8M_{\odot}$, 
$f_{\rm QE}$ is $\approx 750$ Hz for model M1414
and $\approx 700$ Hz for model M1315 according to Eq. (\ref{fqe}).
This implies that the peaks in the Fourier spectrum
appear at $\approx 1.5$ and 2.2 kHz for M1414 and
at $\approx 2.25$ kHz for M1315.
These frequencies will be too high to be detected by the 
first LIGO. However, these quasiperiodic gravitational waves
will be interesting targets for 
resonant-mass detectors and/or specially designed 
advanced interferometers such as the advanced LIGO \cite{KIP}. 

It should be mentioned
that the peak frequencies in the post-Newtonian simulation
\cite{FR2} are smaller than those found in our study for 
a given neutron star mass and radius $\approx 15$ km. 
This may be due to the fact that, in our fully general
relativistic simulation, the gravity is taken into account
correctly and is stronger than that in the post-Newtonian
approximation. 
Consequently, the formed hypermassive neutron star is more compact
and hence the oscillation frequency higher.

The magnitude of the quasiradial oscillation is reflected in 
the amplitude of gravitational waves for the (2,0) mode \cite{SS}. 
In the early phase ($t-r_{\rm obs} \alt P_{t=0}$), 
this mode is dominated by a stationary quadrupole mode
which is not associated with gravitational waves, 
but after a hypermassive neutron star is formed 
it becomes a dominant component. 
Figure \ref{FIG14} shows that the amplitude of the (2,0) mode
for model M1414 is larger than that for M1315 by a factor of $\sim 2$. 
This results from the fact that the amplitude of the quasiradial
oscillation for M1414 is larger than that for M1315. 

The frequency of gravitational waves for the (2,0) mode is within
the sensitive band of kilometer-size laser interferometers as
$\sim f_{\rm QE} \sim 0.7(2.8M_{\odot}/M_0)$ kHz. Although
the amplitude is $\sim 5 \%$ of that of the dominant (2,2) mode,
this mode does not damp soon as indicated in Fig. \ref{FIG6}(a).
Therefore if the cycles are accumulated using a theoretical template,
the effective amplitude may be much larger than that for
one cycle and may be as large as the amplitude of
$R_{22}(\sim M_0/r)$ at $f \sim f_{\rm QE}$. Unfortunately,
it is difficult to exactly estimate the effective magnitude
from the present numerical results of finite duration. However,
gravitational waves of this mode may be an interesting target even for the
first-generation gravitational wave detectors such as the first LIGO.

\subsubsection{Dependence of inspiral waveforms on mass ratios}

From Fig. \ref{FIG15}, we find that the maximum amplitude
is smaller for models of the smaller rest-mass ratios.
According to the quadrupole formula, 
the maximum amplitude for a given total mass
is proportional to $Q_M/(1+Q_M)^2$ 
which is in a small range 0.246--0.25 for $0.8 < Q_M < 1$. 
This suggests that the maximum amplitude would depend weakly 
on the value of $Q_M$. However, the ratios of the maximum amplitude 
for models M1517 and M1418 to that for model M1616 are 0.966 and 
0.909, respectively. This implies that the maximum amplitude is suppressed 
with the decrease of $Q_M$. This results from the fact that 
the tidal effect plays a more important role in the close binaries
of the smaller rest-mass ratios: Since the tidal disruption sets in
at a larger orbital separation for the smaller rest-mass ratios,
the maximum amplitude should be decreased. [Similar
results are also found in Fig. \ref{FIG16}(b) (see below).] 
This property has been reported in 
the Newtonian and post-Newtonian studies, too \cite{RS,FR,FR2}.
According to \cite{FR,FR2}, the suppression factor is proportional 
to $Q_M$ (for a fixed value of $M_*$), 
agreeing with our results approximately. 

Another difference of gravitational waveforms 
between models of equal-mass and unequal-mass binaries can be seen 
in the modes of odd values of $m$ (Figs. \ref{FIG14} and \ref{FIG16}).
For the merger of equal-mass binary neutron stars, the amplitude for
those modes is zero because of the $\pi$-rotation symmetry. On the other hand, 
it is not negligible for the mergers of unequal-mass binaries. 
However, the amplitude is at most 5\% of that of the (2,2) mode.

\subsubsection{Radiated energy and angular momentum}

\begin{figure}[t]
\begin{center}
\epsfxsize=2.75in
\leavevmode
(a)\epsffile{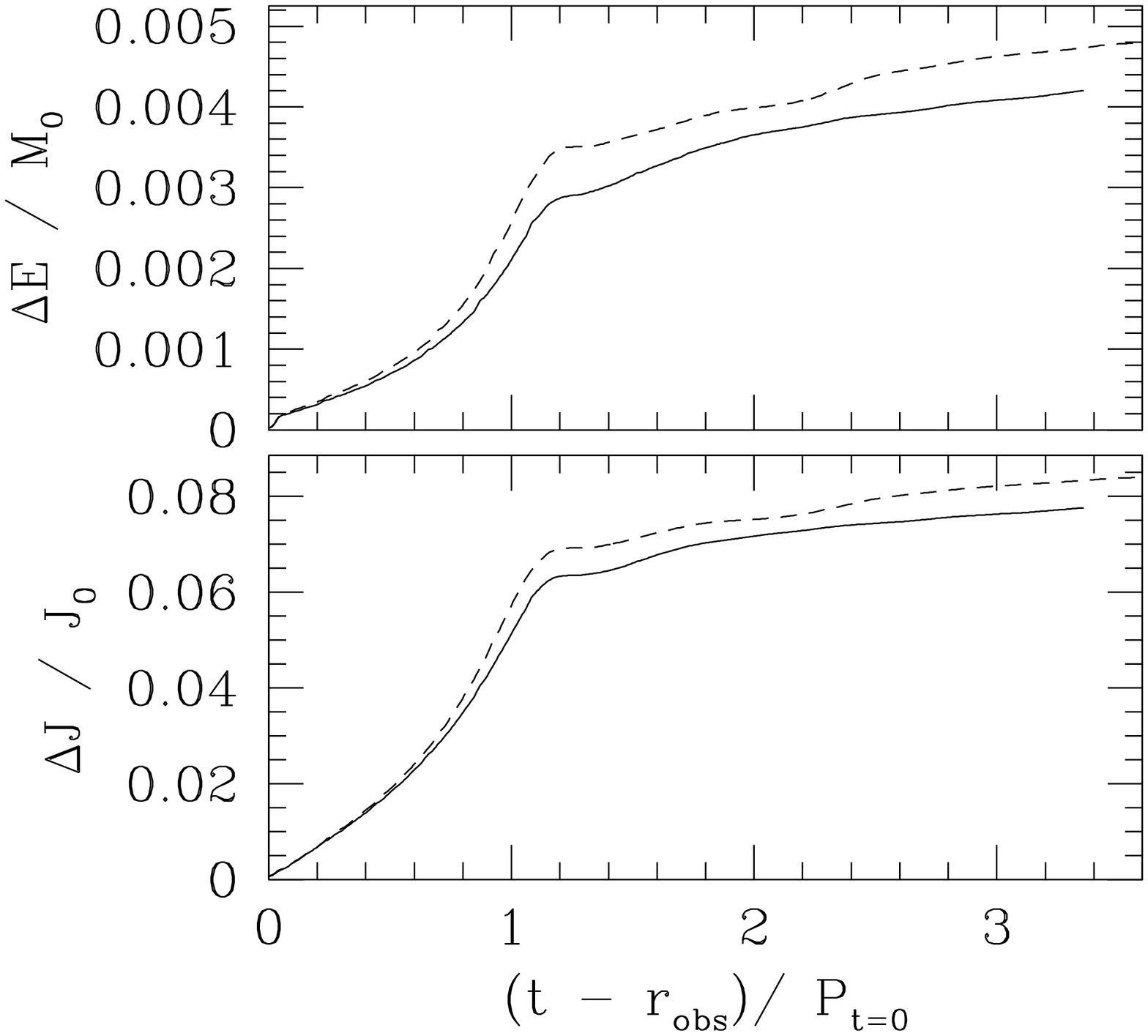}
\epsfxsize=2.75in
\leavevmode
~~~~~(b)\epsffile{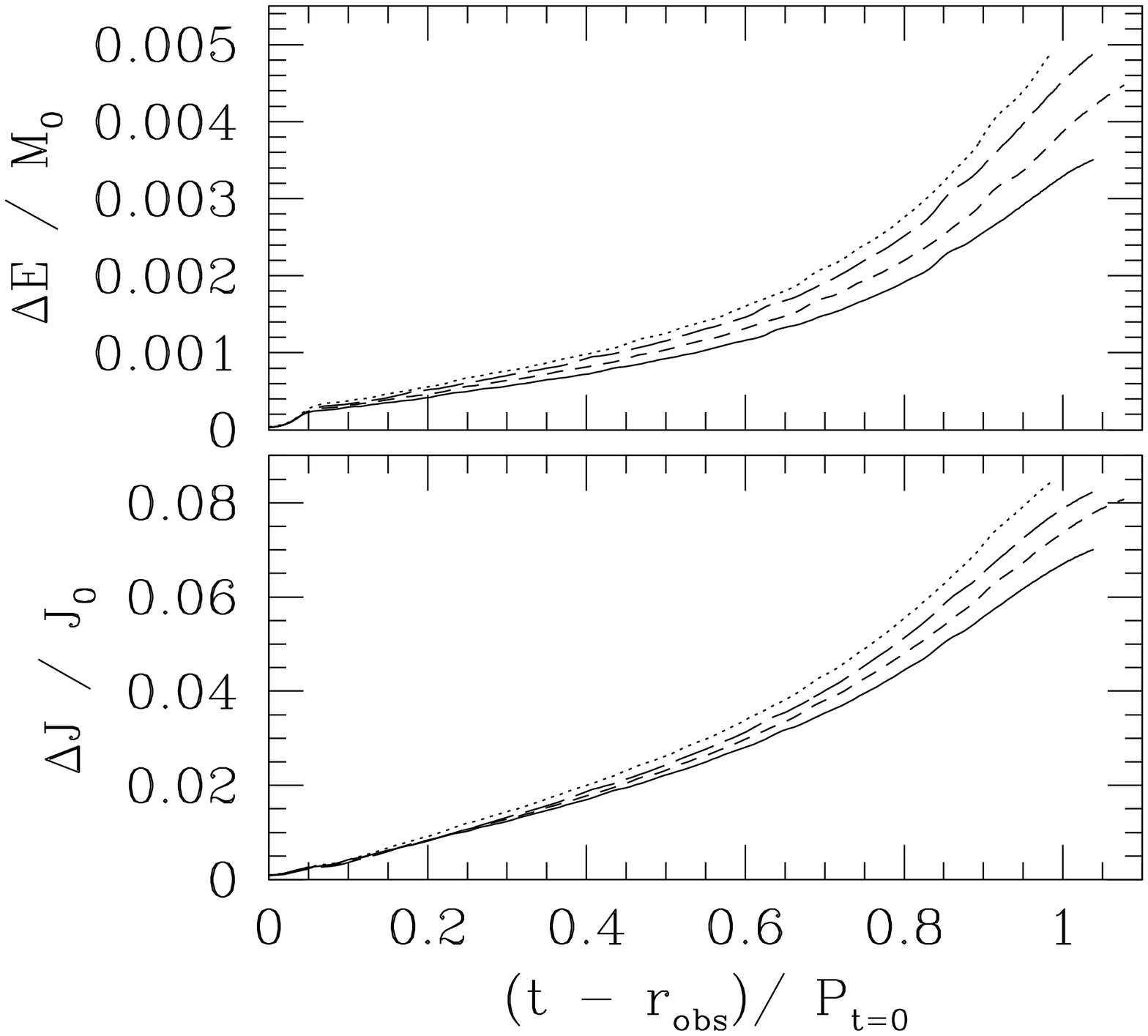}
\end{center}
\vspace{-2mm}
\caption{
The accumulated energy and angular momentum loss by
gravitational radiation as a function of the retarded time
(a) for models M1315 (solid curves) and M1414 (dashed curves), and 
(b) for models M1418 (solid curves), M1517 (dashed curves), 
M1616 (long-dashed curves), and M159183 (dotted curves). 
\label{FIG16}}
\end{figure}

\begin{figure}[t]
\begin{center}
\epsfxsize=2.8in
\leavevmode
(a)\epsffile{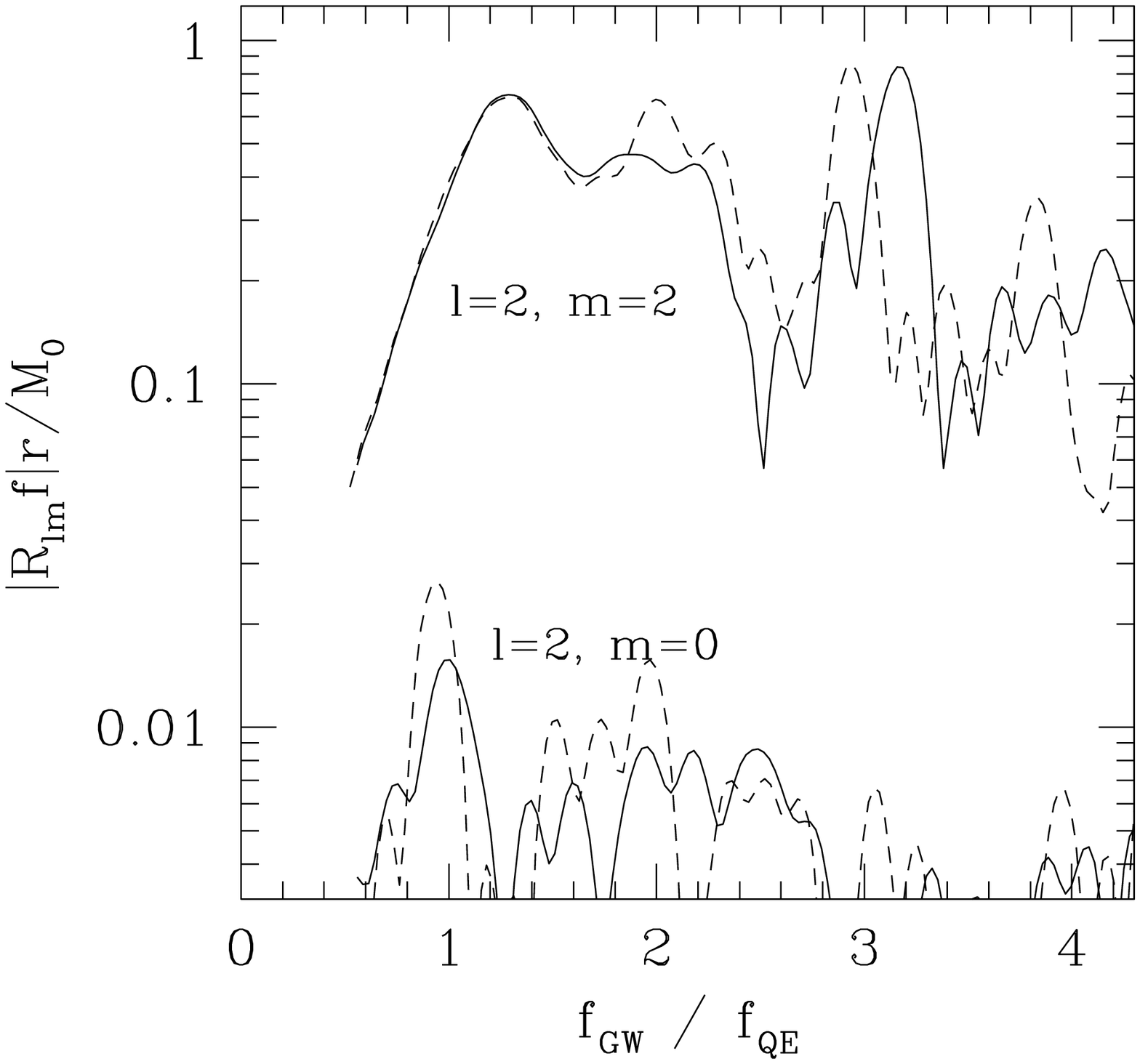}
\epsfxsize=2.8in
\leavevmode
~~~~(b)\epsffile{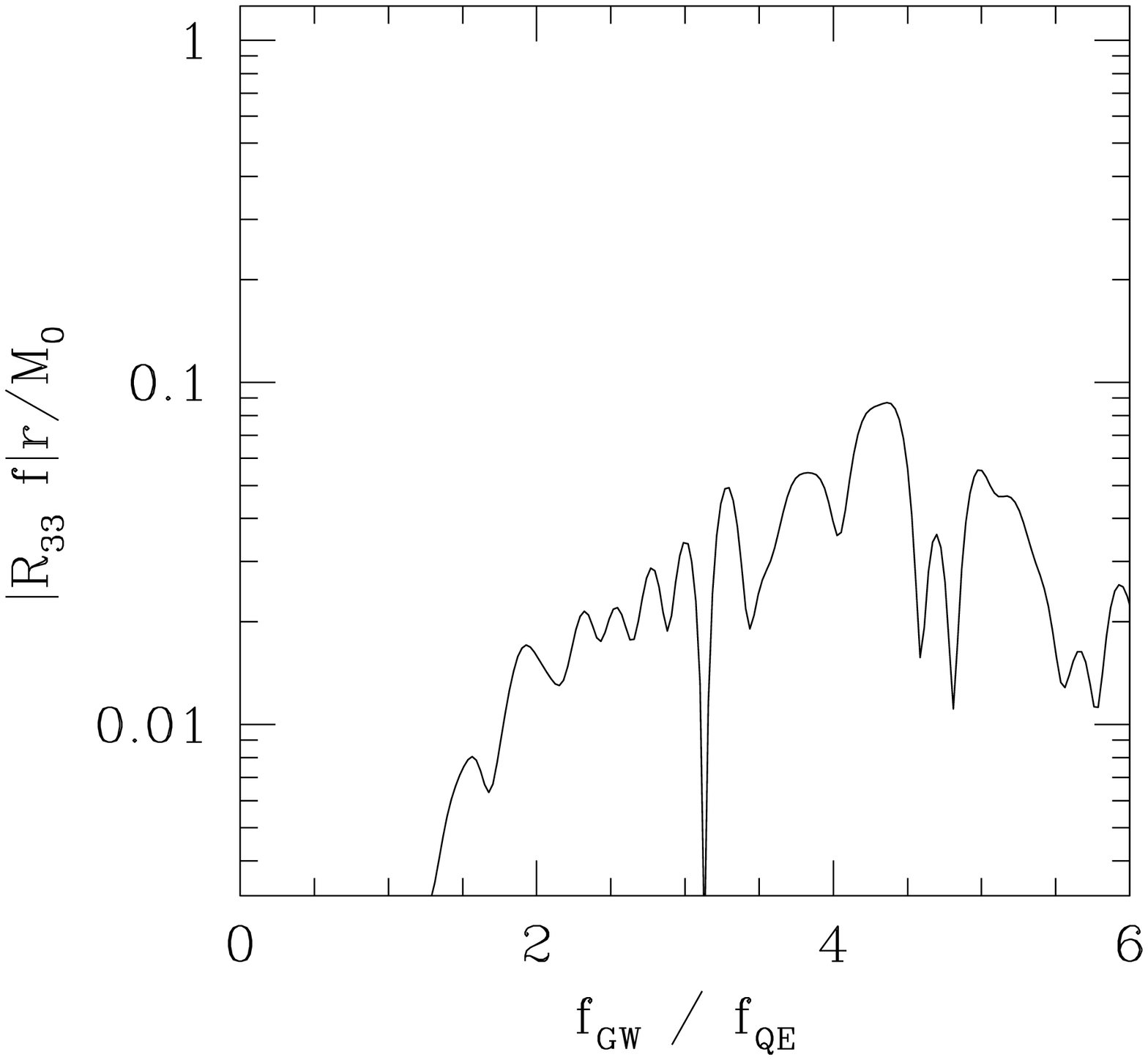}
\end{center}
\vspace{-2mm}
\caption{(a) Fourier spectrum of $l=m=2$ and $l=2,m=0$ 
for models M1414 (dashed curve) and M1315 (solid curve).
(b) Fourier spectrum of $l=m=3$ for model M1315. 
\label{FIG17}}
\end{figure}

Figure \ref{FIG16} shows that in the final inspiral phase 
($t-r_{\rm obs} \alt P_{t=0}$), $\sim 0.3$--0.5\% of the 
initial ADM mass and $\sim 6$--8\% of the initial angular momentum are 
carried away by gravitational radiation. ($M$ does not change much 
but $J$ decreases significantly.) This implies that the nondimensional 
angular momentum parameter $q$ decreases by $\sim 5$--7\% due to the 
gravitational radiation. Since a large fraction of baryon mass
of the system is swallowed into a black hole 
for models M1616, M1517, M1418, and M159183, 
the ADM mass of the black hole should be $\sim M_0$ within a few 
percents error. From this fact, we may expect that the final value of $q$ 
is $\sim 0.95 q_0$ within a few percents error and, hence, it is 
in the range between $0.8$ and 0.9. This value is approximately 
consistent with the value derived from the area of apparent horizons 
computed at the termination of the simulations. 

Figure \ref{FIG16} also indicates that 
the energy and angular momentum loss by gravitational radiation
decrease with the decrease of rest-mass ratios. 
The reason is that
tidal disruption takes place before the orbital separation
becomes as small as the sum of radii of two stars 
for the merger of unequal-mass neutron stars. The orbital separation
at the tidal disruption is larger for the smaller rest-mass ratios 
for a fixed value of the total rest-mass of the system.
This implies that the maximum value of the
compactness of binary orbit is smaller for the smaller rest-mass ratios, 
resulting in that the amount of gravitational radiation becomes smaller. 

For the case of hypermassive neutron star formation, 
the energy and the angular momentum are carried away gradually
due to gravitational radiation emitted by the quasiperiodic
nonaxisymmetric oscillations. 
Since the emission time scale is much longer than the dynamical
time scale, it is impossible to follow the longterm 
evolution of the hypermassive neutron stars to the final state. 
If we assume that the angular momentum is dissipated 
by gravitational waves with the same rate as that at the termination
of the simulation, the angular momentum will become smaller than $0.1J_0$ 
around $t \agt 300P_{t=0}$. Since the hypermassive neutron stars
are supported by the centrifugal force, 
they will collapse to a black hole 
as a result of the angular momentum dissipation within $\sim 1$ sec. 

In the SPH calculations \cite{FR,FR2,ORT}, 
the quasiperiodic oscillations of the hypermassive
neutron star damp in much shorter time than in our numerical results.
If we believe their results, the lifetime of the hypermassive 
neutron stars would be much longer. 
The reason for the discrepancy between our and their results is unclear. 
However, as far as our simulations are concerned,
there is no reason for the damping of the nonaxisymmetric oscillation
in such a short time scale since the emission time scale of 
gravitational waves is much longer than one oscillation period
and other damping processes such as dynamical angular momentum
transfer are unlikely to work efficiently. We suspect that
damping found in previous works may be due to a spurious numerical
dissipation or due to an overestimation of gravitational radiation
damping in the post-Newtonian formalism they adopted.

\subsection{Gravitational radiation reaction}

The ADM mass $M$ and the angular momentum $J$ computed in a finite 
computational domain using Eqs. (\ref{eqm00}) and (\ref{eqj00})
decrease with time because of the gravitational radiation.  
However, conservation laws (\ref{eqm01}) and 
(\ref{eqj01}) should still be satisfied. Here, we demonstrate that they 
are satisfied approximately in the present simulations. 

\begin{figure}[t]
\begin{center}
\epsfxsize=2.75in
\leavevmode
(a)\epsffile{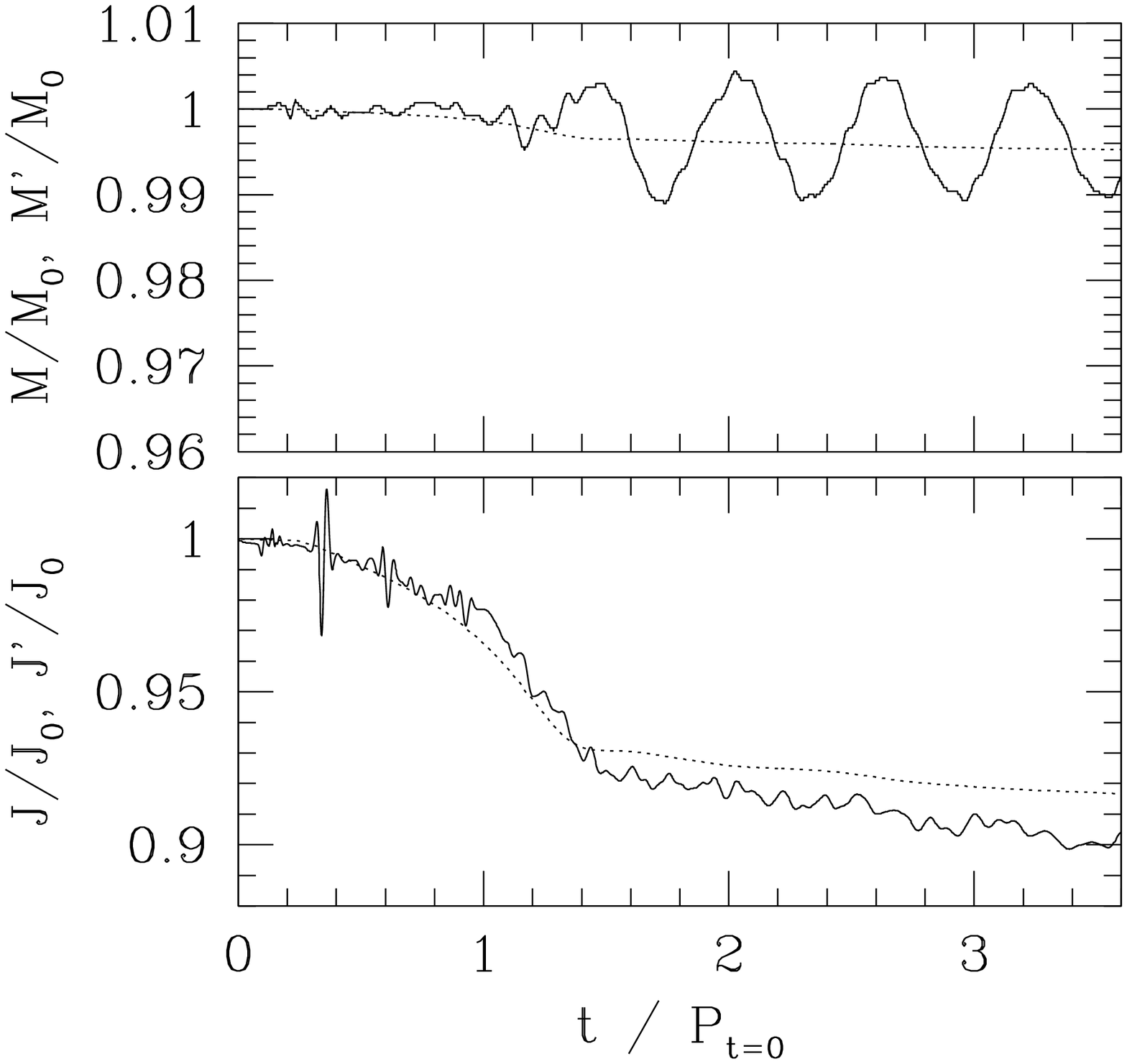}
\epsfxsize=2.75in
\leavevmode
~~~(b)\epsffile{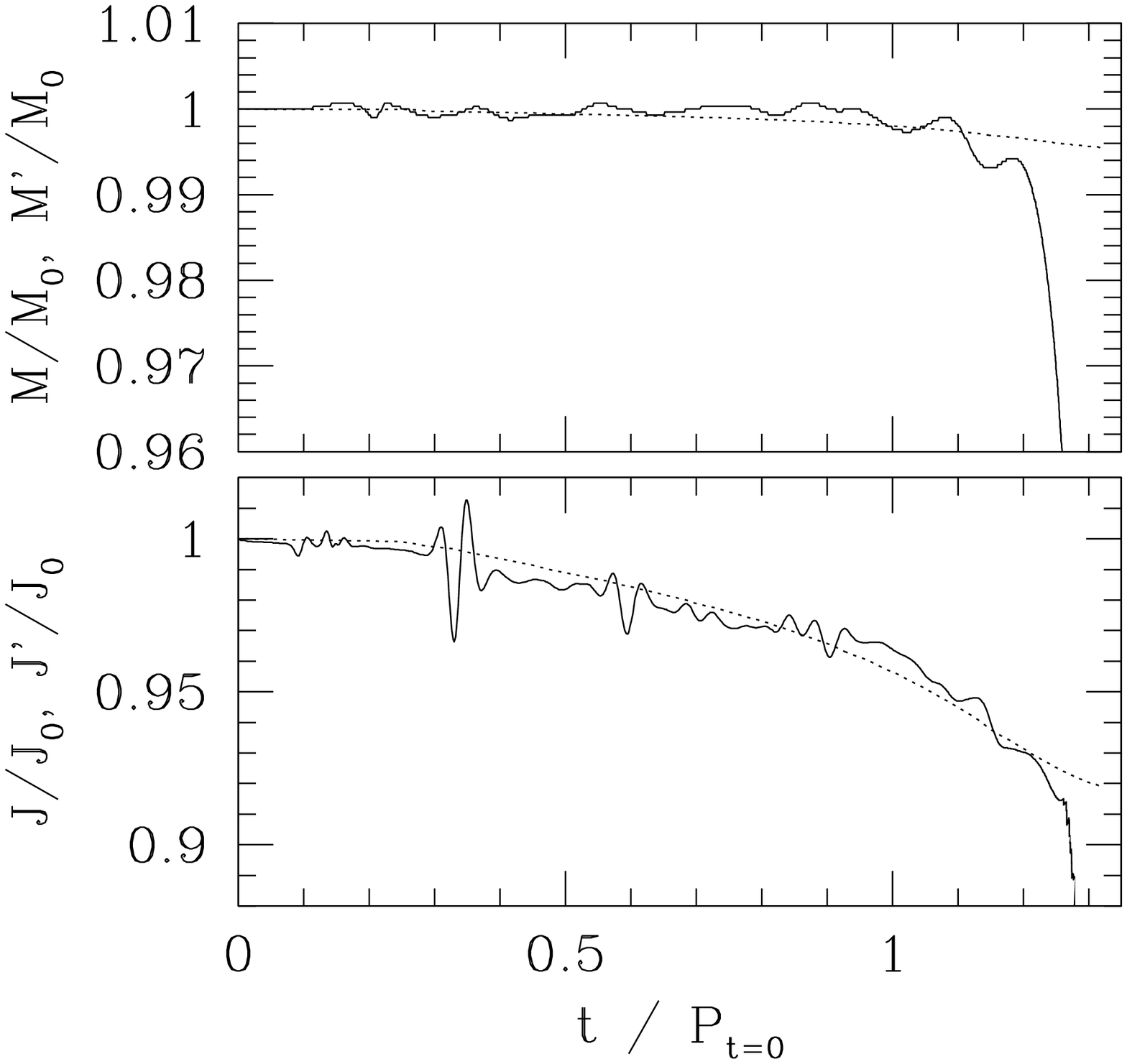}
\end{center}
\vspace{-2mm}
\caption{Evolution of the total energy and angular momentum of
the system calculated by Eqs. (\ref{eqm00}) and (\ref{eqj00})
(solid curves) and that calculated by Eqs. (\ref{eqm02}) and (\ref{eqj02})
(dotted curves) (a) for model M1414 and (b) for model M1517. 
\label{FIG18}}
\end{figure}

Figure \ref{FIG18} shows the time evolution of $M$ and $J$ 
(solid curves) and of the quantities defined 
by the following equations (dotted curves) for models M1414 and M1517:
\beqn
&& M'(t) = M_0 - \Delta E(t),\label{eqm02}\\ 
&& J'(t) = J_0 - \Delta J(t).\label{eqj02} 
\eeqn
The relations $M'(t)=M(t)$ and $J'(t)=J(t)$ are equivalent to 
the conservation of the total ADM mass energy and angular momentum. 
Figure \ref{FIG18} indicates that relations $M=M'$ and $J=J'$ 
are satisfied within $\sim 1\%$ error except for the 
phase in which the merged object collapses to a black hole and, 
as a result, the grid resolution becomes too poor. 

In fully general relativistic simulations, 
the numerical accuracy is restricted by grid resolution and 
by the approximate 
outer boundary conditions imposed 
in a local wave zone. The results presented here
indicate that these errors are 
suppressed within $\sim 1\%$ error in our simulations
in the absence of a black hole. (In the presence of a black hole, the
errors increased to $\sim 10\%$ and the computation crashed.)

The conservation of the angular momentum
which holds approximately in our present simulations 
is a necessary condition for studying the formation of disks and 
a hypermassive neutron star supported 
by centrifugal force, and the final value of $q$ of a black hole. 
The results here indicate the reliability of the numerical results 
on the formation of disks and hypermassive neutron stars, and on 
determination of the final value of $q$ presented in Sec. IV B
and IV C.

\subsection{Calibrations}

\begin{figure}[t]
\begin{center}
\epsfxsize=2.75in
\leavevmode
\epsffile{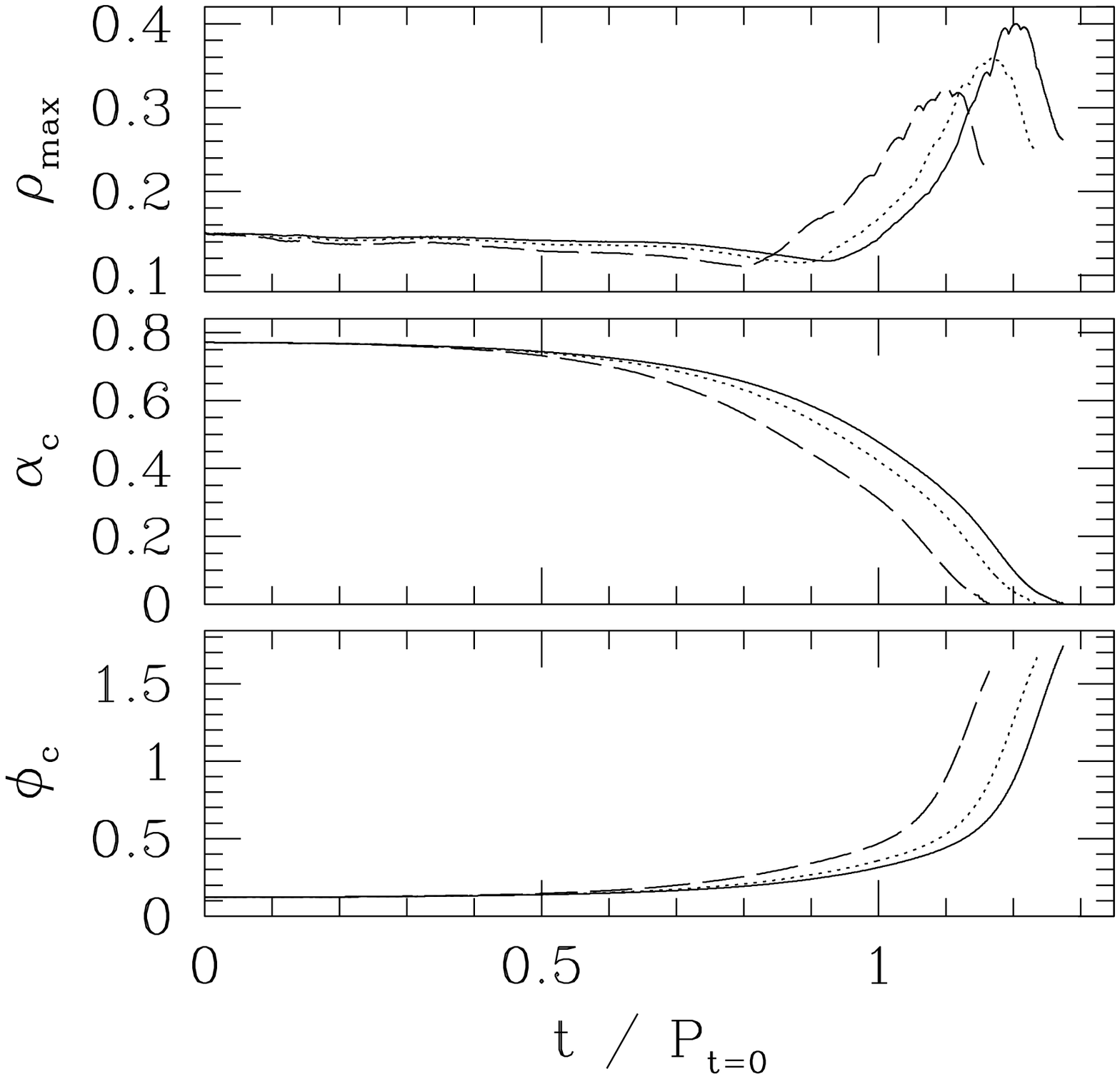}
\epsfxsize=2.75in
\leavevmode
\epsffile{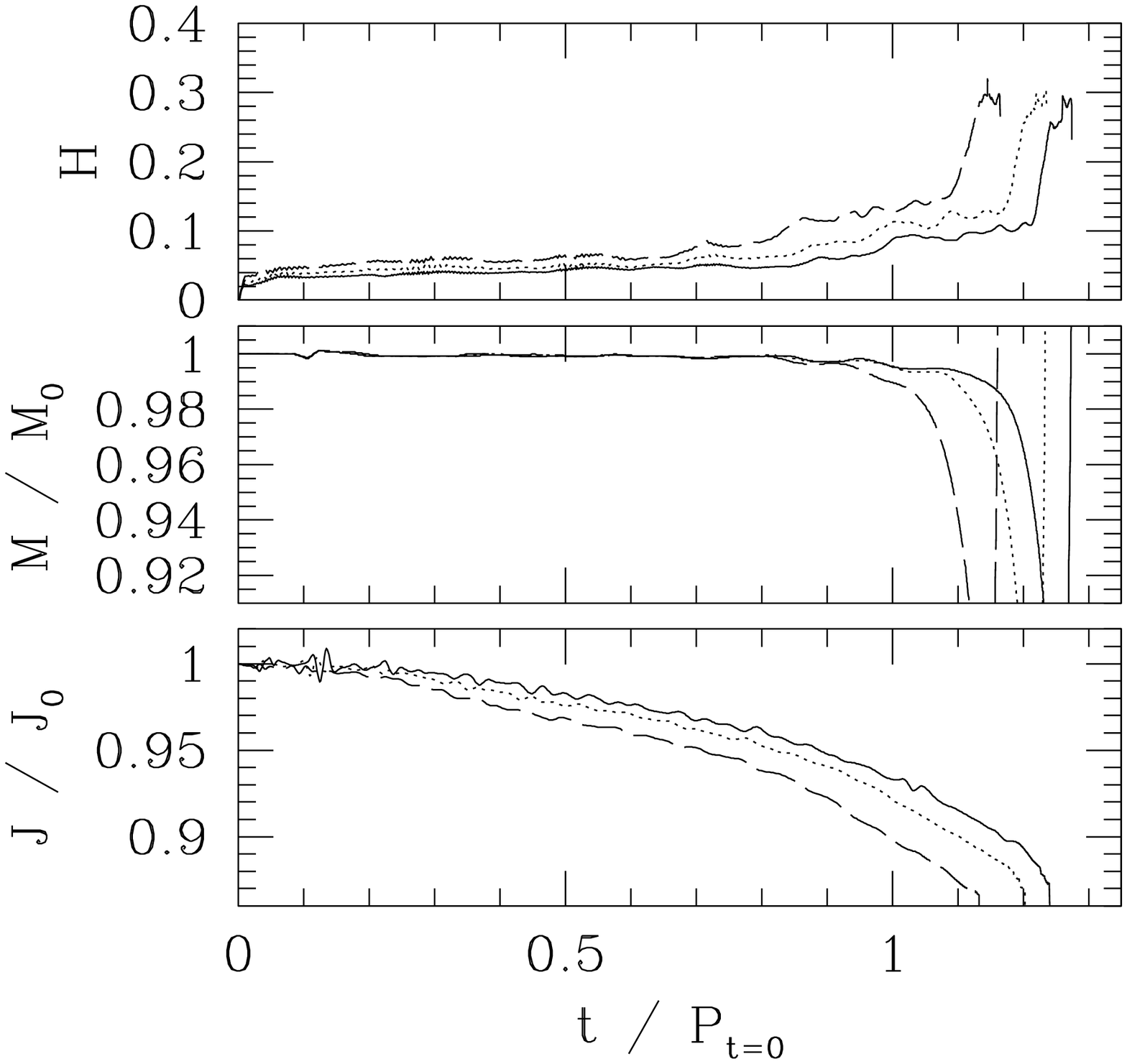}
\end{center}
\vspace{-2mm}
\caption{Time evolution of the maximum density, 
the central values of $\alpha$ and $\phi$, the averaged 
violation of the Hamiltonian constraint ($H$), $M$, and $J$ 
for models M1616-3 (dotted curves), M1616-4 (solid curves),
and M1616-5 (long-dashed curves).
In this figure, effects with regard to the grid resolution are clarified. 
\label{FIG19}}
\end{figure}

\begin{figure}[htb]
\begin{center}
\epsfxsize=2.75in
\leavevmode
\epsffile{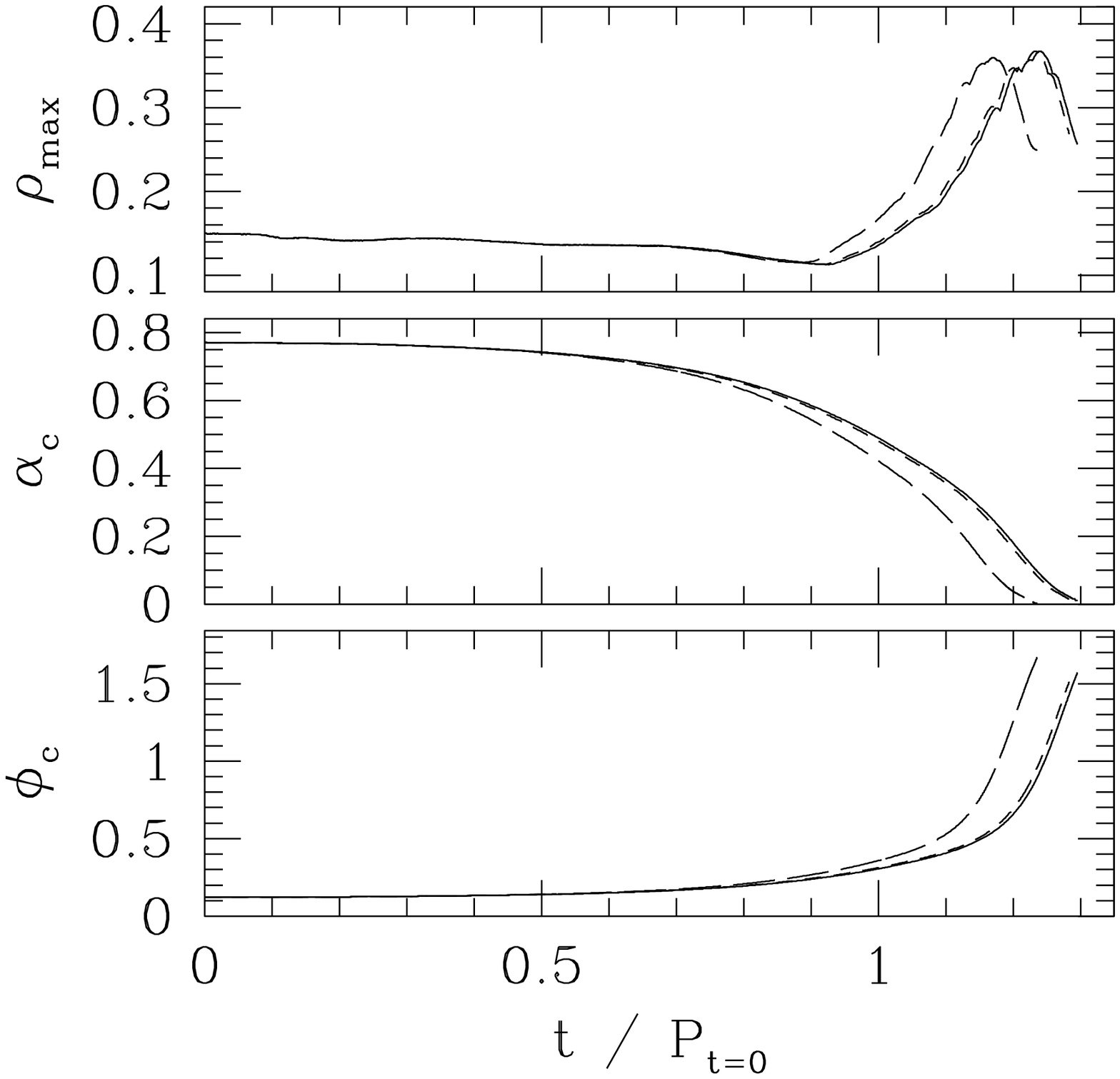}
\epsfxsize=2.75in
\leavevmode
\epsffile{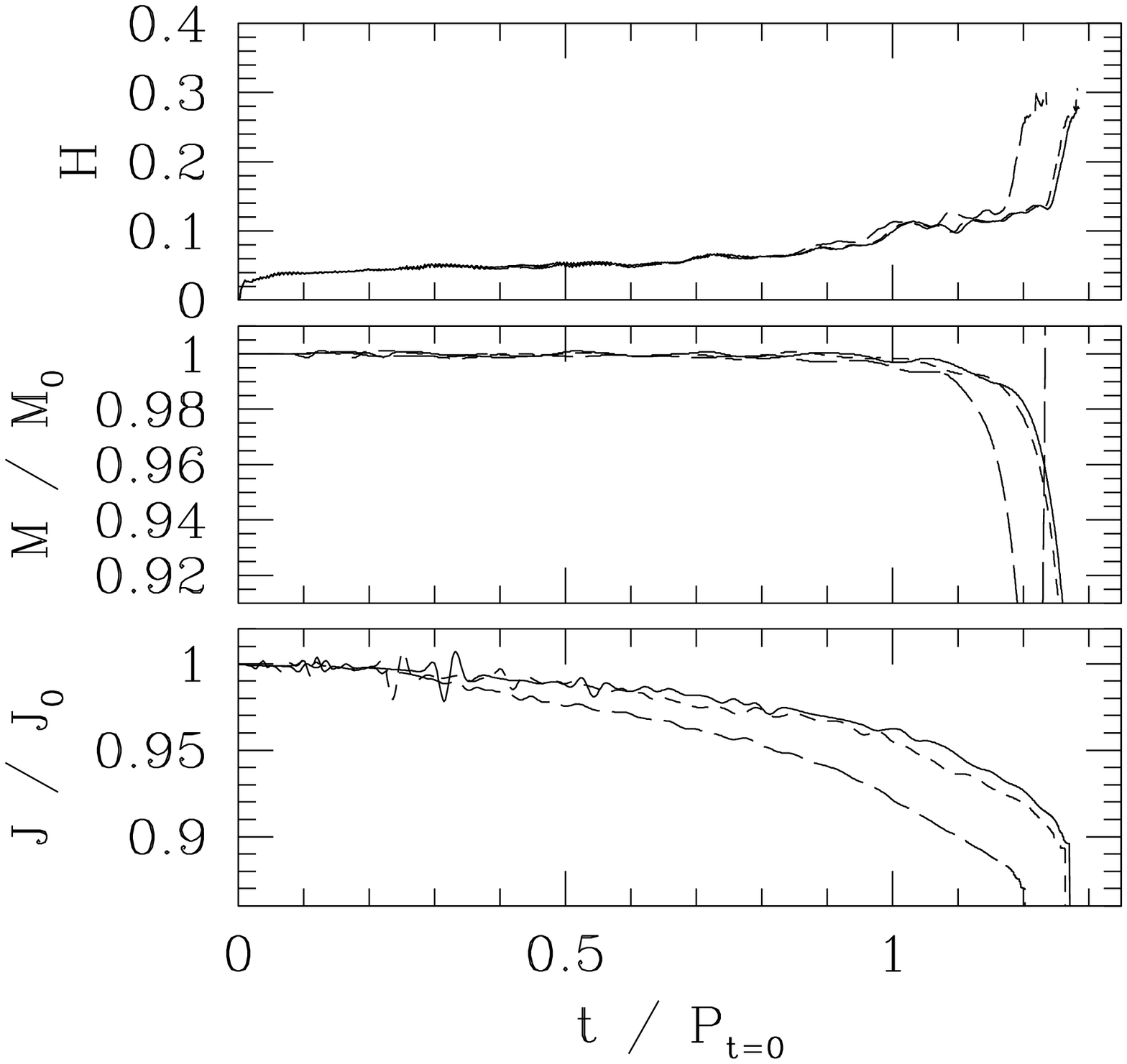}
\end{center}
\vspace{-2mm}
\caption{The same as Fig. \ref{FIG19}, but 
for models M1616 (solid curves), M1616-2 (dashed curves), and
M1616-3 (long-dashed curves). In this figure, effects
with regard to the location of the outer boundaries are clarified. 
\label{FIG20}
}
\end{figure}

\begin{figure}[t]
\begin{center}
\epsfxsize=2.75in
\leavevmode
(a)\epsffile{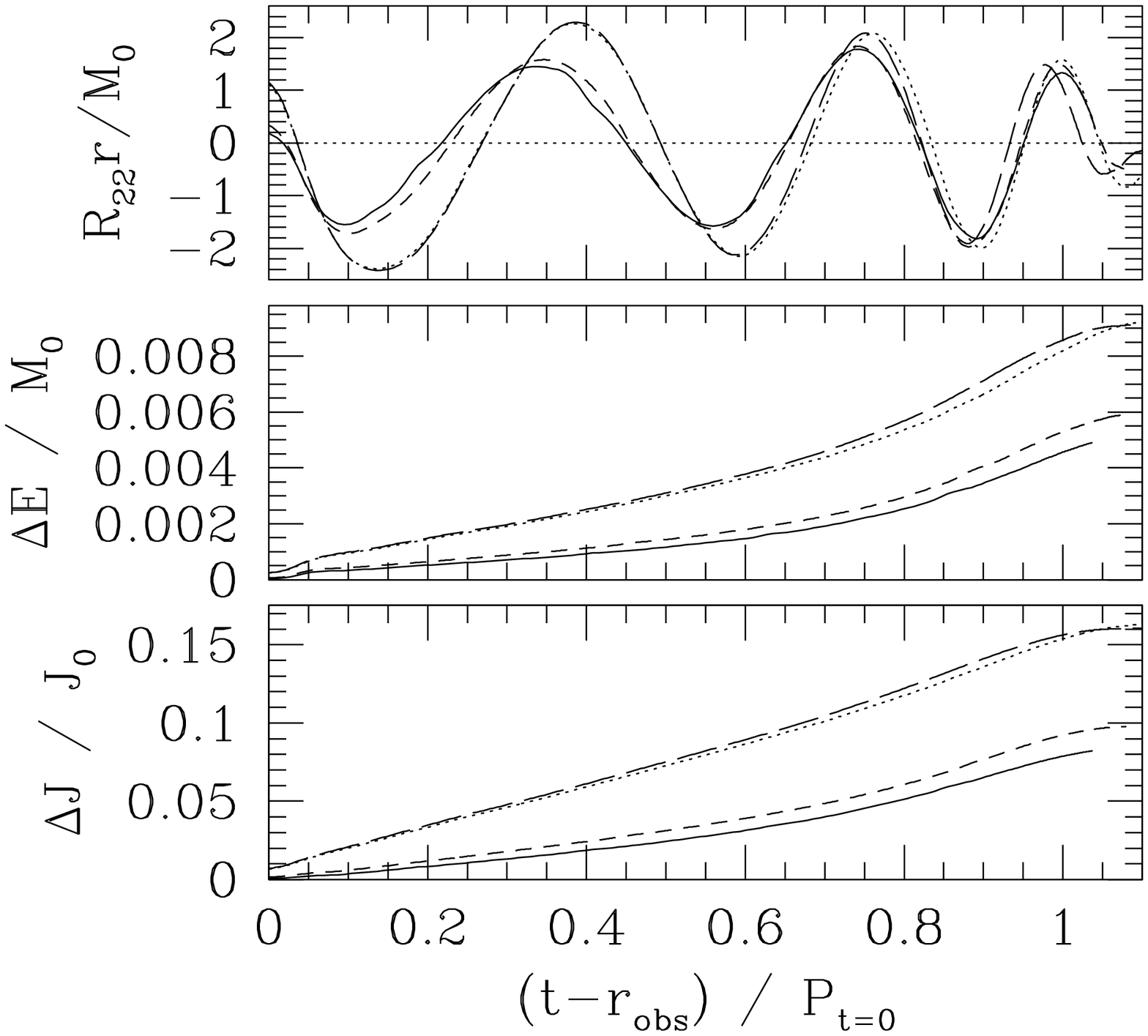}
\epsfxsize=2.75in
\leavevmode
~~~~~(b)\epsffile{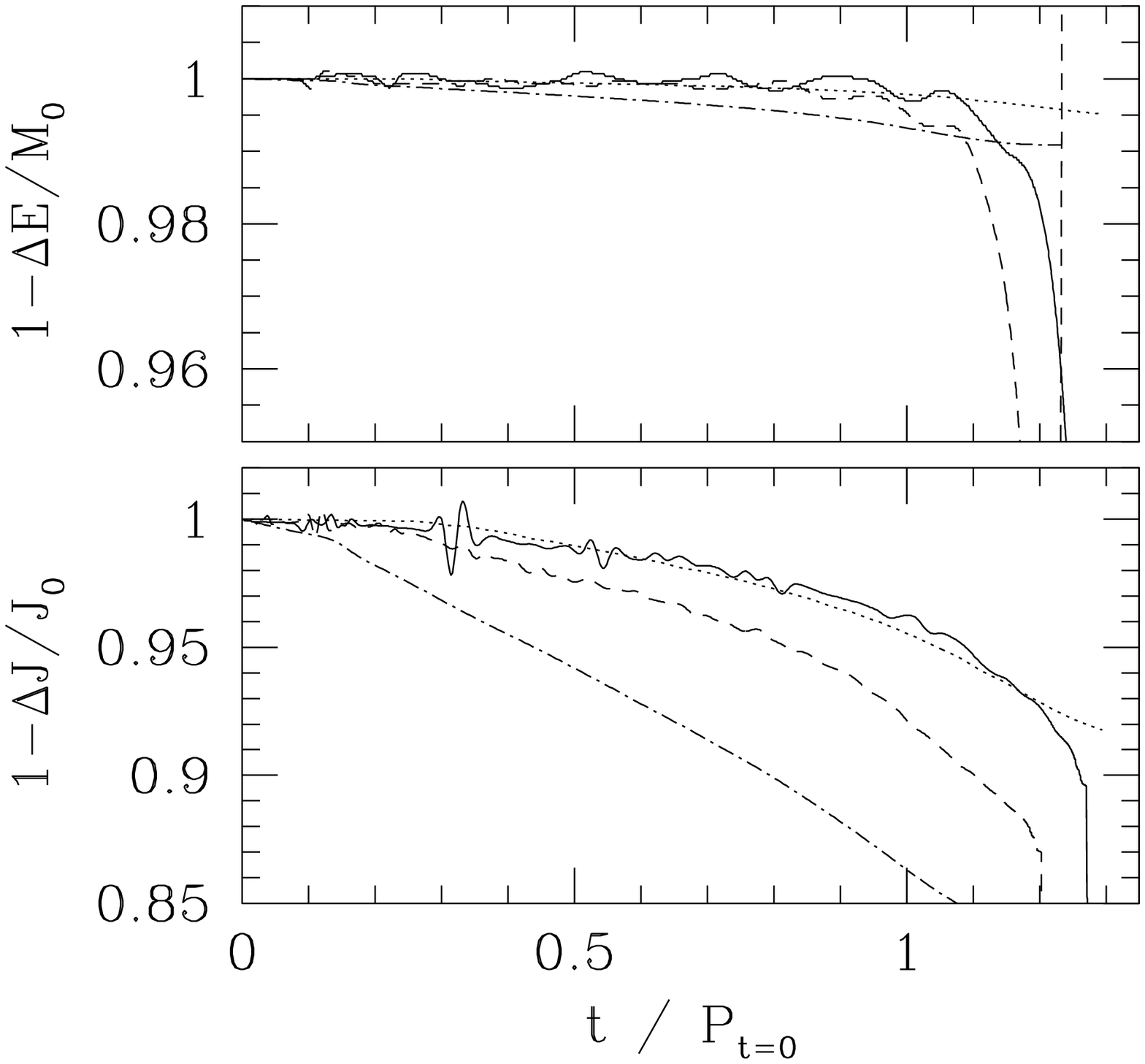}
\end{center}
\vspace{-2mm}
\caption{
(a) Gravitational waveforms ($R_{22+}$), and accumulated energy and
angular momentum of gravitational radiation for models M1616 (solid curves), 
M1616-2 (dashed curves), M1616-3 (long-dashed curves), and
M1616-4 (dotted curves). 
(b) The same as Fig. \ref{FIG18}(b), but for models M1616
(solid and dotted curves) and M1616-3 (dashed and dotted-dashed curves).
The solid and dashed curves denote $M/M_0$ and $J/J_0$, and
the dotted and dotted-dashed curves $1-\Delta E/M_0$ and $1-\Delta J/M_0$. 
\label{FIG21}
}
\end{figure}

Convergence tests were performed employing models M1616 and M1414.
The test simulations were done for additional five models
as listed in Table II. To investigate effects of the location of the
outer boundaries at which approximate boundary conditions were imposed,
the values of $L$ were changed for three levels as 
$L/\lambda_0=0.533$ (M1616), 0.425 (M1616-2), and 0.263 (M1616-3) 
fixing the grid spacing. To see effects with regard to the grid resolution, 
we also performed two additional simulations for models M1616-4 and
M1616-5 in which the location of the outer boundaries 
was the same as that for M1616-3, but the grid spacings were about 5/6 and 
5/4 times, respectively, that for M1616-3.  
A simulation for model M1414-2 was performed to clarify weak dependence
of gravitational waveforms from quasiperiodic oscillations of
a hypermassive neutron star on the value of $L$.

\subsubsection{Convergence test with regard to grid resolution}

Figure \ref{FIG19} shows the evolution of the maximum density, 
the central values of $\alpha$ and $\phi$, 
the averaged violation of the Hamiltonian constraint
$H$ [computed by Eq. (\ref{vioham})], $M$ [computed by Eq. (\ref{eqm00})],
and $J$ [computed by Eq. (\ref{eqj00})] for models M1616-3 (dotted curves),
M1616-4 (solid curves), and M1616-5 (long-dashed curves). 
This figure indicates the dependence of the numerical results 
on the grid resolution for a fixed value of $L$. 
It is found that the convergence of $H$ is at first order. 
A likely reason is as follows: 
Since the vacuum is not allowed in our hydrodynamic implementation,
we have to add an atmosphere of small density outside neutron stars.
In the present work, the density of the atmosphere
is $\sim 10^{-7}$ in units of $\kappa=1$. As a result,
a very steep density gradient appears at the stellar surface. 
In such a region, the transport term of the hydrodynamic 
equations is computed with first-order accuracy 
in space. This effect seems to be non-negligible in 
determining the global order of the accuracy. 

The angular momentum is dissipated and transported
unphysically by numerical effects. 
For the larger grid spacing, the dissipation rate is larger and,
as a result, the duration of the inspiral phase becomes shorter. 
Even in the case of the best resolution (M1616-4), the angular momentum
appears to be dissipated by $\sim 1\%$.
This effect may be the main source for the discrepancy between 
$J$ and $J'$ [see Fig. \ref{FIG18}(a)] in the late
phase $t - r_{\rm obs}\agt P_{t=0}$ for model M1414. 

\subsubsection{Convergence test with regard to $L$}

In Fig. \ref{FIG20}, we show the same figure as that of Fig. \ref{FIG19}
but for models M1616 (solid curves), M1616-2 (dashed curves),
and M1616-3 (long-dashed curves) to make a comparison among
the numerical results with the different values of $L$ and 
a fixed grid spacing. 
It is found that (i) for a smaller value of $L$, the merged 
object collapses earlier, (ii) $H$ depends very weakly on the value of $L$, 
and (iii) the results for models M1616 and M1616-2 are almost identical. 

The reason for (i) is that 
the magnitude of the radiation reaction is overestimated 
with small values of $L$. To explain this effect, 
gravitational waveforms, the radiated energy, and the radiated
angular momentum are shown in Fig. \ref{FIG21}, which indicate that 
the numerical results for models M1616 
and M1616-2 are approximately identical. This implies that
with $L \agt 0.5 \lambda_0$, a convergent result may be achieved. 
On the other hand, with the smaller value of $L < 0.5 \lambda_0$, 
the amplitude of gravitational waves, the 
radiated energy, and the radiated angular momentum are overestimated. 
The radiated energy and angular momentum for model M1616-3 
are about twice as large as those for M1616.
As a result, the orbital separation for M1616-3 decreases more rapidly 
than that for M1616. Moreover, the radiation reaction is not
accurately computed for model M1616-3, so that the 
conservation of the angular momentum ($J+\Delta J=J_0$) is largely
violated [see Fig. \ref{FIG21}(b)]. 
A number of numerical simulations for the binary merger
in full general relativity have been recently performed 
with $L < 0.5\lambda_0$ \cite{bina,marks,illinois}. 
Figure \ref{FIG21}(b) warns that the gravitational
waveforms and the merger process in such numerical simulations 
are not very reliable.

\begin{figure}[t]
\begin{center}
\epsfxsize=2.75in
\leavevmode
\epsffile{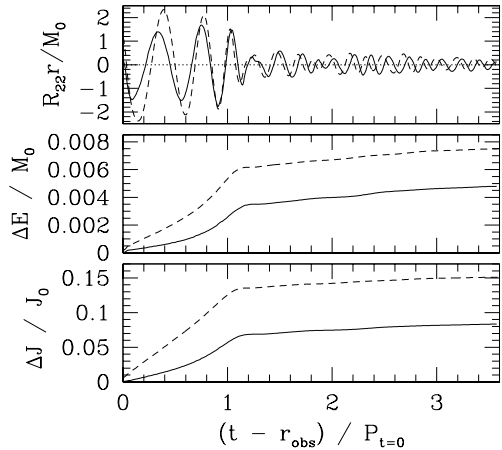}
\end{center}
\vspace{-2mm}
\caption{Gravitational waveforms ($R_{22+}$) and 
accumulated energy and angular momentum 
of gravitational radiation for models M1414 (solid curves) and 
M1414-2 (dashed curves). 
\label{FIG22}}
\end{figure}

Figure \ref{FIG22} is the same figure as that of Fig. \ref{FIG21}(a)  
but for models M1414 (solid curves) and M1414-2 (dashed curves), for 
which the outer boundaries are located at 
$L=0.510\lambda_0$ and $0.252\lambda_0$, respectively. 
For $t - r_{\rm obs} \alt P_{t=0}$, the amplitude of 
gravitational waves, the radiated energy, and the radiated angular momentum
are overestimated for the smaller value of $L$. 
Since the angular momentum is dissipated more rapidly from the
system, the inspiral phase is shorter and
the merger sets in earlier for model M1414-2. 
This results in a phase difference between gravitational waves 
of M1414 and M1414-2. However, for $t - r_{\rm obs} \agt P_{t=0}$,
the amplitude of gravitational waves, the energy luminosity, and 
the angular momentum flux are approximately in agreement between two results 
(besides the slight disagreement in the wave phase). 
This figure shows that quasiperiodic waves emitted from 
oscillating hypermassive neutron stars is calculated accurately
with our choice of $L$, since its wavelength is short enough 
to compute these quantities even for the smaller value of $L$.

\section{Summary}

We performed fully general relativistic simulations 
for the merger of binary neutron stars focusing 
particularly on the unequal-mass case. The following is a summary 
of the scientific results obtained in this paper: 

\begin{itemize}

\item If the total rest-mass of the system is more than 1.7 times of
the maximum allowed rest-mass of spherical neutron stars
a black hole is formed for the $\Gamma$-law equation of state with $n=1$. 
The nondimensional angular momentum parameter of the formed Kerr black hole 
is likely to be in the range between 0.8 and 0.9. 
%%%%%%%%%%%%%%
\item Disk mass around a black hole formed after the merger increases 
with the decrease of rest-mass ratios for a fixed value of the
total baryon rest-mass of binary neutron stars.
It is found that for the rest-mass ratio $\sim 0.85$, 
the disk mass may be several percents of the total mass of the system
if two neutron stars are not very compact.  
%%%%%%%%%%%%%
\item Disk mass around a black hole formed after the merger 
decreases with the increase of the compactness of the system 
for a fixed value of the rest-mass ratio. 
%%%%%%%%%%%%%
\item Shape of the hypermassive neutron stars formed after the merger
depends on the rest-mass ratio of binaries. For the merger 
of equal-mass neutron stars, a hypermassive neutron star of 
a double core is formed. On the other hand, for the merger of 
unequal-mass neutron stars, an asymmetric double core
structure is the outcome. 
%%%%%%%%%%%%
\item In the hypermassive neutron stars formed after the merger, 
both nonaxisymmetric and quasiradial oscillations are excited.
These oscillations induce gravitational radiation. 
%%%%%%%%
\item For the case of hypermassive neutron star formation, the 
characteristic frequency of gravitational waves associated
with nonaxisymmetric oscillations is $\sim 3f_{\rm QE}$, 
which is $\approx 2.2$ kHz assuming that $M_0 \approx 2.8M_{\odot}$.
This value is slightly higher than that found in the post-Newtonian
simulation \cite{FR2}. This is likely due to the fact that
the formed hypermassive neutron star is more compact 
in our simulation in which general relativistic effects
are fully taken into account. 
%%%%%%%
\item The frequency of the peak in the gravitational wave spectrum
associated with the nonaxisymmetric oscillation is higher
for the mergers of the smaller rest-mass ratio
with a given total rest-mass. This reflects the fact that the
formed hypermassive neutron star is more compact for mergers of the 
smaller rest-mass ratios. 
%%%%%%%%%%
\item The amplitude of quasiradial oscillations for hypermassive 
neutron stars is larger for the merger of equal-mass
neutron stars. This is reflected in the 
amplitude of gravitational waves for $R_{20}$
as well as the magnitude of the peak 
at $\sim 2f_{\rm QE}$ of $\hat R_{22}$.
%%%%%%%
\item The characteristic frequency of 
gravitational waves associated with a quasiradial oscillation
is $\sim f_{\rm QE}$. 
The oscillation does not damp quickly. Thus if the cycle of
gravitational waves could be accumulated using a theoretical
template, the effective amplitude may be as large as that of the 
dominant quadrupole component. 
\end{itemize}

The simulations were performed using a new implementation. As a result,
the accuracy of the numerical results is significantly improved.
In particular, we emphasize that the gravitational radiation reaction is
taken into account with a good accuracy in the new implementation. 
We now consider that fundamental parts of the numerical implementation
such as those for Einstein's evolution equation,
general relativistic hydrodynamic equations, 
gauge conditions, and apparent horizon finder are established well 
for simulating spacetimes of no black hole and for 
early growth of formed black holes. 

However, there are still technical issues to be solved.
The following is a list of them: 

\begin{itemize}
%%%%%%%%
\item The black hole forming region does not 
have good resolution in our current computation. 
Consequently, computation crashed soon after formation of
the apparent horizon. 
Obviously, it is necessary to improve the grid resolution 
around the black hole forming region for longer time simulations. 
Since we have to prepare a large computational domain with $L$ 
which is at least half of the wavelength of gravitational waves, using 
restricted computational speed and memory, 
it is desirable to develop numerical techniques such as 
the mesh refinement techniques \cite{AMR} to overcome this problem.
%%%%%%%%
\item Gravitational waveforms are incompletely computed in the 
case of black hole formation, since the computations crash 
soon after the formation of the black holes. 
A straightforward approach to compute such gravitational waves is 
to develop a black hole excision technique \cite{U} 
by which we might be able to continue the simulation for a long time 
duration even after formation of the black holes. 
An alternative approach is to extract gravitational waves 
from a restricted spacetime data set using the so-called
Lazarus technique \cite{BCL}. 
Developing either of two technologies is an issue for the future. 
%%%%%%%%
\item Up until this time, we have performed simulations using 
$\Gamma$-law equations of state and neglecting micro physical effects.
To produce more physical and realistic outputs by 
numerical simulation, it is necessary to take into account 
sophisticated microphysics as done in Newtonian simulations \cite{RJ}. 
%%%%%%%%%%%%%
\item It is desirable to improve the implementation for providing the 
initial conditions. In simulations performed to this time, 
we have used quasiequilibrium states of a conformally flat 
three-metric as the initial conditions for simplicity. 
The conformal flatness approximation becomes a source of 
a certain systematic error when attempting to obtain 
realistic quasiequilibrium states, since the nonconformal part of the 
three-metric is in general nonzero \cite{SU01}. 
As a result, this approximation introduces a systematic error on 
the initial conditions and subsequent merger simulation. 
Since the magnitude of the ignored terms in the conformal flatness 
approximation seems to be small, it is unlikely that this effect 
significantly changes the results obtained in this paper. 
However, this conclusion is not entirely certain. 
To rule out the possibility, it is necessary to perform simulations using 
quasiequilibrium states of generic three geometries as initial conditions.
A few formulations in which the conformal flatness is not assumed
have been proposed recently \cite{SU03a}. 
\end{itemize}

\acknowledgments

We thank S. Mineshige for comments and Arun Thampan for reading the 
manuscript. 
Numerical computations were performed on the FACOM VPP5000 machines 
at the data processing center of NAOJ. 
This work was in part supported by Monbukagakusho 
Grant Nos. 14047207, 15037204, and 15740142.
KT is supported by a JSPS Grant (No. 14-06898).

\newpage
%%%%%%%%%%%%%%%%%%%%%%%%%

\begin{table}[t]
\begin{center}
\begin{tabular}{|c|c|c|c|c|c|c|c|c|c|c|c|c|} \hline
\hspace{-2mm} Model \hspace{-2mm} 
&\hspace{-3mm} $\comp$ \hspace{-3mm} & 
$\rho_{\rm max} $  & $Q_M$ &  $M_*$  & 
\hspace{-3mm} $M_0$ \hspace{-2mm} &  $q_0$  & 
\hspace{-3mm} $\hat P_{\rm t=0}$ \hspace{-2mm} &
\hspace{-3mm} $C_0$  \hspace{-2mm} &  
\hspace{-3mm}$Q_{*}$ \hspace{-2mm} &
\hspace{-3mm}$L/\lambda_0$ \hspace{-2mm} &
\hspace{-3mm} Product \hspace{-2mm} &
\hspace{-2mm} $M_{\rm disk}/M_*$ \hspace{-2mm}\\ \hline\hline
M1414 &0.14 vs 0.14 & 0.118, 0.118 & 1.00
& 0.292 & 0.269 & 0.951 & 193 & 0.102  & 1.62 & 0.510 & NS
& \\ \hline
M1315 &0.13 vs 0.15 & 0.104, 0.134 & 0.901
& 0.292 & 0.269 & 0.961 & 206 & 0.0976 & 1.62 & 0.479 & NS
&\\ \hline
M1616 &0.16 vs 0.16 & 0.151, 0.151 & 1.00
& 0.320 & 0.292 & 0.914 & 158 & 0.116  & 1.78 & 0.533 & BH
& $< 0.2\%$ \\ \hline 
M1517 &0.15 vs 0.17 & 0.133, 0.171 & 0.925
& 0.319 & 0.291 & 0.923 & 169 & 0.111  & 1.77 & 0.507 & BH
& $\alt 2$\% \\ \hline
M1418 &0.14 vs 0.18 & 0.118, 0.195 & 0.855
& 0.317 & 0.290 & 0.933 & 182 & 0.106  & 1.76 & 0.467 & BH 
& $\alt 4$\% \\ \hline
M159183 &0.159 vs 0.183 & 0.149, 0.203 & 0.925
& 0.332 & 0.301 & 0.908 & 156 & 0.118  & 1.84 & 0.498 & BH
& $< 1\%$ \\ \hline
\end{tabular}
\caption{
A list of several quantities for 
quasiequilibria of irrotational binary neutron stars with $n=1$. 
The compactness of each star in isolation $(M/R)_{\infty}$, 
the maximum density for each star, 
the baryon rest-mass ratio $Q_M \equiv M_{*2}/M_{*1}$, 
the total baryon rest-mass,
the ADM mass at $t=0$ ($M_{0}$), 
$q_0=J_0/M_{0}^2$, $\hat P_{t=0}\equiv P_{t=0}/M_{0}$, 
the orbital compactness [$C_0\equiv (M_{0}\Omega)^{2/3}$], 
the ratio of the total baryon rest-mass to 
the maximum allowed mass for a spherical star 
($Q_{*}\equiv M_*/M_{*~\rm max}^{\rm sph}$),
gravitational wavelength in units of $L$ in the maximum
grid number, and the products we found when we stopped simulations,
In the last column, 
the estimated ratio of the disk rest-mass located for $r > 3M_0$
at the termination of the simulation to the total rest-mass 
for the black hole formation case is listed. All quantities are
normalized by $\kappa$ appropriately to be dimensionless: 
The mass, the radius, and the density can be rescaled
to desirable values by appropriately choosing $\kappa$. 
Here, $M_{*~\rm max}^{\rm sph}$ denotes the maximum allowed mass 
of a spherical star ($M_{*~\rm max}^{\rm sph} \approx 0.180$ 
at $\rho_{\rm max} \approx 0.32$ for $n=1$ and $\kappa=1$). 
BH and NS denote ``black hole'' and ``neutron star''. 
}
\end{center}
\end{table}

%%%%%%%%%%%%%%%%%%%%%%%%%

\begin{table}[t]
\begin{center}
\begin{tabular}{|c|c|c|c|c|} \hline
Model & $\Delta / M_0$ & Grid size & $L/\lambda_0$  & $L/M_0$ 
\\ \hline\hline
M1616   & 0.134 & (633,633,317) & 0.533 & 42.2 \\ \hline
M1616-2 & 0.134 & (505,505,252) & 0.425 & 33.7 \\ \hline
M1616-3 & 0.134 & (313,313,157) & 0.263 & 20.8 \\ \hline
M1616-4 & 0.111 & (377,377,189) & 0.263 & 20.8 \\ \hline
M1616-5 & 0.169 & (249,249,125) & 0.263 & 20.8 \\ \hline
M1414   & 0.156 & (633,633,317) & 0.510 & 49.3 \\ \hline
M1414-2 & 0.156 & (313,313,157) & 0.252 & 24.3 \\ \hline
\end{tabular}
\caption{Computational setting for test simulations. 
}
\end{center}
\end{table}


\begin{thebibliography}{99}
\bibitem{foot00}
$M$ and $J$ computed by Eqs. (\ref{eqm00}) and (\ref{eqj00})
in the finite computational domain 
do not have any strict physical meaning such as ADM mass and
angular momentum. This implies that Eqs. 
(\ref{eqm01}) and (\ref{eqj01}) do not have to hold strictly. 
However, if the outer boundaries are located in the local 
wave zone, we could use the linear perturbational treatment for
the spacetime metric. In the linear level, conservation relations
(\ref{eqm01}) and (\ref{eqj01}) seem to hold and, thus,
in the present computation, they should hold approximately.
\end{thebibliography}
\end{document}